\documentclass[11pt]{article}
\usepackage[top=1.1in, left=1in, right=1in, bottom=1.1in]{geometry}
\usepackage{amsmath,amssymb,amsthm,amsfonts,verbatim}
\usepackage{enumerate}
\usepackage{microtype}
\usepackage{url}
\usepackage{algpseudocode}
\usepackage{tikz}
\usetikzlibrary{arrows}
\usepackage{amsmath,amssymb}
\usepackage[justification=centering]{caption}
\usepackage[justification=centering]{subcaption}
\usepackage[english]{babel}
\usepackage{array}
\usepackage{psfrag}
\usepackage{graphicx}
\usepackage{color}
\usepackage{setspace}
\usepackage{hyperref}
\usepackage{algorithm2e}
\usepackage[utf8]{inputenc}
\usepackage{cleveref}
\usepackage{mathrsfs} 
\usepackage{authblk}
\usepackage{threeparttable}
\usepackage[titletoc,title]{appendix}
\usepackage{xcolor}
\usepackage{xfrac}
\usepackage{caption}
\usepackage{bm}

\usepackage{algorithm2e}
\RestyleAlgo{ruled} 
\SetKw{KwInitialization}{Initialization}
\SetKw{KwTheMainBody}{Main Body}

\newcommand{\eref}[1]{equation (\ref{#1})} 
\newcommand{\fref}[1]{Fig. \ref{#1}} 
\newcommand{\xb}{\bm{x}} 
\newcommand{\p}{p(\xb,t)} 
\newcommand{\ccite}[1]{\cite{#1}} 
\newcommand{\etal}{ {\it et al.} } 
\newcommand{\R}{\mathbb{R}}  
\newcommand{\s}{\mathbb{S}}  

\begin{document}


\title{Most probable escape paths in periodically driven nonlinear oscillators
} 
\author[1]{Lautaro Cilenti\thanks{lcilenti@umd.edu}}
\author[2]{Maria K. Cameron\thanks{mariakc@umd.edu}}
\author[1]{Balakumar Balachandran\thanks{balab@umd.edu}}
\affil[1]{\small{Department of Mechanical Engineering, University of Maryland, College Park, MD 20742, USA}}
\affil[2]{\small{Department of Mathematics, University of Maryland, College Park, MD 20742, USA}}                               

\maketitle 

\begin{abstract}
The dynamics of mechanical systems such as turbomachinery with multiple blades are often modeled by arrays of periodically driven coupled nonlinear oscillators. It is known that such systems may have multiple stable vibrational modes, and transitions between them  may occur under the influence of random factors. A methodology for finding most probable escape paths and estimating the transition rates in the small noise limit is developed and applied to a collection of arrays of coupled monostable oscillators with cubic nonlinearity, small damping, and harmonic external forcing. The methodology is built upon the action plot method (Beri et al. 2005) and relies on the large deviation theory, optimal control theory, and  the Floquet theory. The action plot method is promoted to non-autonomous high-dimensional systems, and a method for solving the arising optimization problem with discontinuous objective function restricted to a certain manifold is proposed. The most probable escape paths between stable vibrational modes in arrays of up to five oscillators and the corresponding quasipotential barriers are computed and visualized. The dependence of the quasipotential barrier on the parameters of the system is discussed.
\end{abstract}

\begin{quotation}

Response transitions from one dynamic state to another can occur due to random perturbations in a variety of systems, including mechanical and structural systems. Some examples are sensor arrays, energy harvesters, and rotating machinery.  The aim of the present work is to elucidate these transitions, in particular, when the random perturbations are weak.  To that end, a methodology based on Action Plot Method from Large Deviation Theory and Optimal Control Theory is developed and illustrated for a range of periodically forced systems.  

\end{quotation}

\section{Introduction} \label{IntroductionSection}

Systems consisting of multiple beam-like and plate-like structures including turbomachinery, micro-electromechanical systems (MEMS), and vibration energy harvesters (VEH) can exhibit stable periodic mechanical oscillations that can be unfavorable to structural integrity and/or performance of these systems. In this work, the authors study the effect of random pertubations (noise) in physics-based models of these systems, as noise can induce transitions into and out of undesired stable modes.  Many of these systems can be modeled as coupled Duffing oscillator  \ccite{duffing_erzwungene_1918} arrays with external periodic forcing, $\bm{f}(t)$, and a weak white noise, $\sqrt{\epsilon}\bm{\eta(t)}$, accounting for random influences, as shown below:
\begin{equation}
    \ddot{\bm{x}} +\delta_c \dot{\bm{x}} + \alpha \bm{x} + \beta \bm{x}^3 + \nu \bm{D}_N\bm{x} = \bm{f}(t) + \sqrt{\epsilon}\bm{\eta(t)},\quad \bm{x}\in\mathbb{R}^N.
    \label{eq:mainSDE}
\end{equation}
Here, $N$ is the number of oscillators, 
$\bm{x}$ is the vector of beam tip displacements from their respective equilibrium positions, $\delta_c$ is a damping  related coefficient, $\alpha$ and $\beta$ are scalar constants that render each oscillator monostable, $\bm{x}^3$ denotes the elementwise cube of the vector, $\nu$ is the inter-oscillator coupling coefficient, and $\bm{D}_N$ is the coupling matrix that takes the following form:
\begin{align} 
\bm{D}_1& = 0,\quad \bm{D}_2 = \left[\begin{array}{rr}1&-1\\-1&1\end{array}\right],\label{eq:coupling_matrix}\\
\bm{D}_N &= \left[\begin{array}{rrrr}2&-1&&-1\\-1&2&-1&\\&\ddots&\ddots&\ddots\\-1&&-1&2\end{array}\right],\quad N>2.\notag
\end{align}

\begin{figure}[htbp]
\begin{center}
\centerline{\includegraphics[width=0.5\textwidth]{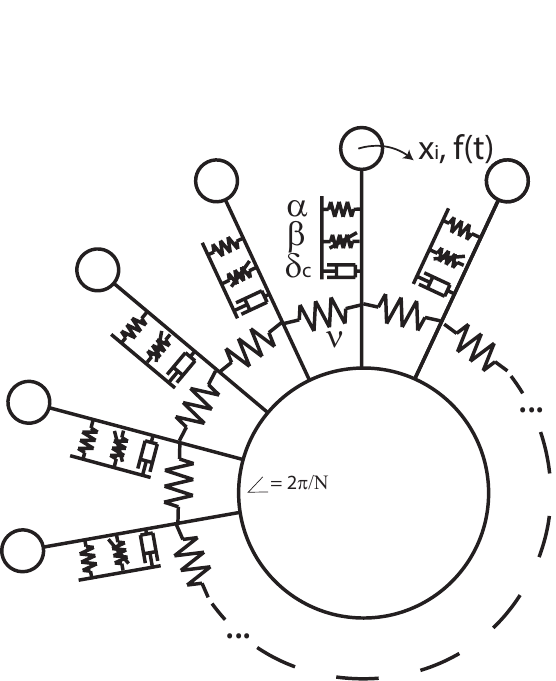}}
\caption{Schematic of  linearly coupled nonlinear spring-- mass-damper systems arranged in a cyclical array.}
\label{Drawing_DuffingArray}
\end{center}
\end{figure}

 It is noted that if the number of oscillators is greater than two, they are coupled in a circular array configuration, as depicted in \fref{Drawing_DuffingArray}. 

Setting the parameter $\epsilon$ in the noise term to zero, the authors obtain the corresponding deterministic ordinary differential equations (ODE) system:
\begin{equation}
    \ddot{\bm{x}} + \delta_c \dot{\bm{x}} + \alpha \bm{x} + \beta \bm{x}^3 + \nu \bm{D}_N\bm{x} = \bm{f}(t),\quad \bm{x}\in\mathbb{R}^N.
    \label{eq:mainODE}
\end{equation}

\begin{figure}[htbp]
\begin{center}
\centerline{\includegraphics[width=0.5\textwidth]{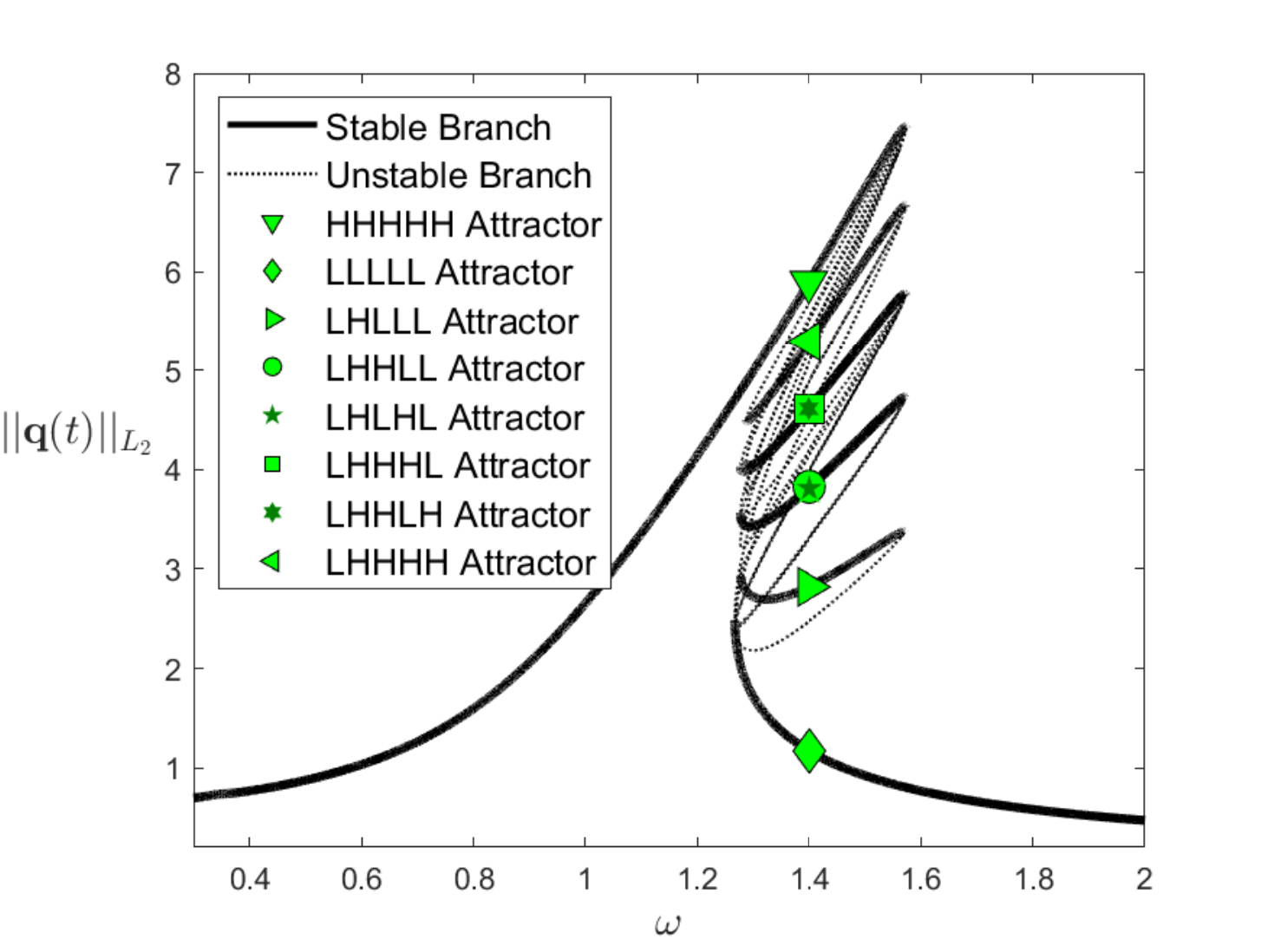}}
\caption{Frequency response of a circular array of $N=5$ coupled forced Duffing oscillators as a function of excitation frequency, $\omega$. At the frequency $\omega = 1.4$, unique attractors of the array are highlighted.  This figure was generated by using the continuation package AUTO2007 \ccite{doedel_auto-07p_2007}. The parameters in the ODE system \eqref{eq:mainODE} were set to $\alpha = 1$, $\beta = 0.3$, $\delta = 0.1$, $\nu = 0.01$,  and the periodic excitation was of the form $f(t) = 0.4 \cos(\omega t)$. Solid (dotted) lines in the figure correspond to stable (unstable) periodic solutions of the system \eqref{eq:mainODE}.   Here, it is of interest to examine how random perturbations influence the transition from one stable solution to another.
}
\label{FiveDuffing_FrequencyResponse}
\end{center}
\end{figure}

It is known that the ODE system \eqref{eq:mainODE} may admit multiple attractors \ccite{papangelo_multistability_2019,grolet_free_2012,dick_intrinsic_2008,ikeda_intrinsic_2013,balachandran_response_2015}; that is, stable periodic solutions that attract other solutions from appropriate open subsets of the initial condition space $(\bm{x}(0),\dot{\bm{x}}(0))$. In \fref{FiveDuffing_FrequencyResponse}, the authors show the response of the system \eqref{eq:mainODE} to the forcing $\bm{f}$ with frequency $\omega$ for a circular array of $N=5$ coupled Duffing oscillators. Each periodic solution $\bm{q}(t)\equiv (\bm{x}(t),\dot{\bm{x}}(t))$ to \eqref{eq:mainODE} is represented by a $L_2$-like norm 
\begin{equation}
    ||\bm{q}(t)||_{L_2} = \sqrt{\int_0^1 \sum_i^{2N} \left(q_i\left(s T \right)\right)^2 ds},
    \label{L2NormDefinition}
\end{equation}
where $T=\sfrac{2\pi}{\omega}$ is the period.

Each attractor is encoded by a sequence of symbols $X_1X_2X_3X_4X_5$, $X_j \in\{\rm{L},\rm{H}\}$, $j = 1,\ldots,5$, where L and H stand, respectively, for ``low amplitude" and ``high amplitude". For example, LHLLL refers to the stable periodic solution where oscillator 2 has high amplitude while the rest of the oscillators have low amplitudes.  From \fref{FiveDuffing_FrequencyResponse}, it is discernible that if the frequency $\omega$ of the external forcing is far from the primary resonance, there exists at least one stable periodic solution.  Furthermore, at a certain range of $\omega_1<\omega<\omega_2$, there are a total of at least 32 periodic attractors corresponding to which each oscillator is in either a low or high amplitude state.  This region with multiple stable solutions is referred to as the hysteresis region of the frequency response. In this region, the state (stable response) of the system is sensitive to initial (prior) conditions. Note that some solution branches in \fref{FiveDuffing_FrequencyResponse} coincide due to circular symmetry or closeness of their representations by $||\bm{q}(t)||_{L_2}$. The phenomenon wherein a smaller subset of the system (one or more oscillators but not all) consists of high amplitude responses is called \emph{localization}. In this work, the authors refer to  \emph{response localization} exclusively as the stable response mode in which only \emph{one} oscillator oscillates with a high amplitude while all of other oscillators oscillate with low amplitudes. 

Response localizations have been demonstrated in experiments of macro-cantilever beam arrays, wherein the beams represent the blades or components of turbomachinery and VEHs \ccite{emad_experimental_2000,kimura_coupled_2009,kimura_experimental_2012,perkins_effects_2016,niedergesas_experimental_2021}. The localized high amplitude oscillations can, on one hand, be destructive in turbomachinery \ccite{papangelo_multistability_2019, kenyon_forced_2002, srinivasan_flutter_1997, castanier_modeling_2006, sever_experimental_2004} and, on the other hand, be beneficial for energy harvesting systems \ccite{gardonio_vibration_2014,lefeuvre_comparison_2006}.

It is remarked that localization in the system governed by \eqref{eq:mainODE} is a nonlinear phenomenon: it would not occur if $\beta = 0$. However, localization may occur also in linear systems due to inhomogeneity \ccite{sever_experimental_2004,ewins_effects_1969,kenyon_forced_2002,yan_vibration_2008,gardonio_vibration_2014}; that is, if $\beta = 0$ and $\alpha$ is replaced with the vector $(\alpha_1,\ldots,\alpha_N)$. 
 In the systems with $N =3$ and $N =5$, here, all stable response localizations have symmetric, anti-phase, neighboring oscillators. 

Adding a stochastic perturbation (noise) to an array of coupled Duffing oscillators; that is,  replacing ODE \eqref{eq:mainODE} with \eqref{eq:mainSDE} can facilitate transitions between various attractors (modes) of these systems \ccite{perkins_effects_2016}.
Therefore, in an application, noise may be used to transition a system out of an undesirable localized mode in a bladed rotor or may help a VEH to get into the desirable high amplitude regime.  


\section{Objectives and Structure}
\begin{figure}[htbp]
\begin{center}
\centerline{\includegraphics[width=0.5\textwidth]{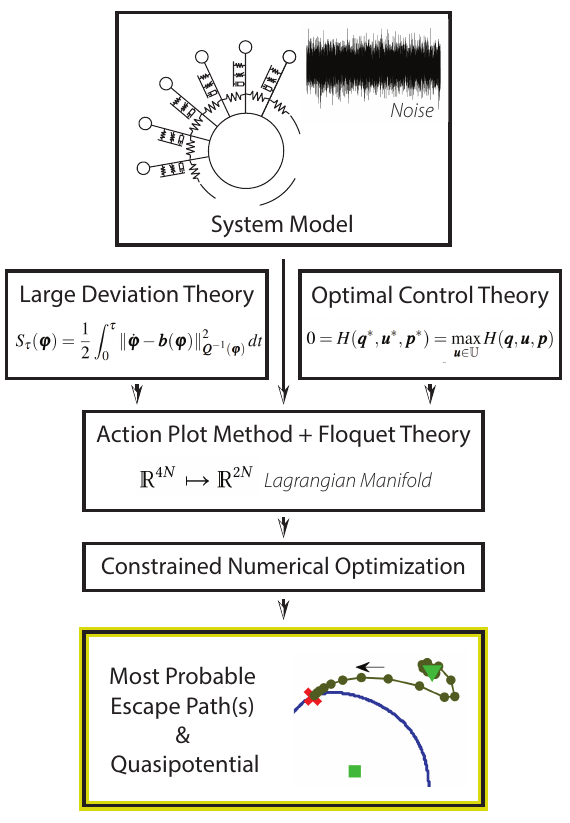}}
    \caption{ A flowchart overview of the main concepts in this work. The authors begin by presenting background on a noise influenced system model, large deviation theory, and optimal control theory. A methodology built on the action plot method combined with the Floquet theory and numerical optimization is used to compute most probable escape paths and the quasipotential barriers of escapes from attractors of the system model.}
\label{SummaryOfEffort}
\end{center}
\end{figure}

The authors achieve the following two objectives here.
\begin{enumerate}
    \item  Conduct a case study on the noise-induced most probable escape paths out of undesired oscillations modes such as response localizations that exist only under the influence of deterministic periodic excitations in circular arrays of nonlinear oscillators.
    \item  Develop a computational tool for accomplishing Objective 1; that is, for finding the quasipotential barriers and most probable escape paths of nonlinear systems with multiple attractors subjected to both deterministic periodic excitations and weak random perturbations.
\end{enumerate} 
By achieving Objective 1, the authors contribute to a highly active area of research and motivate Objective 2, since there is no alternative methodology to compute these escape paths and quasipotential barriers for these high dimensional systems with low damping and deterministic time-dependent external excitations to the best of the authors' knowledge.

The focus of this work is on the weak white noise limit; that is the limit $\epsilon \to 0$.   In a nonlinear system with multiple attractors,  the probability of a transition between its attractors is nonzero no matter how small the white noise intensity is \ccite{freidlin_random_1998}. 
The possibility of these rare events is interesting when the timescale of deterministic dynamics is different from another timescale of interest.  
With regard to turbomachinery, one can imagine blades that  may vibrate with sub-second periods, but the timescale of continuous operation of the system can range from hours to years. Given this duration of operation, rare events become relevant. 
Furthermore, the asymptotic estimates for the weak noise limit often remain relevant for a broad range of noise intensities.

The current work relies on four building blocks:
\begin{enumerate}
    \itemsep0em 
    \item The authors use Freidlin's and Wentzell's large deviations theory (LDT) \cite{freidlin_random_1998} to define the action functional (FW action) whose minimizers are the most probable escape paths from basins of  attractors  of \eqref{eq:mainODE}.  The corresponding minimal values of the action, which are the values of a certain function called the \emph{quasipotential}, serve as a quantitative measure for the difficulty of escape from an attractor and the consequent transition. 
    \item The optimal control theory and Pontryagin's maximum principle \cite{bardi_optimal_2008} are used to identify the set of paths called \emph{Hamiltonian paths}, which are the candidates for being the minimizers of the FW action. As a result, the problem of minimization of the FW action reduces to an optimization problem on the set of initial conditions for the Hamiltonian paths.
    \item The action plot method \cite{beri_solution_2005} provides a nice recipe for cutting the dimensionality of this optimization problem by a factor of two through linearization of the system and approximation of  the \emph{Lagrangian manifold} containing the desired minimizing initial conditions. 
    \item Following Ref. \cite{lin_quasi-potential_2019}, the authors use the Floquet theory for linear ODEs with periodic coefficients \cite{chicone_1999} to adapt this recipe for periodic attractors of non-autonomous stochastic differential equations (SDEs) \eqref{eq:mainSDE}. 
\end{enumerate}
The use of these building blocks is illustrated via the flow chart in \fref{SummaryOfEffort}.


The resulting optimization problem for the optimal set of initial conditions for the Hamiltonian paths
is still very challenging because the objective function is discontinuous and this function's minima lie next to its ``cliffs". In the case of a $2$-dimensional system, i.e., only one oscillator, this problem can be solved graphically as it has been done in references \ccite{beri_solution_2005,chen_noise_2016}.  For the higher dimensional problems considered in this work, the authors propose an optimization algorithm  \emph{stochastic unit vector (SUV)} for finding the minimizers of the FW action.

First, the developed methodology is applied to a single Duffing oscillator to quantify transitions between this system's low amplitude and high amplitude attractors. This example is highly instructive as it is visual and illuminates the technical difficulties for $N>1$. Next, the authors compute the most probable escape paths between attractors of arrays consisting of two, three, and five coupled Duffing oscillators. Finally, the dependencies of the quasipotential values characterizing the difficulty of escapes from the basins of attractors on the external forcing frequency $\omega$ and inter-oscillator coupling coefficient $\nu$ are visualized. These plots are helpful in validating the obtained results.

 The rest of the paper is organized as follows.  In Section \ref {sec:related_work}, the authors discuss related work in the cases of relatively large noise intensity and the weak noise limit.  In Section \ref{Background}, they lay out fundamental background from large deviations theory and optimal control theory. The proposed methodology is detailed in \ref{Methodology}.
   Results obtained for a single forced Duffing oscillator, two coupled  forced Duffing oscillators, and two circular  arrays of $N=3$ and  $N=5$ Duffing oscillators are described in Sections \ref{OneDuffingResults}, \ref{TwoDuffingResults}, and \ref{NDuffingResults}, respectively. 
The paper is concluded with a discussion in Section \ref{Discussion}.  
 Details and parameter values for the computations conducted in this work are found in Appendix A.  Table \ref{tab:NotationTable} in Appendix A is provided as a nomenclature reference for a reader.

%

\section{Related work}
\label{sec:related_work}
Nonlinear oscillators with external periodic forcing and added noise have been a subject of active research for at least the last two decades. 

\subsection{Relatively large noise}
Experimental and numerical studies conducted by using methods limited to \emph{sufficiently large noise intensity} have yielded a qualitative understanding of the dynamics of these systems. Approximations to the stationary probability density response of mono and coupled second-order nonlinear systems were constructed in references  \ccite{von_wagner_calculation_2000,von_wagner_double_2002,martens_calculation_2012,forster_approximate_2018}.
The method based on the construction of the \emph{prehistory probability distributions} \ccite{dykman_optimal_1992} was applied to a single chaotic oscillator to find the most probable escape paths from basins of its attractors \ccite{luchinsky_optimal_2002}.
The average dynamics of the nonlinear systems was approximated by deriving a series of nonlinear moment equations from the Fokker-Planck equation and integrating them \ccite{ramakrishnan_energy_2010,perkins_noise-enhanced_2012,perkins_noise-influenced_2013}, and by means of the cell mapping method  \ccite{xu_global_2003,xu_stochastic_2004}. Transient probability density functions for forced mono and coupled Duffing oscillators were generated by utilizing new developments in the path integral method \ccite{yu_numerical_2004, kumar_modified_2010,narayanan_numerical_2012,cilenti_transient_2021}. A comparison between simulations of circular arrays of six and more oscillators revealed no fundamental differences with regards to the transitions from a response localization to another stable mode \ccite{balachandran_dynamics_2022}. 

Experimental studies of the effect of noise on response localizations \ccite{perkins_effects_2016,agarwal_influence_2018,alofi_coupled_2021} have demonstrated that likelihood of transitions from one stable mode to another depends on the deterministic excitation frequency of the forced system.  Although all of these studies  provide relevant insights into the probabilistic behavior of the considered  stochastic nonlinear mechanical systems, the methodologies used in the earlier studies are not amenable for studying the dynamics in the weak noise limit.

\subsection{Weak noise limit}
The present work is devoted to the study of quasipotential barriers and the most probable escape paths (MPEPs) from basins of various attractors of arrays of periodically forced Duffing oscillators in the limit of the white noise intensity tending to zero. Central to the developed methodology is the \emph{action plot method} introduced in the work of Beri \etal \ccite{beri_solution_2005} for finding MPEPs in two-dimensional nongradient SDEs. An important advantage of the action plot method is that it is suitable for systems with inertia and external forcing. For example, Chen \etal \ccite{chen_noise_2016} used this method to find a MPEP for a chaotic oscillator \ccite{luchinsky_optimal_2002} and validated it by matching it with the computed prehistory probability distribution. The action plot method implies a graphical solution to the optimization problem that arises -- this is the reason for its name -- and hence this approach in its original form \ccite{beri_solution_2005} is limited to low dimensional systems (single oscillator).  The main methodological contribution of the present work is the extension of the action plot method to high dimensional systems with periodic forcing and the development of a minimization technique for solving the optimization problem for initial conditions for the MPEP. 

In addition, the authors would like to highlight a recent work \ccite{chao_tao_2021} that offers an asymptotic technique for computing the quasipotential barriers in a bistable multidimensional oscillator with small noise and small periodic forcing based on an asymptotic expansion of the FW action in powers of the scaling parameter used for the forcing. While this approach is feasible for high dimensions, it is not suitable for the oscillator arrays studied in the present work as they are monostable in the absence of the periodic forcing. Moreover, the periodic forcing is not assumed to be small in this work.

%


\section{Background} 
\label{Background}

\subsection{Large Deviations Theory}

 In this section, the authors provide a brief background from large deviations theory, which is used in this work to define the quasipotential.

The Freidlin-Wentzell large deviations theory \cite{freidlin_random_1998} primarily deals with \emph{autonomous} SDEs with \emph{non-degenerate} noise
\begin{equation}
    d\bm{q} = \bm{b}(\bm{q}) dt + \sqrt{\varepsilon} \, \bm{G}(\bm{q}) \, d\bm{W}(t),
    \label{B_SDE_GEN}
\end{equation}
where $d{\bf W}(t)$ is the standard Brownian motion.  The drift field  $\bm{b}$ and the diffusion matrix $\bm{Q}(\bm{q}) = \bm{G}\bm{G}^\top$ are assumed to be continuously differentiable.

The work performed by random perturbations can be quantified with the action functional $S_\tau(\bm{\varphi})$. When $\bm{Q}$ is invertible; that is, the noise in \eref{B_SDE_GEN} is non-degenerate, the Freidlin-Wentzell action is defined on the set of absolutely continuous paths for $\bm{\varphi}$ by \ccite{freidlin_random_1998}
\begin{equation}
  S_\tau(\bm{\varphi}) = \frac{1}{2} \int_0^\tau \left\|\dot{\bm{\varphi}}- \bm{b}(\bm{\varphi})\right\|^2_{\bm{Q}^{-1}(\bm{\varphi})}  dt,
  \label{B_ActionFunctional}
\end{equation}

where  $\|\cdot\|^2_{\bm{Q}^{-1}}$ denotes the inner product $\langle \cdot ,\bm{Q}^{-1} \cdot \rangle$. One takeaway from \eqref{B_ActionFunctional} is that the action is zero if the path $\bm{\varphi}$ follows a deterministic trajectory of the corresponding ODE system
\begin{equation}
    \label{eq:B_ODE}
    \dot{\bm{q}} = \bm{b}(\bm{q}). 
\end{equation} 
The probability for a stochastic trajectory of SDEs \eqref{B_SDE_GEN} to follow a small $\delta$-tube surrounding a given absolutely continuous path $\bm{\varphi}$ scales with $\epsilon$ as $\exp\{-S_\tau(\bm{\varphi})/\epsilon\}$ (see Refs. \ccite{freidlin_random_1998,dembo_1998}).

The \emph{quasipotential  $ U_{\mathcal{A}}(\bm{q})$ at a point $\bm{q}$ with respect to an attractor} $\mathcal{A}$ of \eqref{eq:B_ODE} is defined as the infimum of the action taken over all possible absolutely continuous paths starting at $\mathcal{A}$ and ending at $\bm{q}$ and all travel times $\tau$:
\begin{equation}
  U_{\mathcal{A}}(\bm{q}) =  \inf_{\bm{\varphi},\tau} \{  S_\tau(\bm{\varphi}) \, |\, \bm{\varphi}(0) \in \mathcal{A},~ \bm{\varphi}(\tau) = \bm{q}\}.
  \label{B_quasipotential}
\end{equation}
Note that this infimum is achieved at $\tau = \infty$ as it takes infinitely long time to escape from the attractor in the weak noise limit. Furthermore, if the attractor is not a point attractor but a limit cycle or a more complex object, the infimum is achieved via an escape path of infinite length, as it makes infinitely many revolutions in neighborhood of the attractor while escaping from it. This minimizing path is called the \emph{minimum action path} or \emph{MAP}. 

Let $D$ be a domain with a smooth boundary $\partial D$ lying in the basin of $\mathcal{A}$.
The quasipotential allows us to estimate the expected exit time $\tau_{exit}$ from $D$ up to the exponential order \ccite{freidlin_random_1998,dembo_1998}:
\begin{equation}
    \lim_{\varepsilon \to 0} \varepsilon \textrm{ log } E\left[\tau_{exit} \right] = \inf_{\bm{q} \in \partial D} U_{\mathcal{A}}(\bm{q})
    \label{B_ExpectedWaitTime}
\end{equation}
Therefore, in the limit $\epsilon\rightarrow 0$, the system escapes the domain $D$
after a waiting time logarithmically equivalent to $\exp\left\{- \inf_{\bm{q} \in \partial D} U_{\mathcal{A}}(\bm{q})/\epsilon\right\}$. Moreover, the escape path lies in a small tube surrounding the minimizer of the FW action \eqref{B_ActionFunctional}, i.e. the MAP, with probability tending to one as $\epsilon\rightarrow 0$. Consequently, this MAP is often referred to as the \emph{most probable escape path (MPEP)}. 
 In the present work, the authors study the escape from the basin of attraction of $ \mathcal{A}$. Therefore $D$ is chosen to be the basin of attraction and $\partial D$ is the basin boundary.

The generalization of these results of  LDT for systems with inertia has been addressed only partially in the mathematical literature, to the best of the authors' knowledge. This has been done for the dynamics governed by $\ddot{\bm{x}} + \delta_c \dot{\bm{x}} + \bm{b} (\bm{x}) = \sqrt{\varepsilon} \bm{\dot{W}}(t)$ for the case where $\bm{b}$ is either uniformly Lipschitz \ccite{chen_smoluchowskikramers_2005} or a gradient of a smooth potential function \ccite{freidlin_2004} $\bm{b} = -\nabla \bm{V}$. The conjecture that similar results for the MPEPs and the expected exit times hold for oscillators of the form \eqref{eq:mainSDE} has been in physical literature since at least 1979 -- see reference \cite{dykman_1979}. This conjecture is rooted in Feynman's path integral theory \cite{feynman_1965}. A rigorous proof of an estimate for the exit time similar to \eqref{B_ExpectedWaitTime} for the dynamics governed by SDEs \eqref{eq:mainSDE} is yet to be given to the best of the authors' knowledge. 

Finally, it worth mentioning that there is strong numerical evidence suggesting that the MPEP for a nonlinear oscillator with external periodic forcing is the MAP obtained by matching the MAP with the plot of the prehistory probability distribution \ccite{chen_noise_2016}. 

\subsection{Definitions for periodically driven nonlinear oscillators}
Now consider SDE \eqref{eq:mainSDE} with periodic forcing $\bm{f}(t)$ of period $T$  and the corresponding ODE \eqref{eq:mainODE}. There is an important difference between a periodic solution of an autonomous ODE and a periodic solution of the time-dependent or non-autonomous ODE \eqref{eq:mainODE}. Any time shift of a periodic solution of an autonomous ODE gives another periodic solution of the same ODE. In contrast, only time shifts of periodic solutions to \eqref{eq:mainODE} that are multiples of the period result in other periodic solutions to \eqref{eq:mainODE}. Therefore, every point $(\bm{x},\bm{v})$ of a periodic solution to \eqref{eq:mainODE} is associated with a particular phase $\theta:=t\thinspace{\rm mod}\thinspace T$. Hence, it is natural to make ODE \eqref{eq:mainODE} autonomous by introducing the phase variable 
$$
\theta:=t\thinspace{\rm mod}\thinspace T \in \mathbb{S}^1_T\equiv \mathbb{S}^1(T/{2\pi}),
$$
where $\mathbb{S}^1_T$ is a circle of radius ${T}/{2\pi}$. The resulting autonomous ODE on the manifold $\mathbb{R}^{2N}\times\mathbb{S}^1_T$ is:
\begin{align}
    \dot{\bm{x}} &= \bm{v},\notag\\
    \dot{\bm{v}} & = -\delta_c \bm{ v} -\bm{K}(\bm{x}) + \bm{f}(\theta),\label{AP_ODE1}\\
    \dot{\theta} & = 1,\notag\\
    \bm{x}&\in\mathbb{R}^N,\quad\bm{v}\in\mathbb{R}^N,\quad \theta\in\mathbb{S}^1_T.\notag
\end{align}
Here, the notation $\bm{K}(x): = \alpha \bm{x} + \beta \bm{x}^3 + \nu \bm{D}_N\bm{x}$ is introduced for brevity and convenience of generalization. The attracting periodic trajectories of the ODE \eqref{eq:mainODE} are the attractors of \eqref{AP_ODE1}. 

The corresponding SDE \eqref{eq:mainSDE} is made autonomous in the same way:
\begin{equation}
\begin{split}
        & d\xb   = \bm{v} \, dt \\ 
         & d\bm{v}   = 
    \left(- \delta_c \bm{v} - \bm{K} (\bm{x}) + \bm{f}(\theta)  \right)dt
    + \sqrt{\varepsilon} \,  d{\bm{W}}(t) \\ 
        & d\theta   = dt \\ 
        & \bm{x} \in\mathbb{R}^N,\quad\bm{v}\in\mathbb{R}^N,\quad \theta\in\mathbb{S}^1_T. \\ 
\end{split}
    \label{B_SDE_Autonomous}
\end{equation}

The action functional for SDE \eqref{B_SDE_Autonomous} on the set of absolutely continuous paths $\{[\bm{\varphi}(t),\theta(t)]~|~t_0\le t\le \tau\}\subset\mathbb{R}^N\times\mathbb{S}^1_T$ is defined as
\begin{equation}
    \begin{split}
  S_{[t_0,\tau]} ([\bm{\varphi},\theta])
  & = \frac{1}{2} \int_{t_0}^{\tau} ||\ddot{\bm{\varphi}} +  \delta_c \dot{\bm{\varphi}} + \bm{K} (\bm{\varphi}) - \bm{f}(\theta)||^2    dt \\
  \end{split}
  \label{B_ActionFunctionalDegenerate}
\end{equation}
 According to Dykman's and Krivoglaz's conjecture \cite{dykman_1979}, which is consistent with Feyman's path integral theory \cite{feynman_1965}, and supported by a strong numerical evidence \cite{chen_noise_2016},  the minimizers of \eqref{B_ActionFunctionalDegenerate} have the following significance. Let $[\bm{\varphi}^{\ast}(t),\theta^{\ast}(t)]$, $t_0\le t\le \tau$, be a minimizer of \eqref{B_ActionFunctionalDegenerate} amongst all paths satisfying $[\bm{\varphi}(t_0),\theta(t_0)] = [\bm{\varphi}^{\ast}(t_0),\theta^{\ast}(t_0)]$ and 
$[\bm{\varphi}(\tau),\theta(\tau)] = [\bm{\varphi}^{\ast}(\tau),\theta^{\ast}(\tau)]$.
Then the probability for a trajectory of \eqref{B_SDE_Autonomous}  starting at $[\bm{\varphi}^{\ast}(t_0),\theta^{\ast}(t_0)]$ and ending at $[\bm{\varphi}^{\ast}(\tau),\theta^{\ast}(\tau)]$ to follow a small tube around a path connecting these points is maximized if this path a minimizer of \eqref{B_ActionFunctionalDegenerate}. Moreover, this probability scales with the parameter $\epsilon$ in \eqref{B_SDE_Autonomous} as $\exp\{-S_{[t_0,\tau]} ([\bm{\phi}^{\ast},\theta^{\ast}])/\epsilon\}$. Therefore, in order to find the most probable escape path from a basin of attractor $\hat{\bm{A}}_0$ of \eqref{AP_ODE1}, one should minimize \eqref{B_ActionFunctionalDegenerate} with respect to all paths and all times $t_0\in[0,T)$ and $\tau\in[0,\infty]$ satisfying the boundary conditions $[\bm{\varphi}(t_0),\theta(t_0)] \in\hat{\bm{A}}_0$, $[\bm{\varphi}(\tau),\theta(\tau)]\in\partial B_0$ where $B_0$ denotes the basin of $\hat{\bm{A}}_0$.

\subsection{Pontryagin's Maximum Principle}

Once the action functional (the cost functional) is known, one can use optimal control theory \ccite{bardi_optimal_2008,kirk_optimal_2004} to simplify the problem of minimizing it. The relevant ODE with an $N-dimensional$ control function $\bm{u}(t)$ is given by
\begin{align}
    \dot{\bm{x}} &= \bm{v},\notag\\
    \dot{\bm{v}} & = -\delta_c \bm{ v} -\bm{K}(\bm{x}) + \bm{f}(\theta) + \bm{u}(t),\label{AP_control}\\
    \dot{\theta} & = 1,\notag\\
    \bm{x}&\in\mathbb{R}^N,\quad\bm{v}\in\mathbb{R}^N,\quad \theta\in\mathbb{S}^1_T.\notag
\end{align}
Expressing $\bm{u}$ from the equation for $\dot{\bm{v}}$ and comparing it to the action functional \eqref{B_ActionFunctionalDegenerate} we obtain the cost functional
\begin{equation}
  S(\bm{u}) =  \frac{1}{2}\int_{t_0}^{\tau}\|\bm{u}\|^2 \, dt.
  \label{B_PerformanceMeasure}
\end{equation}
The optimal control problem consists in finding a function $\bm{u}$ that minimizes the cost functional subject to some user-chosen constraints. 

Suppose that ODEs \eqref{AP_ODE1} admit $n_a>1$ attractors, 
$\hat{\bm{A}}_k$, $0\le k\le n_a-1$.  The authors denote the basin of $\hat{\bm{A}}_k$ by $B_k$ and the basin boundary by $\partial B_k$, $0\le k\le n_a-1$.  Index 0 is reserved for the initial attractor, from which the most probable escape path emerges. 
Thus, the constraints are set as
\begin{equation}
    \label{AP_BC1}
    (\bm{x}(t_0),\bm{v}(t_0),\theta(t_0))\in \hat{\bm{A}}_0,\quad
    (\bm{x}(\tau),\bm{v}(\tau),\theta(\tau))\in\partial B_0.
\end{equation}

The optimal control problem 
\begin{align}
    &S(\bm{u})=\frac{1}{2}\int_{t_0}^{\tau}\|\bm{u}\|^2 \, dt\rightarrow \min\label{eq:our_problem}\\
    &\text{subject to \eqref{AP_control} and \eqref{AP_BC1}}\notag
\end{align} 
will be solved with the aid of \emph{Pontryagin's Maximum Principle} \ccite{bardi_optimal_2008} whose basic form is the following. 
Suppose a controlled system is governed by a system of ODEs
\begin{equation}
    \label{genODE}
    \dot{\bm{q}} = \bm{b}(\bm{q},\bm{u}).
\end{equation}
Let the cost functional be 
\begin{equation}
    \label{gencost}
     S(\bm{u})=\int_{t_0}^{t_1}L(\bm{q},\bm{u})\,dt.
\end{equation}
Then, the \emph{control theory Hamiltonian} is defined by
\begin{equation}
    \label{genHam}
    H(\bm{q},\bm{u},\bm{p}): = \bm{p}^\top \bm{b}(\bm{q},\bm{u}) - L(\bm{q},\bm{u}),
\end{equation}
where $\bm{p}$ is a costate variable. \emph{Pontryagin's  maximum principle} states that
if $\bm{u}^{\ast}(t)$ is the minimizer of \eqref{gencost} 
and $\bm{q}^{\ast}(t)$ is the corresponding optimal trajectory of \eqref{genODE} then there exists a function $\bm{p}^\ast(t)$, that, together with $\bm{q}^{\ast}(t)$ satisfies the Hamilton canonical equations \ccite{bardi_optimal_2008}
\begin{equation}
\label{Heq}
         \left( \begin{array}{c} \dot{\bm{q}}  \\ \dot{\bm{p}}  \end{array} \right) =
  \left( \begin{array}{r}  \nabla_{\bm{p}} H  \\ - \nabla_{\bm{q}} H \end{array} \right). 
  \end{equation}
The costate variables $\bm{p}$ are the momenta in the context of classical mechanics \ccite{arnold_1978} but they  do not have that physical meaning in the current problem. 

 Furthermore, according to Pontryagin's maximum principle, the Hamiltonian is identically equal to zero and achieves its maximum along the optimal trajectory  $(\bm{q}^*,\bm{p}^*)$ of \eqref{Heq}:
 \begin{equation}
    0 = H(\bm{q}^*,\bm{u}^*,\bm{p}^*) =\max_{\bm{u}\in \mathbb{U}} H(\bm{q},\bm{u},\bm{p}),
  \label{B_Pontragin}
\end{equation}
 where $\mathbb{U}$ is the set of admissible controls.

  For the control problem \eqref{eq:our_problem}, the state and costate variables are 
 $$
 \hat{\bm{q}} \equiv
 \left(\begin{array}{c}\bm{q}\\\theta\end{array}\right) = 
 \left(\begin{array}{c}\bm{x}\\\bm{v}\\\theta\end{array}\right)
 \quad{\rm and}\quad 
 \hat{\bm{p} }\equiv 
 \left(\begin{array}{c}\bm{p}\\p_\theta\end{array}\right) = 
 \left(\begin{array}{c}\bm{p}_{\bm{x}}\\\bm{p}_{\bm{v}}\\p_\theta\end{array}\right).
 $$
The optimal control $\bm{u}^{\ast}$ can be found using the fact that the Hamiltonian achieves its maximum at the optimal control $\bm{u}^*$. Therefore
\begin{align}
0=&\nabla_{\bm{u}} H(\hat{\bm{q}},\bm{u}^{\ast},\hat{\bm{p}})  \notag\\
=& \nabla_{\bm{u}} \left(\bm{p}_{\bm{x}}^\top \bm{v} + 
\bm{p}_{\bm{v}}^\top\left[-\delta_c \bm{v}-\bm{K}(\bm{x})+\bm{f}(\theta) +\bm{u}^{\ast}\right]\right. \notag \\
&\left.+p_{\theta} - \frac{1}{2}\|\bm{u}^{\ast}\|^2 \right)\notag \\
 =& \bm{p}_{\bm{v}} - \bm{u}^{\ast}.
\end{align}
 Hence $\bm{u}^{\ast} = \bm{p}_{\bm{v}}^{\ast}$ and the optimal path satisfies the Hamiltonian dynamical system
\begin{equation}
  \left( \begin{array}{c} \dot{\bm{x}} \\ \dot{\bm{v}} \\\dot{\theta} \\ \dot{\bm{p_x}} \\ \dot{\bm{p_v}} \\\dot{p}_{\theta} \end{array} \right)
  =
  \left( \begin{array}{c}  \bm{v} \\ 
    - \delta_c \bm{v} - \bm{K} (\bm{x}) + \bm{f}(\theta) + \bm{p_v} \\ 
    1 \\
    \nabla_{\bm{x}} \left(\bm{p_v}^\top  \bm{K} (\bm{x})\right) \\
    - \bm{p_x} + \delta_c \bm{p_v}
    \\-\bm{p}_{\bm{v}}^\top\bm{f}'(\theta)
        \end{array} 
    \right).
  \label{AP_Ham1}
\end{equation}

Any attractor $\hat{\bm{A}}_k$ of the ODEs \eqref{AP_ODE1}  can be mapped onto a saddle-type invariant set $\bar{\bm{A}}_k$ of the corresponding Hamiltonian ODEs \eqref{AP_Ham1} by including the costate variables $\hat{\bm{p}}$ and setting them to zero. Since the first boundary condition in \eqref{AP_BC1} is set at the attractor $\hat{\bm{A}}_0$, 
the corresponding trajectory of \eqref{AP_Ham1} passing through $(\hat{\bm{q}}(t_0),\bm{0})$  will stay forever at the corresponding saddle-type invariant set $\bar{\bm{A}}_0$. On the other hand, a whole family of trajectories called the \emph{Hamiltonian paths} emanates from $\bar{\bm{A}}_0$ at time $t_0=-\infty$ and constitutes the $(2N+1)$-dimensional unstable manifold of $\bar{\bm{A}}_0$ called the \emph{Lagrangian manifold}.  This will be clear from the structure of the Jacobian matrix obtained in Section \ref{AP_BC}.

Therefore, the optimal control problem \eqref{eq:our_problem} is reduced to finding the initial condition for the optimal Hamiltonian path at some finite moment in time. This is done by surrounding the attractor $\hat{\bm{A}}_0$ with a small neighborhood and searching for the initial conditions for the optimal Hamiltonian path in the intersection of the boundary of this neighborhood and the Largangian manifold. This construction is elaborated in Section  \ref{AP_BC}. 



\section{Methodology}
\label{Methodology}
The methodology is developed for SDEs of the form  \eqref{B_SDE_Autonomous} where the external forcing $\bm{f}(t)$ is periodic. It is assumed that the corresponding ODEs \eqref{AP_ODE1} admit $n_a\ge 2$ attractors $\hat{\bm{A}}_0$, $\ldots$, $\hat{\bm{A}}_{n_a-1}$ with basins $B_0$, $\ldots$, $B_{n_a-1}$, and these \emph{attractors are periodic orbits}. The intersections of basin boundaries $\partial B_k$ and $\partial B_l$ are denoted by $\partial B_{kl}$, $0\le k,l\le n_a-1$. 

Let the system governed by SDEs \eqref{B_SDE_Autonomous} be initially located near attractor $\hat{\bm{A}}_0$. The implication of Pontryagin's maximum principle is that the minimizer of the action functional \eqref{B_ActionFunctionalDegenerate} is a Hamiltonian path given by \eqref{AP_Ham1}. As soon as the Hamiltonian path reaches the basin boundary $\partial B_0$, the authors reset and fix the costates $\bm{p_x}$ and $\bm{p_v}$ to zero. From that point, the path continues a trajectory of \eqref{AP_ODE1} restricted to basin boundary $\partial B_0$.
This trajectory approaches an attractor $S$ for the dynamics restricted to  $\partial B_0$. For the systems studied in this work, $S$ is always a saddle cycle. Let $S\equiv S_{0k}$ belongs to $\partial B_{0k}$ separating basins of $\hat{\bm{A}}_0$ and $\hat{\bm{A}}_k$, $k\neq 0$. Then, one can complete the transition path from $\hat{\bm{A}}_0$ to $\hat{\bm{A}}_k$ by adding a trajectory going from $S_{0k}$ to $\hat{\bm{A}}_k$. Since the contribution to action (13) along any trajectory is zero, the total action along the transition path from $\hat{\bm{A}}_0$ to $\hat{\bm{A}}_k$ is equal to the action along the Hamiltonian path. The authors' numerical experiments suggest (see Figs. \ref{OneDuffing_H_L2MPEP},
\ref{OneDuffing_L_L2MPEP},
\ref{TwoDuffing_HH_L2MPEP},
\ref{TwoDuffing_LL_L2MPEP},
\ref{TwoDuffing_HL_L2MPEP},
\ref{ThreeDuffing_LHL_L2MPEP})
that the maximum likelihood transition path from $\hat{\bm{A}}_0$ to $\hat{\bm{A}}_k$ is a concatenation of  $(i)$ a Hamiltonian path that approaches  $S_{0k}$ that is infinitely long, $(ii)$ the saddle periodic orbit $S_{0k}$, and $(iii)$ a trajectory of \eqref{AP_ODE1} starting infinitely close to $S_{0k}$ and approaching $\hat{\bm{A}}_k$. The key challenge is that the initial condition for the optimal Hamiltonian path is unknown.

\subsection{The Lagrangian Manifold} 
\label{AP_BC}
The action plot method \ccite{beri_solution_2005} utilizes Pontryagin's maximum principle to reduce the problem of finding the most probable escape path (MPEP) to the problem of finding the initial condition for the Hamiltonian path and offers an elegant way to simplify this problem. This is done by linearizing the Hamiltonian system near an invariant set corresponding to an attractor of the original ODE and approximating the unstable Lagrangian manifold by a linear relationship between $\hat{\bm{q}}(0)$ and $\hat{\bm{p}}(0)$. The construction of the Lagrangian manifold adapted for the case of periodic forcing is detailed below.  The Floquet theory \ccite{chicone_1999} for linear ODEs with periodic coefficients is used to facilitate finding the unstable manifold. The latter is inspired by Ref. \ccite{lin_quasi-potential_2019}.

The equations for $\theta$ and $p_{\theta}$ in the Hamiltonian ODE \eqref{AP_Ham1} can be omitted, as they do not affect the other equations. The resulting non-autonomous Hamiltonian ODE system is given by
\begin{equation}
  \left( \begin{array}{c} \dot{\bm{x}} \\ \dot{\bm{v}} \\ \dot{\bm{p_x}} \\ \dot{\bm{p_v}} \\ \end{array} \right)
  =
  \left( \begin{array}{c}  \bm{v} \\ 
    - \delta_c \bm{v} - \bm{K} (\bm{x}) + \bm{f}(t) + \bm{p_v} \\ 
    \nabla_{\bm{x}} \left(\bm{p_v}^\top  \bm{K} (\bm{x})\right) \\
    - \bm{p_x} + \delta_c \bm{p_v}
        \end{array} 
    \right).
  \label{AP_Ham2}
\end{equation}
The periodic trajectories of \eqref{AP_Ham1} are naturally mapped onto periodic trajectories of \eqref{AP_Ham2} by dropping of the components $\theta$ and $p_{\theta}$. Let ${\hat{\bm{A}}}_0=:(\bm{x}(t),\bm{v}(t),\theta(t))$ be an attractor of the ODE system \eqref{AP_ODE1}. It corresponds to a saddle-type periodic trajectory $\bar{\bm{A}}_0$ of the Hamiltonian dynamical system \eqref{AP_Ham1}. In turn, $\bar{\bm{A}}_0$ corresponds to a periodic solution $\tilde{\bm{A}}_0: = (\bm{x}(t),\bm{v}(t),\bm{0},\bm{0}))$ of the non-autonomous Hamiltonian dynamical system \eqref{AP_Ham2}. The authors' goal is to find the set of the initial conditions for the Hamiltonian paths forming the Lagrangian manifold. To do so, one needs to linearize ODEs \eqref{AP_Ham2} near the periodic solution $\tilde{\bm{A}}_0$. The Jacobian of the right-hand side of \eqref{AP_Ham2} is
 \begin{equation}
     \begin{split}
        &J(\bm{x},\bm{v},\bm{p}_{\bm{x}},\bm{p}_{\bm{v}},t_0) \\
      = & \left( \begin{array}{cccc}  \bm{0} & \bm{I}  & \bm{0} & \bm{0}  \\ 
         -\bm{J}_{\bm{K}}(\bm{x}) & -\delta_c \bm{I}  & \bm{0} & \bm{I}  \\ 
         \nabla_{\bm{x}} \nabla_{\bm{x}}\left[\bm{p}_{\bm{v}}^\top \bm{K}(\bm{x})\right] &\bm{0}  & \bm{0} & \bm{J}_{\bm{K}}(\bm{x})^\top  \\ 
          \bm{0} &  \bm{0} & -\bm{I}  & \delta_c \bm{I}  \\ 
         \end{array} \right), \\  
     \end{split}
     \label{AP_Jacobian}
 \end{equation}
 and its value at $(\hat{\bm{x}}(t),\hat{\bm{v}}(t),\bm{0},\bm{0}))$ is
\begin{equation}
     \begin{split}
         &J(\hat{\bm{x}},\hat{\bm{v}},\bm{0},\bm{0},t_0)\\
         =&  \left( \begin{array}{cccc}  \bm{0} & \bm{I}  & \bm{0} & \bm{0}  \\
         -\bm{J}_{\bm{K}}(\hat{\bm{x}}) & -\delta_c \bm{I}  & \bm{0} & \bm{I}  \\ 
         \bm{0}  &\bm{0}  & \bm{0} &  \bm{J}_{\bm{K}}(\hat{\bm{x}})^\top   \\         %
          \bm{0} &  \bm{0} & -\bm{I}  & \delta_c \bm{I}  \\ 
         \end{array} \right) \equiv J_{\bm{\tilde{A}}_0}(t_0).
         \end{split}
     \label{AP_LinearizedSystem}
 \end{equation}
 In equations \eqref{AP_Jacobian} and \eqref{AP_LinearizedSystem}, $\bm{J}_{\bm{K}}$ denotes the Jacobian matrix of the vector field $\bm{K}$. The linear ODE with periodic coefficients of period $T$ approximating \eqref{AP_Ham2} near $\tilde{\bm{A}}_0$ is of the form 
 \begin{equation}
     \label{AP_ODElinper}
     \dot{\bm{z}} = J_{\bm{\tilde{A}}_0}(t_0) \bm{z}.
 \end{equation}
 Equation \eqref{AP_ODElinper} has a fundamental solution matrix $\bm{\Psi}(t)$.
 According to Floquet's theorem \ccite{chicone_1999}, the monodromy matrix
 \begin{equation}
    \label{monodromy}
    \bm{\Psi}(t_0+T)\bm{\Psi}(t_0)^{-1}
\end{equation}
 provides a coordinate change that results in the Poincar\'e mapping for the corresponding linearized autonomous system with $\theta_0 = t_0 \thinspace{\rm mod}\thinspace T \in \mathbb{S}^1_T$.  The eigenvalues of the monodromy matrix, called \emph{characteristic multipliers}, are intrinsic to the ODE, as they remain invariant for all $t_0\in\mathbb{R}$. The corresponding matrix of eigenvectors is a periodic function of $t_0$.  The structure of the matrix $J_{\bm{\tilde{A}}_0}(t_0)$ implies that
  the monodromy matrix has $2N$ eigenvalues less than 1 (stable) and $2N$ eigenvalues greater than 1 (unstable). The $4N\times 2N$ matrix functions $\bm{\Phi}^s_{\theta_0}$ and $\bm{\Phi}^u_{\theta_0}$ are formed, respectively, by the stable and unstable eigenvectors. At each fixed $\theta_0\in[0,T]$, the span of $\bm{\Phi}^u_{\theta_0}$ is the linear unstable manifold.
Since ODE \eqref{AP_ODElinper} describes the time evolution of the first-order perturbation to the periodic solution $\tilde{\bm{A}}_0$, the Lagrangian manifold near $\tilde{\bm{A}}_0$ is approximated by the sum of trajectories of the Hamiltonian ODE \eqref{AP_Ham2} passing through the set of initial conditions
 \begin{equation}
     \label{AP_IC}
     \left\{\tilde{\bm{A}}_0(\theta_0) + \hat{\gamma}  \bm{\Phi}^u_{\theta_0} \bm{c} \,~\vline~\,\bm{c}\in\mathbb{R}^{2N},\theta_0 \in \mathbb{S}^1_T \right\},
 \end{equation}
 where $\hat{\gamma}>0$ is a scaling parameter. It is worth noting that the choice of $\hat{\gamma}$ would be arbitrary if \eref{AP_Ham2} were linear, and $\theta_0$ could be set to zero if \eref{AP_Ham2} were autonomous. However, since \eqref{AP_Ham2} is nonlinear and nonautonomous, 
 $\hat{\gamma}| \bm{c} |$ needs to be small and all $\theta_0\in[0,T)$ should be considered.
 
 Any escape path in $\mathbb{R}^{2N}\times\mathbb{S}^1$ leaving $\hat{\bm{A}}_0$ must cross through a torus that surrounds $\hat{\bm{A}}_0$. For each fixed $\theta_0 \in \mathbb{S}^1_T$, the point $\hat{\bm{A}}_0(\theta_0)$ can be completely surrounded by a $2N-1$-dimensional sphere embedded in $\mathbb{R}^{2N}$. This allows one to further reduce the optimization space as follows. Consider the sphere of radius $\gamma$:
 \begin{equation}
     \begin{split}
         \mathbb{S}_{\gamma}^{2N-1} = \{\bm{q}_\gamma \in \R^{2N} \textrm{ such that } |\bm{q}_\gamma| = \gamma\}.
     \end{split}
     \label{AP_qgamma}
 \end{equation}
 The values of $\bm{q}_\gamma$ that are in the Lagrangian manifold must satisfy
  \begin{equation}
      \begin{split}
          \bm{A}_0(\theta_0)  + \hat{\gamma} \bm{Z}_{\bm{q},\theta_0} \bm{c} & = \bm{A}_0(\theta_0) + \bm{q}_\gamma , \\
      \end{split}
     \label{AP_qgammaandlagrangianmanifold}
 \end{equation}
 where $\bm{A}_0(\theta_0)$ consists of $\bm{x}$- and $\bm{v}$-components of  $\tilde{\bm{A}}_0(\theta_0)$ and the matrix, $\bm{Z}_{\bm{q},\theta_0} $ is a sub-matrix of dimension $2N\times2N$ extracted from $\bm{\Phi}^u_{\theta_0} $ such that
  \begin{equation}
        \left[\begin{matrix}
        \bm{Z}_{\bm{q},\theta_0} \\
        \bm{Z}_{\bm{p},\theta_0}
        \end{matrix}\right]
        \equiv
      \bm{\Phi}^u_{\theta_0}.
     \label{AP_UnstableEigenSubMatrices}
 \end{equation}
 It follows from \eqref{AP_qgammaandlagrangianmanifold} that 
    \begin{equation}
        \hat{\gamma}\bm{c}  =  \bm{Z}_{\bm{q},\theta_0}^{-1} \bm{q}_\gamma,
     \label{AP_cCoefficient}
 \end{equation}
 which can be used to solve for the initial costate perturbations:
  \begin{equation}
      \begin{split}
          \bm{p}_{\gamma,\theta_0} & =  \bm{Z}_{\bm{p},\theta_0} \hat{\gamma} \bm{c}  \\
          & =  \bm{Z}_{\bm{p},\theta_0} \bm{Z}_{\bm{q},\theta_0}^{-1} \bm{q}_\gamma.
      \end{split}
     \label{AP_pgamma}
 \end{equation}
 With the aid of \eqref{AP_qgamma} and \eqref{AP_pgamma}, one can define the space of initial conditions as 
  \begin{equation}
     \label{AP_ICReduced}
    W \equiv \left\{\tilde{\bm{A}}_0(\theta_0) +       \left[\begin{matrix}
        \bm{q}_\gamma\\
         \bm{p}_{\gamma,\theta_0}
        \end{matrix}\right]   \,~\vline~\,\bm{q}_\gamma\in\mathbb{S}^{2N-1}_\gamma,\theta_0 \in \mathbb{S}^1_T \right\},
 \end{equation}
 which is now a $2N$-dimensional manifold. 

 In this work, the authors have used $\gamma = 10^{-20}$ for $N=1$, a single oscillator, and $\gamma = 10^{-15}$ for the $N>1$ oscillators. 
 Note that it is not worth it to pick $\gamma$ too small as  there is a trade-off between the error in the approximation of the Lagrangian manifold and the error accumulating due to numerical integration of Hamiltonian paths.

\subsection{The Escape Time} \label{EscapeTimeSection}
The escape cost from the basin of $\hat{\bm{A}}_0$ for a Hamiltonian path $(\bm{q},\bm{p})$ defined by \eqref{AP_Ham2} and the initial condition $(\bm{q}_0,\bm{p}_0)$  of the form  \eqref{AP_ICReduced} 
is found by numerical integration of
the action \eqref{B_ActionFunctionalDegenerate} until the moment of time $\tau(\bm{q}_{\gamma},\theta_0)$ when the path reaches the basin boundary $\partial B_0$:
\begin{equation}
    \label{AP_EscapeTime}
    \tau(\bm{q}_\gamma,\theta_0) = \inf_{t\in \mathbf{R}}\{\bm{q}(t)\in\partial B_0,~\|\bm{q}(\theta_0)\| = \gamma\}.
\end{equation}
The  escape  time for most  Hamiltonian  paths  is  finite when radius $\gamma$ in \eqref{AP_ICReduced} is positive, except for when the path is a heteroclinic orbit. In fact, the results presented in this paper suggest that it is exactly the case that the most probable escape paths approach attractors as $t\rightarrow -\infty$ and approach saddle cycles on basin boundaries as $t\rightarrow +\infty$.

\subsection{Optimization and Implementation} 
\label{OptimizationSection}
\begin{figure}[htbp]
\begin{center}
\centerline{\includegraphics[width=0.5\textwidth]{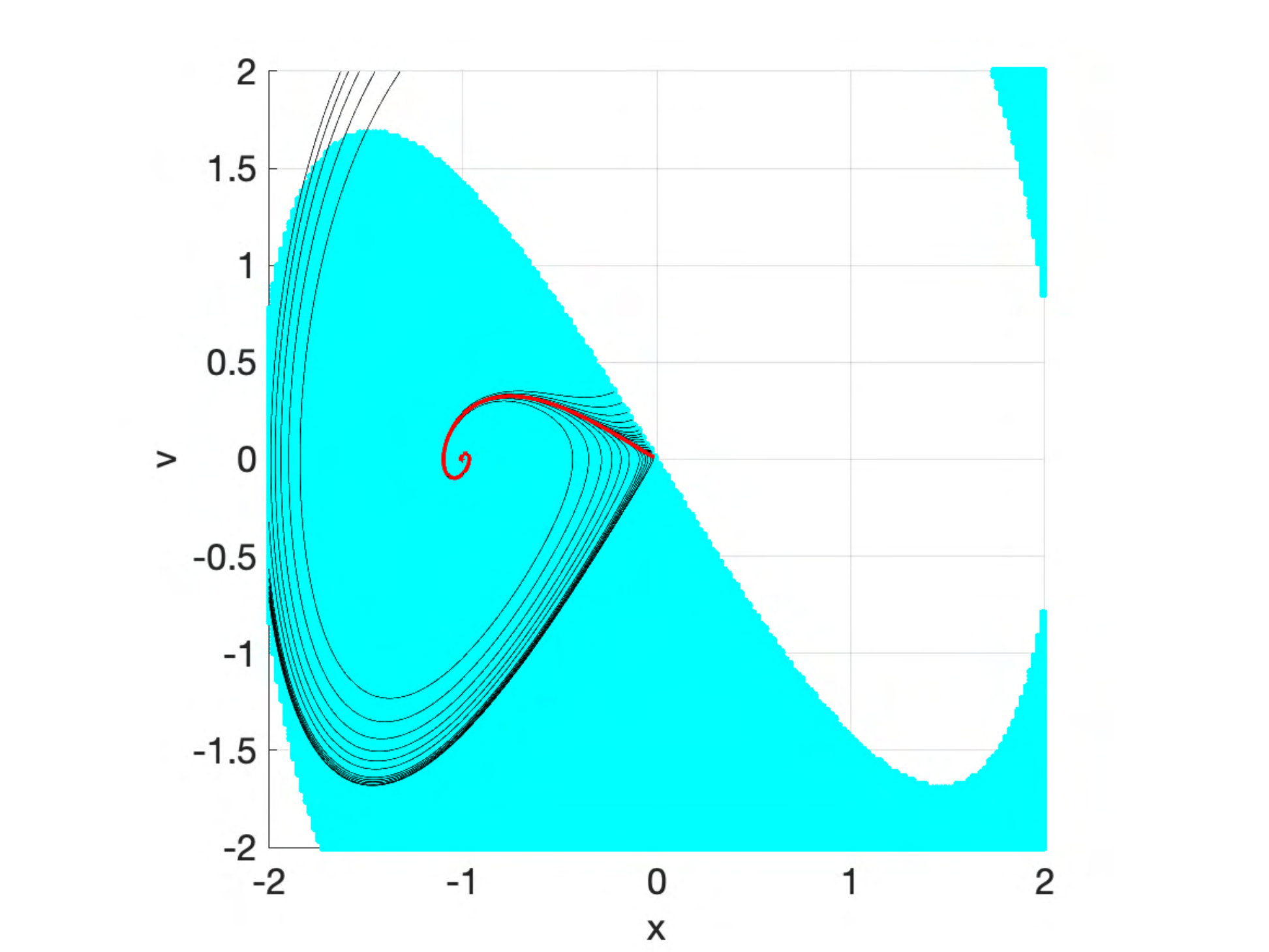}}
\caption{An illustration for the origin of discontinuities in the cost function \eqref{AP_Cost} on the example of a single bistable Duffing oscillator $\ddot{{x}} + \dot{{x}}-{{x}}+{x}^3 = \sqrt{\epsilon}{\eta}(t)$. This oscillator has stable equilibria at $(\pm1,0)$ and a saddle at the origin. The basin of $(-1,0)$ is shaded cyan. The optimal Hamiltonian escape path and a collection of other Hamiltonian paths with close initial conditions are shown in red and black, respectively. It is apparent that overshooting initial conditions results in a much longer escape route and a sharp increase of escape cost.}
\label{fig:bistableDuffing}
\end{center}
\end{figure}

 \begin{algorithm}[b]
\caption{Stochastic Unit Vectors}\label{alg:SUV}
\KwInitialization{
Select an initial point $\bm{x}_0$ on a $d$-dimensional manifold $\mathcal{M}$ and an initial step size $\sigma$. Set $i=0$. Select the minimal step size $\sigma_{\min}$.
}
\\
\KwTheMainBody\\
\While {$\sigma\ge \sigma_{\min}$ }
{
 {\bf 1:} Sample a random unit vector $\bm{e}_1\in T_{\bm{x}_i}\mathcal{M}$ (the tangent space of $\mathcal{M}$ at $\bm{x}_i$) from a uniform distribution. \\
 {\bf 2:} Complete  $\bm{e}_1$ to an orthonormal basis $[\bm{e}_1,\bm{e}_2,...,\bm{e}_d] $ in the tangent space $T_{\bm{x}_i}\mathcal{M}$. \\
 {\bf 3:} Consider a collection of points 
 $Y = \{ \, \bm{y}_{j} =  \bm{x}_{i} + \sigma \, \bm{e}_j , \,    j = [1, 2 ,..., d], $\\
 $\quad \quad \quad  \bm{y}_{j} = \bm{x}_{i} - \sigma \, \bm{e}_j, \quad   j = [d+1, d+2 ,..., 2d]\}$. \\
 {\bf 4:} Calculate the cost $C(\bm{y}_{j})$ at each $\bm{y}_j$, $1\le j\le 2d$. \\ 
 {\bf 5:}  
  \eIf {$\min_{1\le j\le 2d} C(\bm{y}_{j}) \geq  C(\bm{x}_{i})$ }
  {Remain at $\bm{x}_i$ and reduce the step size: $\bm{x}_{i+1} = \bm{x}_{i},~ \sigma  = \sigma/2.$} 
 {  Set $\tilde{\bm{x}} = \arg\min_{1\le j\le 2d} C(\bm{y}_{j})$ and project $\tilde{\bm{x}}$ onto the manifold $\mathcal{M}$: $\bm{x}_{i+1}={\sf Proj}_{\mathcal{M}}(\tilde{\bm{x}})$. } 
 {\bf 6:} $i = i+1$. 
 }
\end{algorithm}

Based on the argument presented in Sections \ref{AP_BC} and \ref{EscapeTimeSection}  the authors reduce the optimization problem \eqref{eq:our_problem} for the most probable escape path (MPEP) to
\begin{equation}
    \begin{split}
        \label{AP_Cost}
    &  C(\bm{q}_\gamma,\theta_0) = \frac{1}{2} \int_{\theta_0}^{\tau(\bm{q}_\gamma,\theta_0) }||\bm{p}_{\bm{v}}(t)||^2    dt \rightarrow \min\\
    \end{split}
\end{equation}
where $\bm{p_v} \equiv \bm{u}$ follows the dynamics of \eqref{AP_Ham2} with initial conditions defined by $\eqref{AP_ICReduced}$. The integral time bounds are defined by $\eqref{AP_ICReduced}$ and \eqref{AP_EscapeTime}.  The authors minimize \eqref{AP_Cost} with respect to both $\bm{q}_\gamma$ and $\theta_0$ in the manifold $\mathcal{M}:=\s^{2N-1}_\gamma\times\s^{1}_T$. The contribution of $\theta_0$ to the minimizers is discussed in Section \ref{OneDuffingResults}.

A set of $n_{\rm init}$ initial points $\bm{q}_\gamma$ and $\theta_0$ is sampled uniformly on the manifold $\mathcal{M}:=\s^{2N-1}_\gamma\times\s^{1}_T$. The corresponding initial conditions for the Hamiltonian paths are defined according to \eqref{AP_ICReduced}. Next, each Hamiltonian path \eqref{AP_Ham2} together with its cost functional \eqref{B_ActionFunctionalDegenerate} is integrated until a crossing of the basin boundary is detected. The corresponding moment of time is $\tau(\bm{q}_\gamma,\theta_0)$. The found values of the cost functional are first sorted in the ascending order, and then $n_s<n_{\rm init}$ initial points  $\bm{q}_\gamma$ and $\theta_0$ corresponding to $n_s$ smallest values of the cost functional are selected as the candidates for local minimizers. The values of $n_{\rm init}$ and $n_s$ are found in Table \ref{tab:ComputationalCostTable} in the appendix.

The cost function \eqref{AP_Cost} is discontinuous and its minima lie at the bottoms of its ``cliffs" -- see Figure \ref{OneDuffing_PhaseQuasipotential} ahead depicting the escape cost for a single oscillator. Such behavior of the cost function can be understood as follows.
Consider a sequence of Hamiltonian paths with initial conditions $(\bm{q}_{\gamma}^j,\theta_0^j)$ escaping from the basin of the attractor $\hat{\bm{A}}_0$ whose escape points approach a saddle cycle lying at the boundary of $B_0$ and whose escape costs $C(\bm{q}_{\gamma}^j,\theta_0^j)$ decrease. The goal of numerical minimization of \eqref{AP_Cost} is to  generate a sequence of Hamiltonian paths converging to a limiting path that has minimal escape cost and approaches the saddle cycle infinitely long. Suppose the optimizer ``overshoots". Then the path will fail to exit the basin of $\hat{\bm{A}}_0$ near the saddle and proceed to an extra loop causing a jump in the escape cost. This phenomenon is illustrated on a simple example of a bistable Duffing oscillator $\ddot{{x}} + \dot{{x}}-{{x}}+{x}^3 = \sqrt{\epsilon}{\eta}(t)$ in Figure \ref{fig:bistableDuffing}. 

\begin{figure}[htbp]
\begin{center}
\centerline{\includegraphics[width=0.5\textwidth]{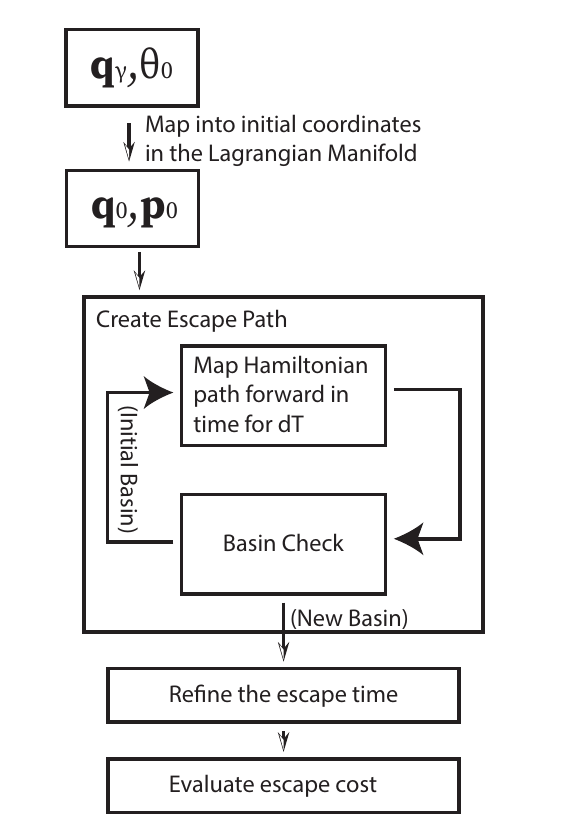}}
\caption{A flow chart of the process of evaluating the cost function given an input $\bm{q}_\gamma$ and $\theta_0$. A detailed version of the process shown here is described in the pseudocode of Algorithm 2.}
\label{fig:Algorithm2}
\end{center}
\end{figure}

\begin{algorithm}[t]
\caption{The Cost Function: Step 4 in Algorithm \ref{alg:SUV}}\label{alg:CostFunction}
{\bf Input:} Initial conditions $\bf{q}_\gamma, \theta_0$ \\ 
\bf{Initialization} \\
\bf{ i:} \textnormal{Define $t_i \gets \theta_0$, $i \gets 0$} \\ 
\bf{ ii:} \textnormal{Define initial condition $(\bm{q}(t_{i}),\bm{p}(t_{i})) \in W$ using $\bf{q}_\gamma, \theta_0$} \\ 
\bf{ iii:} \textnormal{Let {\tt index0} be the index of the initial attractor.} \\ 
\bf{ iv:} \textnormal{Define a function, ${\tt Aindex}(\bm{q}(t))$, that returns $m$ if $\|{\bm{q}(t) - \bm{A}_m(t) }\| \leq R \quad \forall m = [0, 1, ..., n_a-1]$ or {\tt noindex} otherwise. Here, $R$ is a small radius and $\bm{A}_m(t)$ is the position of attractor $m$ at time $t$.} \\ 
\bf{ v:} \textnormal{Define the system's differential equations as $ODE_S$} \\ 
\bf{ vi:} \textnormal{Define the system's corresponding Hamiltonian differential equations as $ODE_H$} \\ 
$\quad$ \\
\bf{Escape via Integration} \\
\While{\emph{ ${\sf index} == 0$}}
{
 {\bf 1:} \textnormal{ $i \gets i + 1, \, t_i \gets t_{i-1} + dT$} \\
 {\bf 2:} \textnormal{Evaluate $(\bm{q}(t_{i}),\bm{p}(t_{i}))$ by integrating $ODE_H$ forward in time from $t_{i-1}$ to $t_i$  with initial condition $(\bm{q}(t_{i-1}),\bm{p}(t_{i-1}))$.}  \\
 {\bf 3:} \textnormal{$j \gets 0, \, \check{t}_j \gets t_i, \, \check{\bm{q}}(\check{t}_j) \gets \bm{q}(t_{i})$} \\
{\bf 4:} \While{\emph{${\tt Aindex}(\check{\bm{q}}(\check{t}_j)) == {\tt noindex}$}}
{
    {\bf a:} \textnormal{ $j \gets j + 1, \, \check{t}_j \gets \check{t}_{j-1} + dT$} \\
    {\bf b:} \textnormal{Evaluate $\check{\bm{q}}(\check{t}_{j})$ by integrating $ODE_S$ forward in time from $\check{t}_{j-1}$ to $\check{t}_{j}$  with initial condition $\check{\bm{\bm{q}}}(\check{t}_{j-1})$.}  \\
}
{\bf 5:} \textnormal{${\tt index} \gets {\tt Aindex}(\check{\bm{q}}(\check{t}_j))$} \\ 
}
$\quad$\\
\bf{Refine the Escape Time} \\
\bf{i:} \textnormal{Define upper and lower time bounds for the escape time: $\tau_H = t_i, \tau_L = t_{i-1}$} \\
\bf{ii:}  $d\tau \gets \tau_H-\tau_L$  \\ 
\While{ \emph{$d\tau > \epsilon$}}
{
\bf{1:} $\tau \gets (\tau_L+\tau_H)/2$  \\ 
{\bf 2:} \textnormal{$j \gets 0, \, \check{t}_j \gets \tau, \, \check{\bm{q}}(\check{t}_j) \gets \bm{q}(\tau)$} \\
{\bf 3:} \While{${\tt Aindex}(\check{\bm{q}}(\check{t}_j)) =={\tt noindex}$ }
{
    {\bf a:} \textnormal{ $j \gets j + 1, \, \check{t}_j \gets \check{t}_{j-1} + dT$} \\
    {\bf b:} \textnormal{Evaluate $\check{\bm{q}}(\check{t}_{j})$ by integrating $ODE_S$ forward in time from $\check{t}_{j-1}$ to $\check{t}_{j}$  with initial condition $\check{\bm{q}}(\check{t}_{j-1})$.}  \\
}
{\bf 4:} \eIf {
\emph{${\tt Aindex}(\check{\bm{q}}(\check{t}_j)) == {\tt index0}$}
}
{
 \textnormal{$\tau_L \gets \tau$}  
}
{
 \textnormal{$\tau_H \gets \tau$}    
} 
\bf{5:} $d\tau \gets \tau_H-\tau_L$  \\ 

}
$\quad$\\
\bf{Evaluate the Cost} \\
\textnormal{ Evaluate the cost by integrating \eqref{AP_Cost} along the path $(\bm{q}(t),\bm{p}(t))$
on the time interval $\theta_0\leq t \leq \tau$.}
\end{algorithm}

A variety of gradient-based and gradient-free methods and their combinations have been explored as optimizers for  \eqref{AP_Cost}. The gradient of the cost function has been approximated by finite differences. 

The gradient-based methods include the gradient descent and the Fletcher-Reeves nonlinear conjugate gradient methods projected on the manifold $\mathcal{M}=\s^{2N-1}_\gamma\times\s^{1}_T$.  The gradient-free techniques that the authors have used are the Nelder-Mead simplex method and an originally developed \emph{stochastic unit vectors} method (SUV) summarized in the pseudocode in Algorithm \ref{alg:SUV}. While gradient-based methods are more efficient while stepping within smooth regions of the cost function than Nelder-Mead or SUV, they are unable to crawl along the edges of the bottoms of the cliffs. Hence, as a proximity of a cliff is detected, a gradient-free method continues the search. The authors have found that there is no significant advantage in starting minimization with a gradient-based technique. Furthermore, the brute-force SUV algorithm turned out to work the most robustly and reliably on the problems addressed in this work. 

 As with all gradient free methods that cannot guarantee convergence to a global minimum, it is best to validate the minimizer with a secondary approach. The authors increase the probability of finding a global minimum by first sampling the cost function space with a large number of initial points prior to beginning the optimization as mentioned earlier in this section. The authors validate minimizers in this work by checking that they approach saddle cycles, and that the resulting quasipotential values follow a trend with respect to parameters such as frequency and coupling. Since the optimization and the initial sampling are stochastic, the authors also validate the minimizers by running the optimization multiple times with different randomly picked initial conditions. 

%

The other challenge of numerical minimization of the cost function \eqref{AP_Cost} for two or more oscillators, i.e., when the dimension $d$ of the manifold $\mathcal{M}$ is four or higher, is that the basins of attraction must be sampled online during optimization as it is impossible to store high-resolution data of the basins of attraction or the basin boundaries for $d\ge 4$. Only the attractors are determined offline and stored. 
Therefore, Step 4 of Algorithm \ref{alg:SUV}; that is, the calculation of the exit cost  that involves the integration of Hamiltonian paths and the detection of crossing basin boundary requires elaboration. 
Each Hamiltonian path $(\bm{q}(t),\bm{p}(t))$ traced by the minimization algorithm is checked at equispaced moments of time $t_i$, $i =1,2,\ldots$ for having left or not the basin of the initial attractor. This is done by integrating the original unperturbed ODE \eqref{eq:mainODE} starting from the point $\bm{q}(t_i)=(\bm{x}(t_i),\dot{\bm{x}}(t_i))$ at time $t_i$ forward in time. If it is found that that $(\bm{q}(t_i),t_i{\rm \, mod \,} T)$ does not belong to the basin of the initial attractor, tracing of the Hamiltonian path is stopped and a refinement procedure is implemented to determine the escape time $\tau(\bm{q}_\gamma,\theta_0)\in (t_{i-1},t_i)$ more precisely. The evaluation of exit cost is detailed in Algorithm \ref{alg:CostFunction} and illustrated in \fref{fig:Algorithm2}. An important parameter in this Algorithm is $dT$, the time interval between any two consecutive check times $t_{i}$ and $t_{i+1}$.
If $dT$ is too small, the cost function can take a long time to evaluate. If $dT$ is too large, the unstable dynamics of the Hamiltonian system blow up during the time span, which increases the integration time and can cause the algorithm to miss the initial escape from the basin. It was found that a good default value for $dT$ is half-period, $T/2$. However,  when the basins are shallow or small, a smaller value of $dT$ should be used. The code is set up to decrease $dT$ and restart automatically if a blow-up of a Hamiltonian path is detected. The code is written in MATLAB. The ODE solver {\tt ode45} with absolute and relative error tolerances of $10^{-8}$ is used for numerical integration.

The location of the attractors and the frequency response diagrams in sections \ref{OneDuffingResults},\ref{TwoDuffingResults}, and \ref{NDuffingResults} were computed by using a combination of the shooting method and a continuation package. The shooting method was used to generate a  stable solution of the system at one excitation frequency. The continuation package AUTO2007 \ccite{doedel_auto-07p_2007} was used to continue the solution branch for different frequency values. The use of the continuation procedure allows for determining the stable solutions branches, the unstable solution branches, the Floquet multipliers of each solution, and the bifurcation points with respect to the excitation frequency. This additional information is useful for visualization and understanding of the behavior of the system, but not necessary to solve the optimization problem for the most probable escape path.


\section{Forced Duffing Oscillator} \label{OneDuffingResults}
\begin{figure}[htbp]
\begin{center}
\centerline{\includegraphics[width=0.5\textwidth]{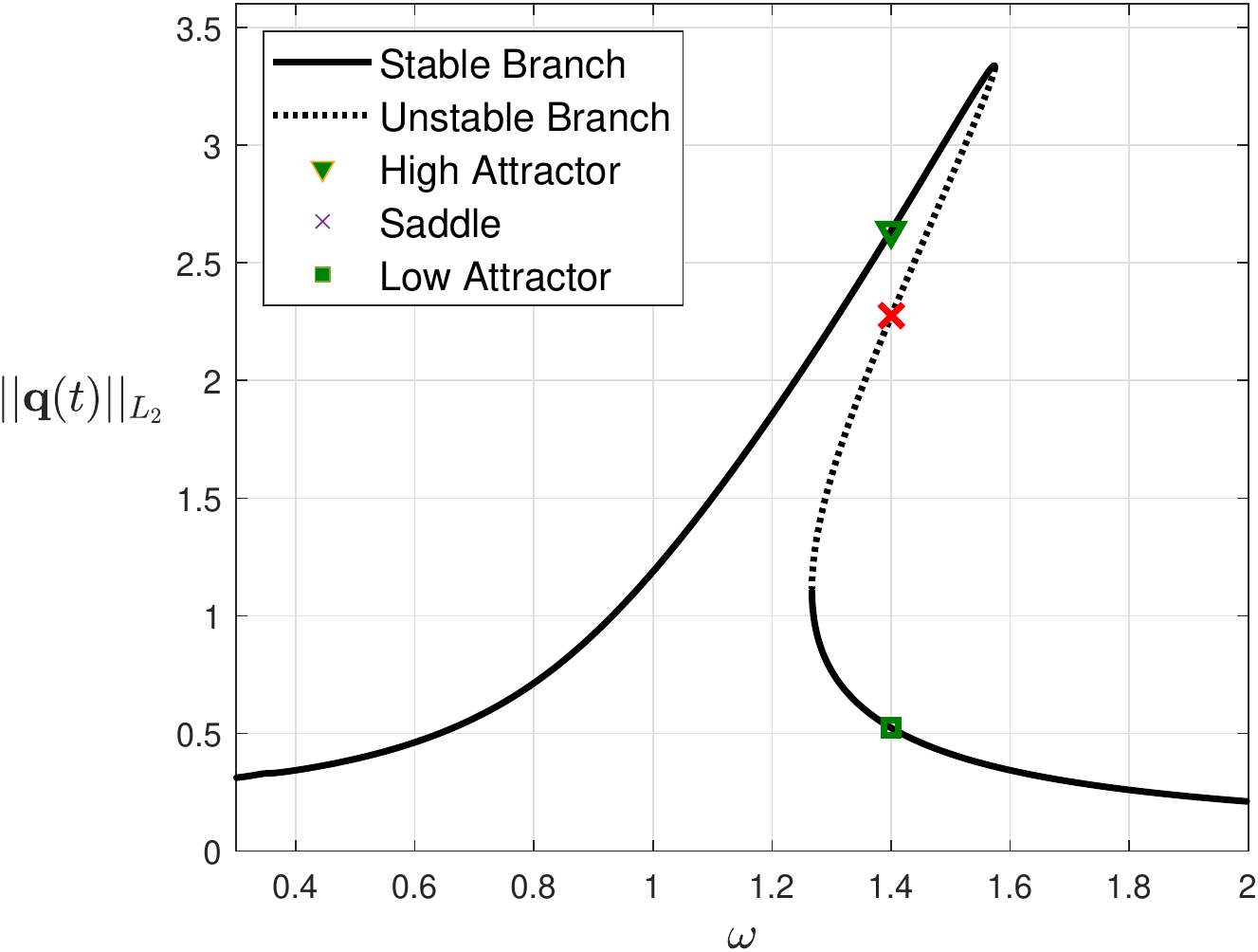}}
\caption{Frequency response of a forced Duffing oscillator as a function of excitation frequency, $\omega$. Attractor and saddle locations are highlighted at the frequency of $\omega = 1.4$. This figure was generated by using the continuation package AUTO2007 \ccite{doedel_auto-07p_2007}. }
\label{OneDuffing_FrequencyResponse}
\end{center}
\end{figure}
\begin{figure}[htbp]
\begin{center}
\centerline{\includegraphics[width=0.5\textwidth]{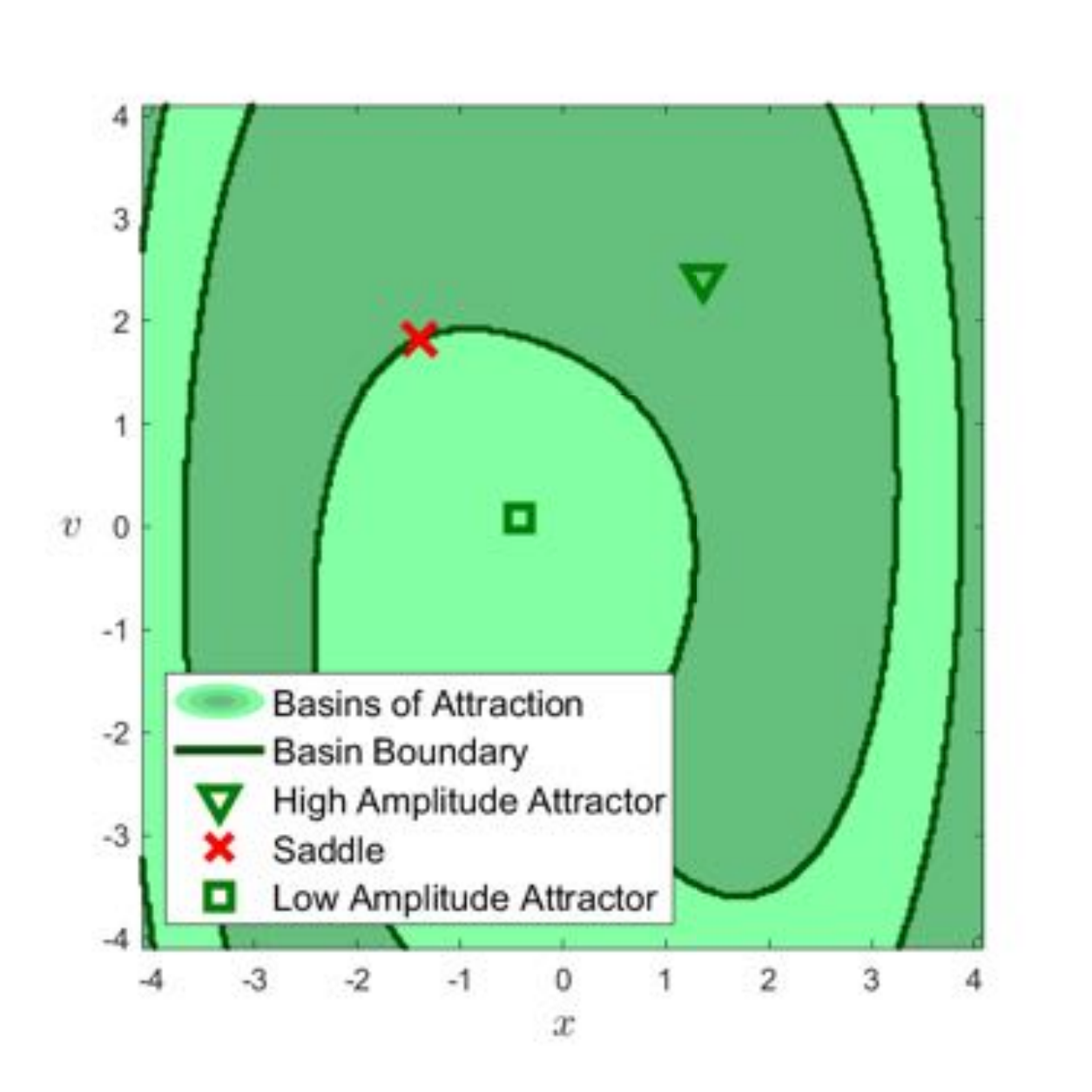}}
\caption{Poincar\'e Map of the Basins of Attraction at the frequency of $\omega = 1.4$. }
\label{OneDuffingBasins}
\end{center}
\end{figure}

This section is devoted to a detailed analysis of noise-driven transitions between the high amplitude and low amplitude attractors of a single monostable Duffing oscillator with external harmonic forcing. The methodology introduced in the previous section is applied to find the MPEPs. The case of a single oscillator is instructive and visual, as its study illuminates the difficulties of the problem and explains the authors' technical decisions. 

The dynamics of the Duffing oscillator studied in this section is governed by SDE \eqref{eq:mainSDE} with $N=1$. 
The values of the parameters chosen are $\alpha = 1, \, \beta = 0.3, \, \delta = 0.1,$  and $\, F(t) = 0.4 \text{ cos}(\omega t) $. The frequency response of this oscillator is shown in \fref{OneDuffing_FrequencyResponse}. 
As the frequency $\omega$ of the external force increases, the dynamics of the Duffing oscillator undergoes bifurcations at $\omega_1 \approx 1.27$  and at $\omega_2 \approx 1.57$.
At $0 <\omega <\omega_1$, at least one periodic solution exists and its $L_2$-norm is an increasing function of $\omega$. At $\omega = \omega_1$, a pair of a stable and unstable periodic solutions are born. As a result, there exist two stable periodic solutions for $\omega_1<\omega<\omega_2$, one of which with the smaller $L_2$-norm is born at $\omega=\omega_1$, while the other with higher $L_2$-norm one continues to exist from lower values of $\omega$. They will be referred to as the low amplitude attractor and  the high amplitude attractor, respectively. At $\omega = \omega_2$, the high amplitude periodic solution annihilates with the unstable periodic solution born at $\omega=\omega_1$. Only the low amplitude attractor exists for $\omega>\omega_2$, and its $L_2$-norm decreases as $\omega$ increases. 

For the frequency of $\omega = 1.4$, the Poincar\'e map of the periodic solutions  and the basins of attractions of each attracting solution at $\theta = 0$ is shown in \fref{OneDuffingBasins}.

The  autonomous Hamiltonian system corresponding to this oscillator is given by
\begin{equation}
    \left(\begin{array}{c} \dot{q}_1\\ \dot{q}_{2}\\ \dot{p}_{1}\\ \dot{p}_{2} \\ \dot{\theta} \end{array}\right) = 
    \left(\begin{array}{c} q_{2}\\ -\beta\,{q_{1}}^3-\alpha \,q_{1}+p_{2}-\delta_c \,q_{2} + F \cos(\omega \theta)\\ p_{2}\,\left(3\,\beta\,{q_{1}}^2+\alpha\right)\\ \delta_c \,p_{2}-p_{1} \\ 1\end{array}\right),
    \label{OneDuffing_HamiltonianRHS}
\end{equation}
where $(q_1,q_2,p_1,p_2,\theta)\in\mathbb{R}^4\times\mathbb{S}^1_T$. The action functional for this system is:
\begin{equation}
    \begin{split}
            S(\varphi,\theta) &=  \frac{1}{2}  \int_{\theta_0}^{\tau} \,p_2^2\, dt \\
            & = \frac{1}{2}  \int_{\theta_0}^{\tau} \,\left(\beta\,\varphi^3+\alpha \,\varphi+\mathrm{\ddot{\varphi}}+\delta_c \,\mathrm{\dot{\varphi}} - F \cos(\omega \theta)\right)^2 dt.
    \end{split}
    \label{OneDuffing_ActionFunctional}
\end{equation}
The Lagrangian manifold for this systems is three-dimensional, and the set $W$ of initial conditions for the Hamiltonian paths is, respectively, two-dimensional: $(\bm{q}_\gamma,\theta_0)\in \mathbb{S}^{1}_{\gamma}\times \mathbb{S}^1_T$.

\begin{figure}[htbp]
\begin{center}
\centerline{\includegraphics[width=0.5\textwidth]{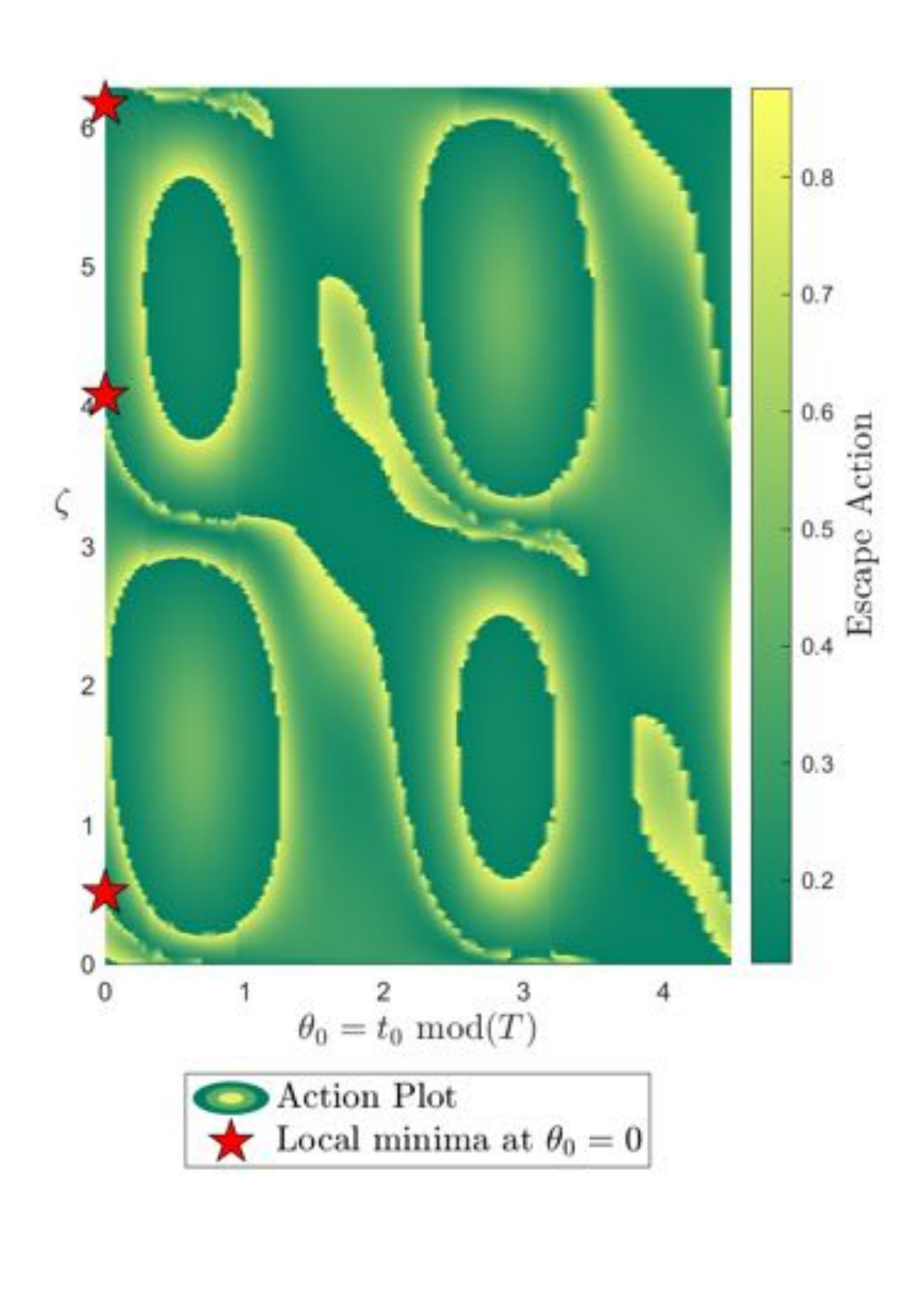}}
\caption{The cost \eqref{OneDuffing_ActionFunctional} to escape from the high amplitude attractor as a function of $\zeta$ and the initial time $\theta_0 = t_0\thinspace{\rm mod}\thinspace T$. The variable $\zeta$ is the angle around a circle that completely surrounds the initial attractor at time $\theta_0$. Three lowest local minima at $\theta_0=0$ are marked with red stars. They correspond to the minimum escape cost up to numerical errors. 
}
\label{OneDuffing_ActionPlotWithPhase}
\end{center}
\end{figure}

\begin{figure}[htbp]
\begin{center}
\centerline{\includegraphics[width=0.5\textwidth]{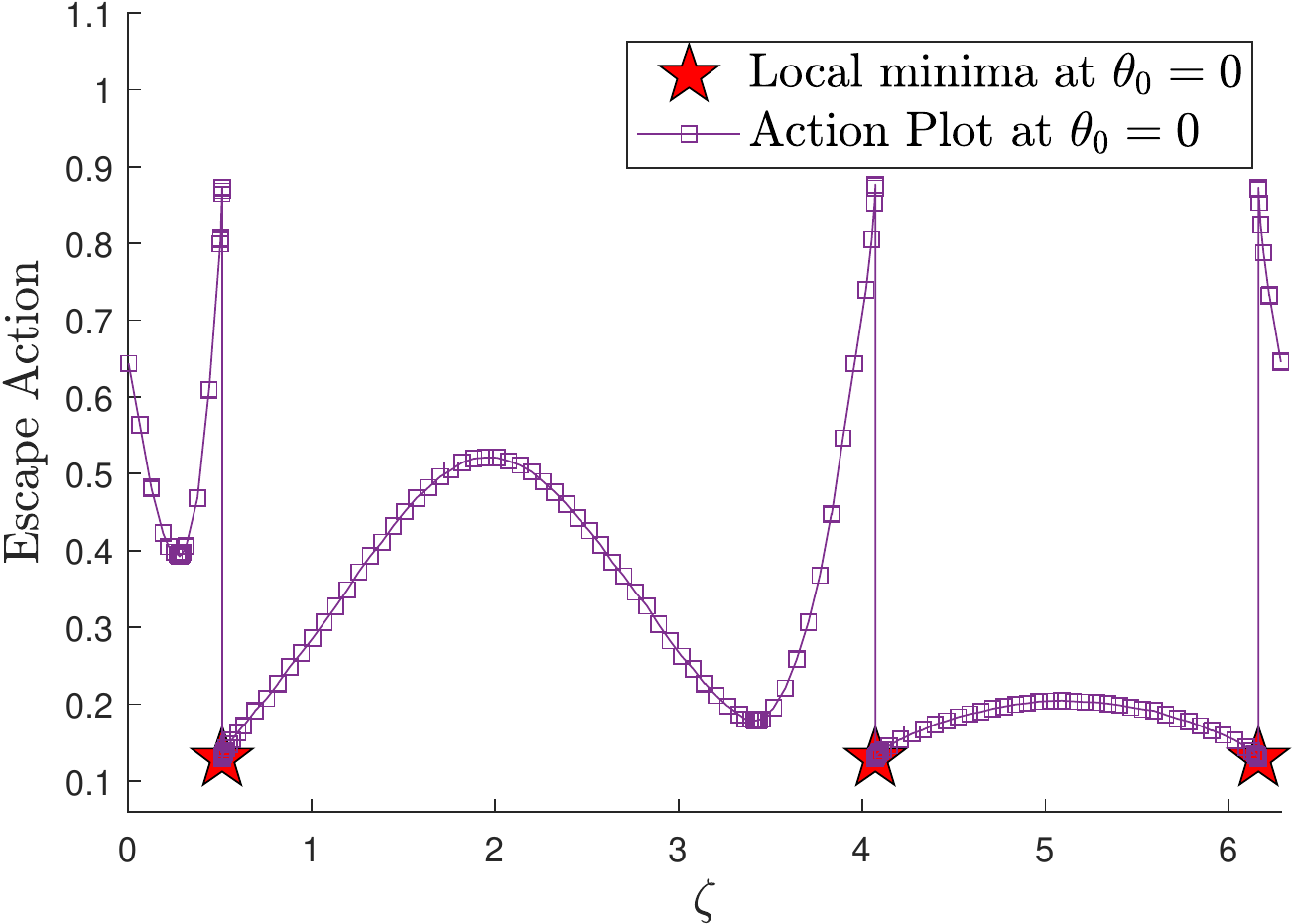}}
\caption{The cost to escape from the high amplitude attractor as a function of $\zeta$ for fixed  phase $\theta_0 = 0$. The variable $\zeta$ is the angle around a circle that completely surrounds the initial attractor at $\theta_0 $. Note that the minimizers of the cost function arise immediately next to discontinuities.}
\label{OneDuffing_ZetaEscapeAction}
\end{center}
\end{figure}

\begin{figure}[htbp]
\begin{center}
\centerline{\includegraphics[width=0.5\textwidth]{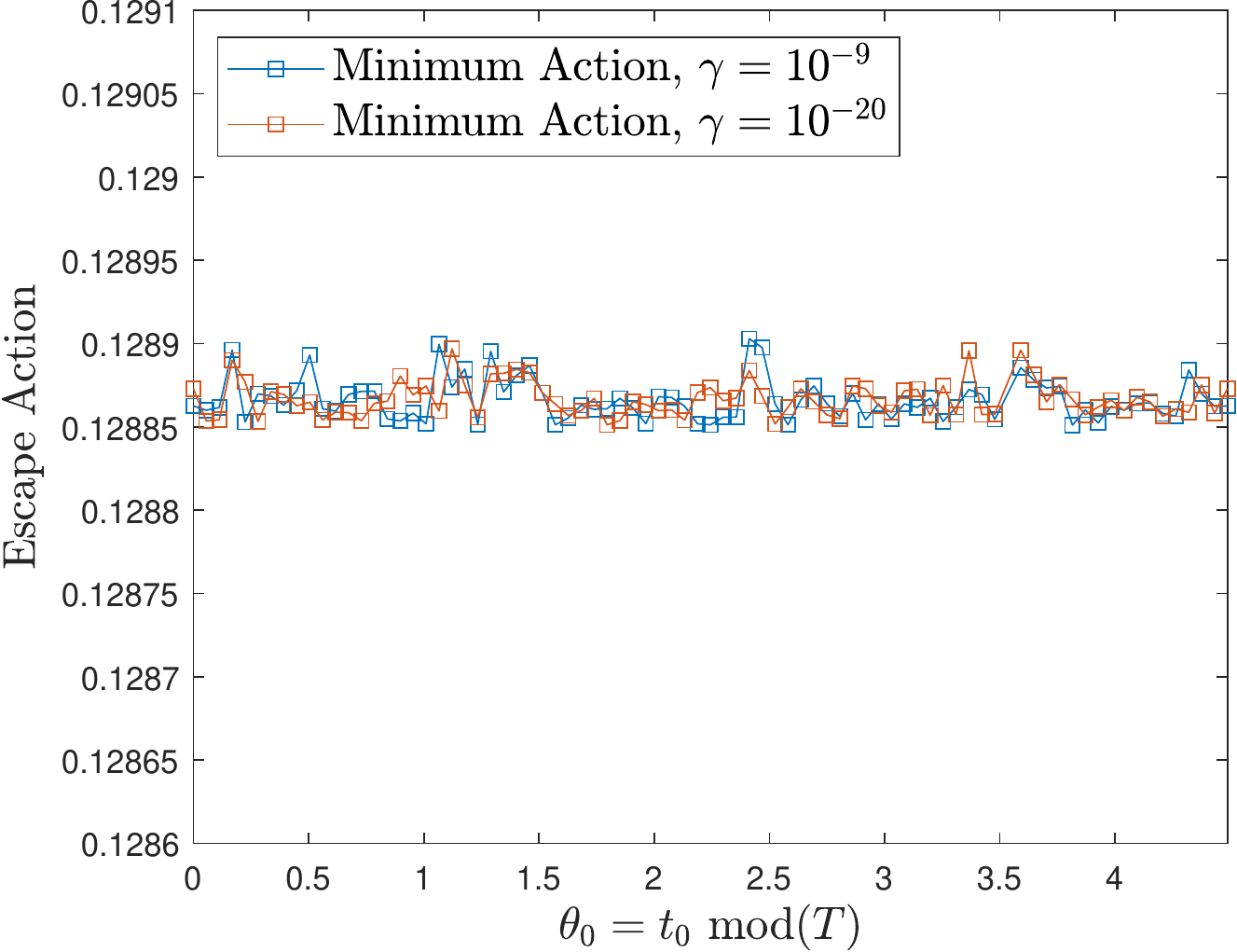}}
\caption{{The escape action from the high amplitude attractor is minimized with respect to $\zeta$ at each initial phase $\theta_0$.  Notice that with some small error tolerance, the minimum action does not change with $\theta_0$.}}
\label{OneDuffing_PhaseQuasipotential}
\end{center}
\end{figure}

The escape cost from the high amplitude attractor to the low amplitude attractor  is displayed in \fref{OneDuffing_ActionPlotWithPhase} as a function of $(\zeta,\theta_0)\in\mathbb{S}^{1}_{\gamma}\times \mathbb{S}^1_T$. The variable $\zeta$ is the angle around the  circle $ \mathbb{S}^{1}_{\gamma}$. \fref{OneDuffing_ActionPlotWithPhase} is used to elucidate the complexity of the optimization problem. Discontinuities of the cost function form  ``cliffs". The cost function has multiple local minima located at the bottom of the cliffs. Furthermore, the cost function along the curves delineating the bottom of the cliffs is nearly constant, as one can see in Figs. \fref{OneDuffing_ActionPlotWithPhase}, \ref{OneDuffing_ZetaEscapeAction}, and \ref{OneDuffing_PhaseQuasipotential} -- its variation along these curves is comparable to the numerical errors. Therefore, there are whole families of escape paths with values of cost functional matching up to numerical errors. {Hence, MPEPs are non-unique at least up to the available numerical accuracy.}

\begin{figure}[htbp]
\begin{center}
\centerline{\includegraphics[width=0.5\textwidth]{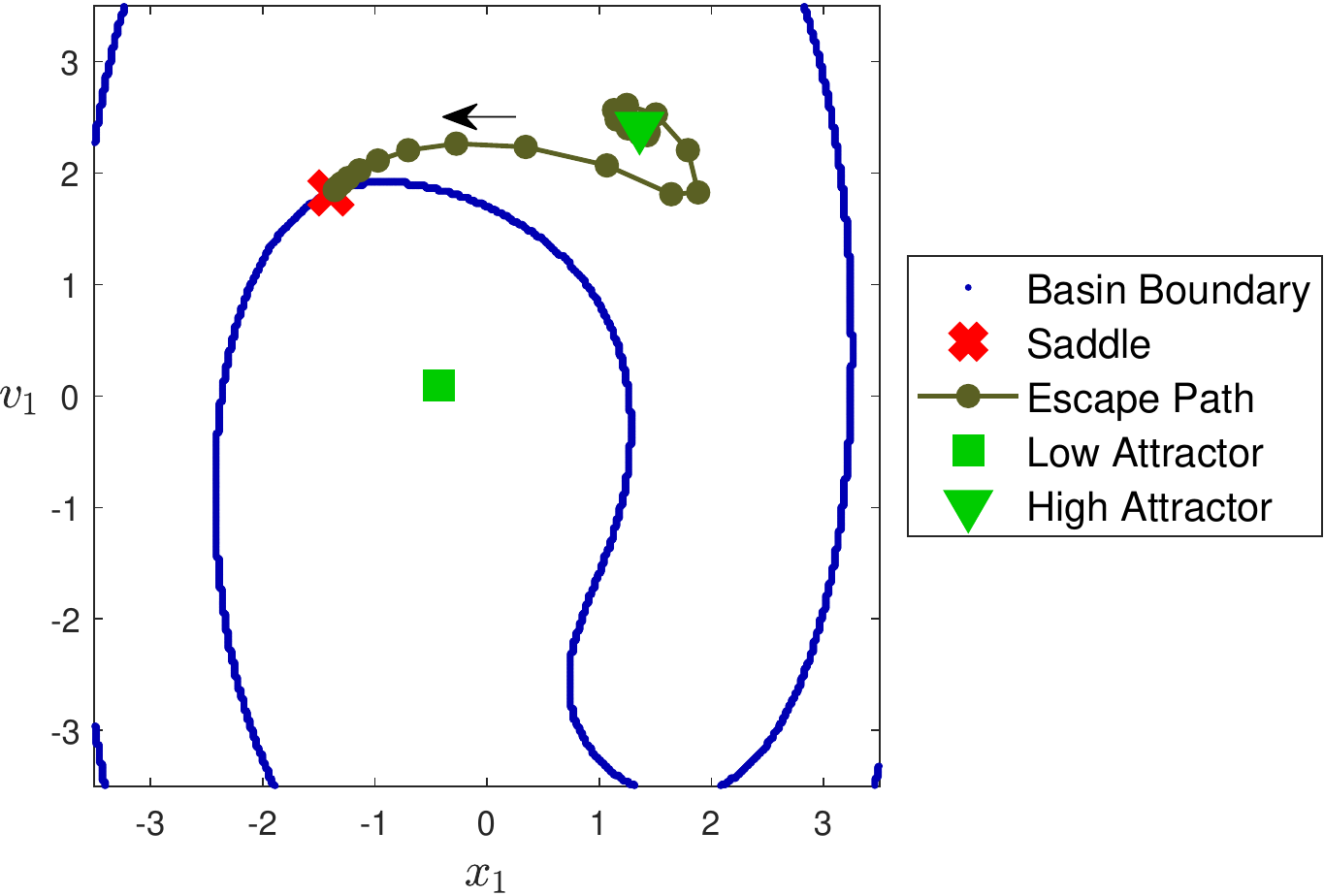}}
\caption{The Poincar\'e section of one MPEP that escapes the high amplitude attractor. This escape path begins at the attractor, rotates clockwise in the Poincar\'e section, and ends at the saddle.}
\label{OneDuffing_H_MPEP}
\end{center}
\end{figure}

The Poincar\'e section of one of the MPEPs corresponding to the global minimum of the cost functional \eqref{OneDuffing_ActionFunctional}, specifically, the one corresponding to the initial condition marked with the bottom red star in \fref{OneDuffing_ActionPlotWithPhase}, is shown in \fref{OneDuffing_H_MPEP}. Here and throughout the rest of the paper, the Poincar\'e sections all correspond to the phase angle $\theta = 0$. The solid marks are the intersections of the MPEP with the plane $\theta = 0$. The MPEP goes through one revolution between each pair of consecutive points of its Poincar\'e section. The line segments connecting the marks are added only to indicate the ordering of these intersections -- they are \emph{not} the projections of the MPEP segments onto the plane $\theta = 0$. 

\begin{figure}[htbp]
\begin{center}
\centerline{\includegraphics[width=0.5\textwidth]{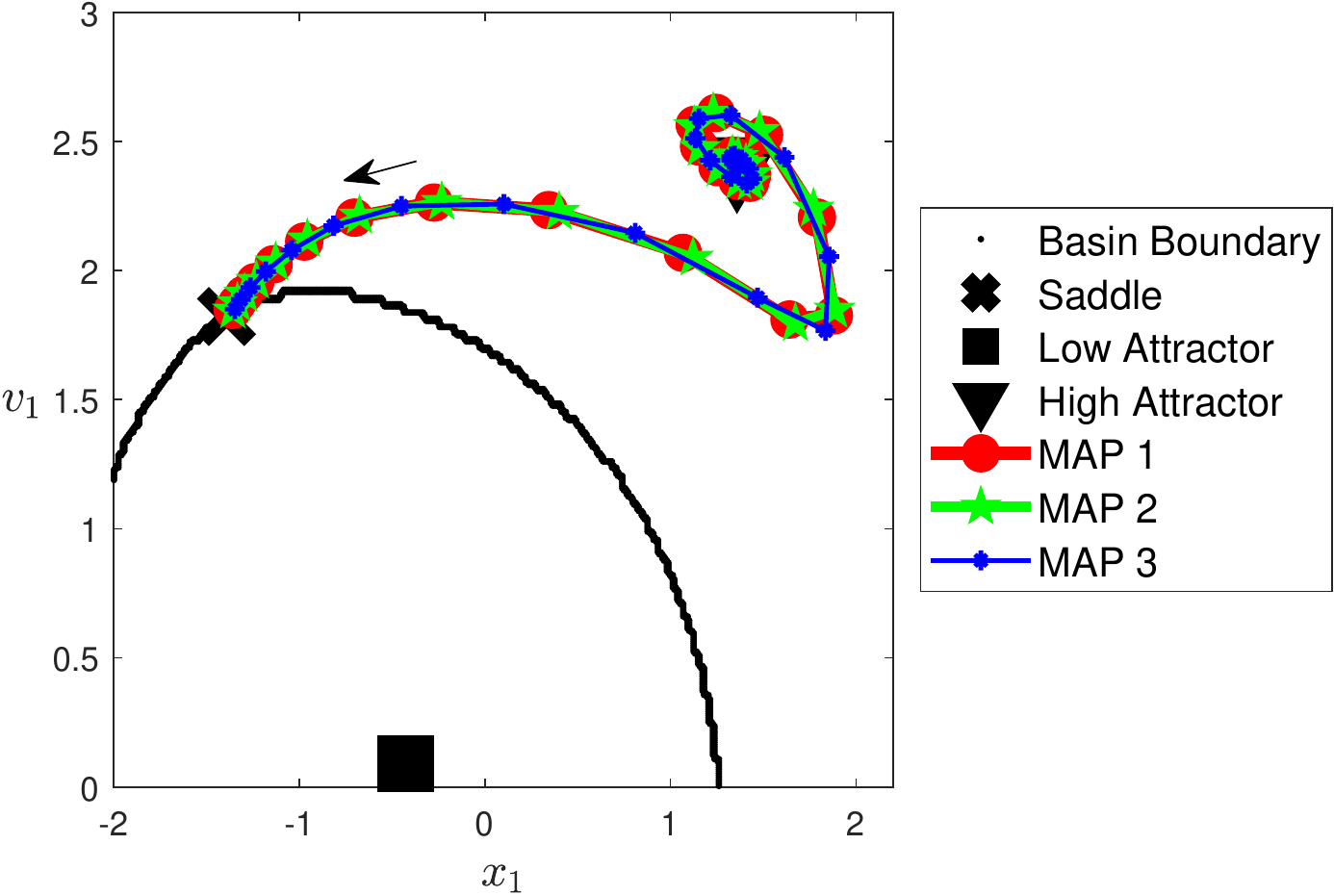}}
\caption{The Poincar\'e sections of three MPEPs, which escape the high amplitude attractor with $\theta_0 = 0$, are plotted on top of each other. Notice that these escape paths all overlap in phase space, but cross the Poincar\'e section at different times. These MPEPs and all other MPEPs of this system share the same quasipotential, share this same characteristic in phase space, and have the same probability of occurrence.}
\label{OneDuffing_H_MPEP_All}
\end{center}
\end{figure}

\begin{figure}[htbp]
\begin{center}
\centerline{\includegraphics[width=0.5\textwidth]{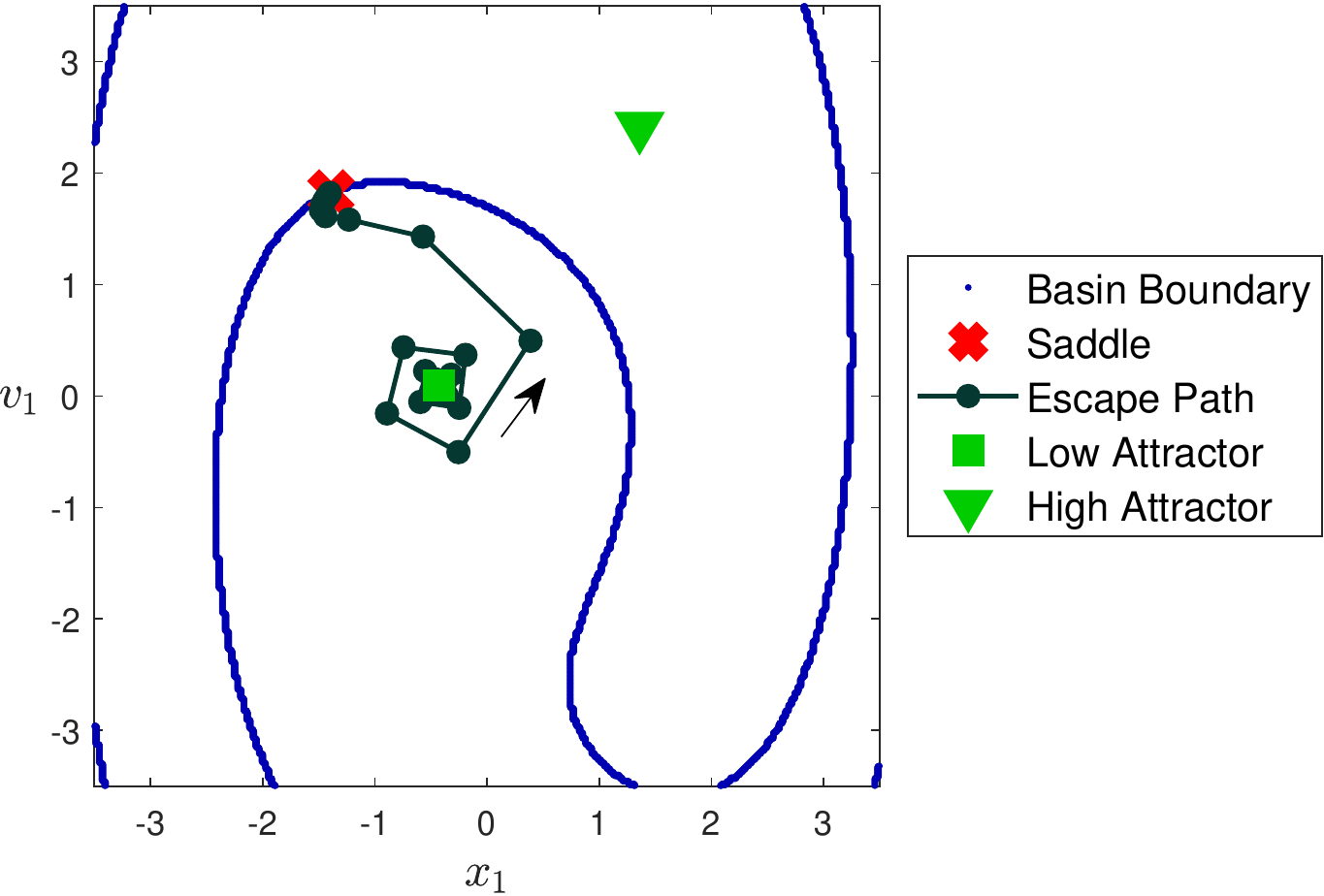}}
\caption{The Poincar\'e section of a MPEP that escapes the low amplitude attractor. This escape path begins at the attractor, rotates counter-clockwise in the Poincar\'e section, and ends at the saddle.}
\label{OneDuffing_L_MPEP}
\end{center}
\end{figure}

The Poincar\'e section of the MPEPs defined by the initial conditions corresponding to the three red stars in \fref{OneDuffing_ActionPlotWithPhase} are shown in \fref{OneDuffing_H_MPEP_All}.
These MPEPs escape from the high amplitude attractor and approach the saddle cycle lying in the basin boundary of the high and low amplitude attractors. 
The Poincar\'e section of a MPEP going from the low amplitude attractor to the saddle cycle is shown in \fref{OneDuffing_L_MPEP}. Figures \ref{OneDuffing_H_MPEP_All} and \ref{OneDuffing_L_MPEP} suggest that \emph{all  MPEPs} have infinitely long escape from the attractors and an infinitely long approach to the saddle cycles lying in the basin boundaries.

\begin{figure}[htbp]
\begin{center}
\centerline{\includegraphics[width=0.5\textwidth]{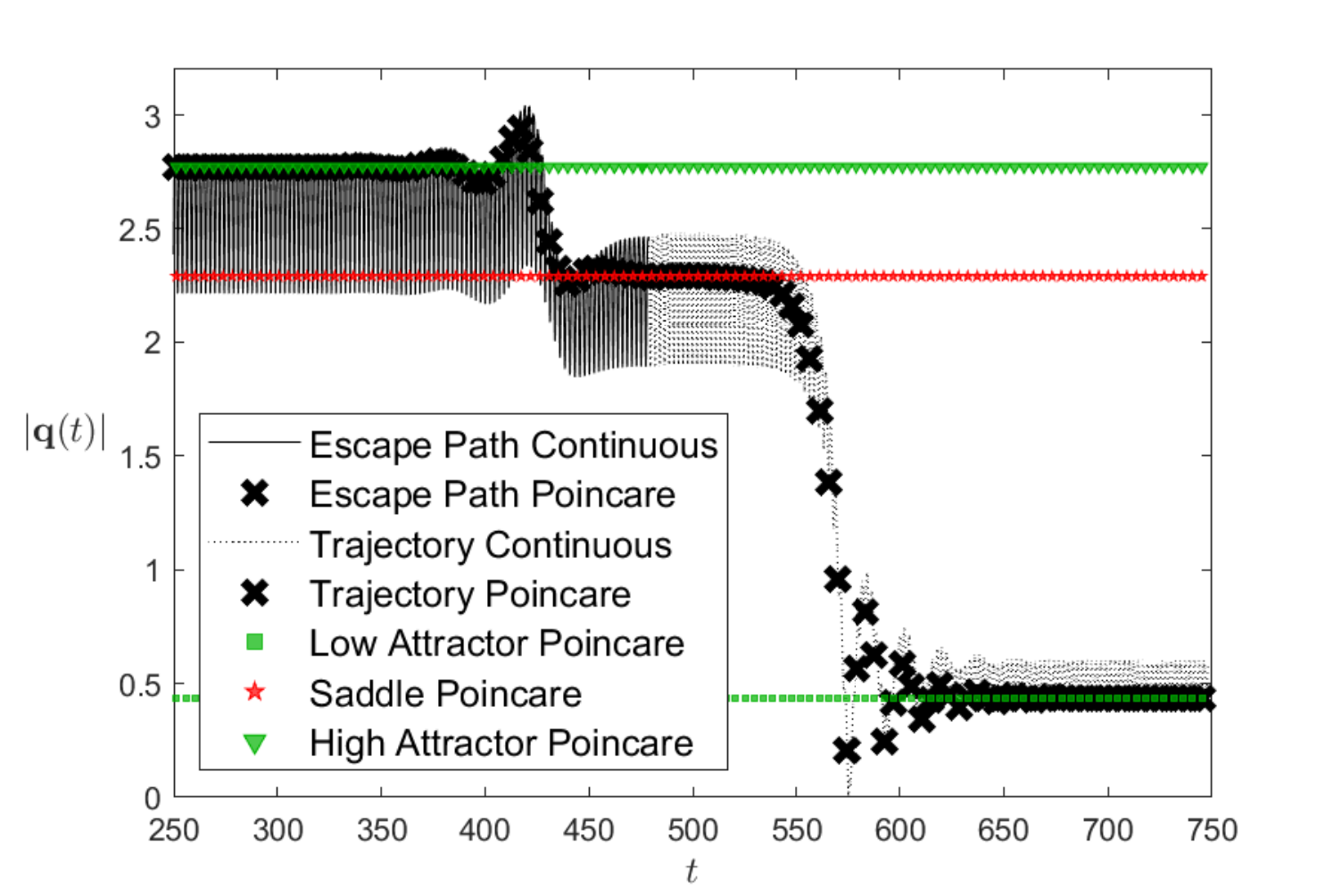}}
\caption{Euclidean norm, $|\bm{q}(t)| = \sqrt{\sum_i^{2N} q_i\left(t\right)^2}$, at all time steps of a MPEP that escapes the high amplitude attractor (solid) and of an unperturbed trajectory  that descends from the saddle to the low amplitude attractor (dotted).  }
\label{OneDuffing_H_L2MPEP}
\end{center}
\end{figure}
\begin{figure}[htbp]
\begin{center}
\centerline{\includegraphics[width=0.5\textwidth]{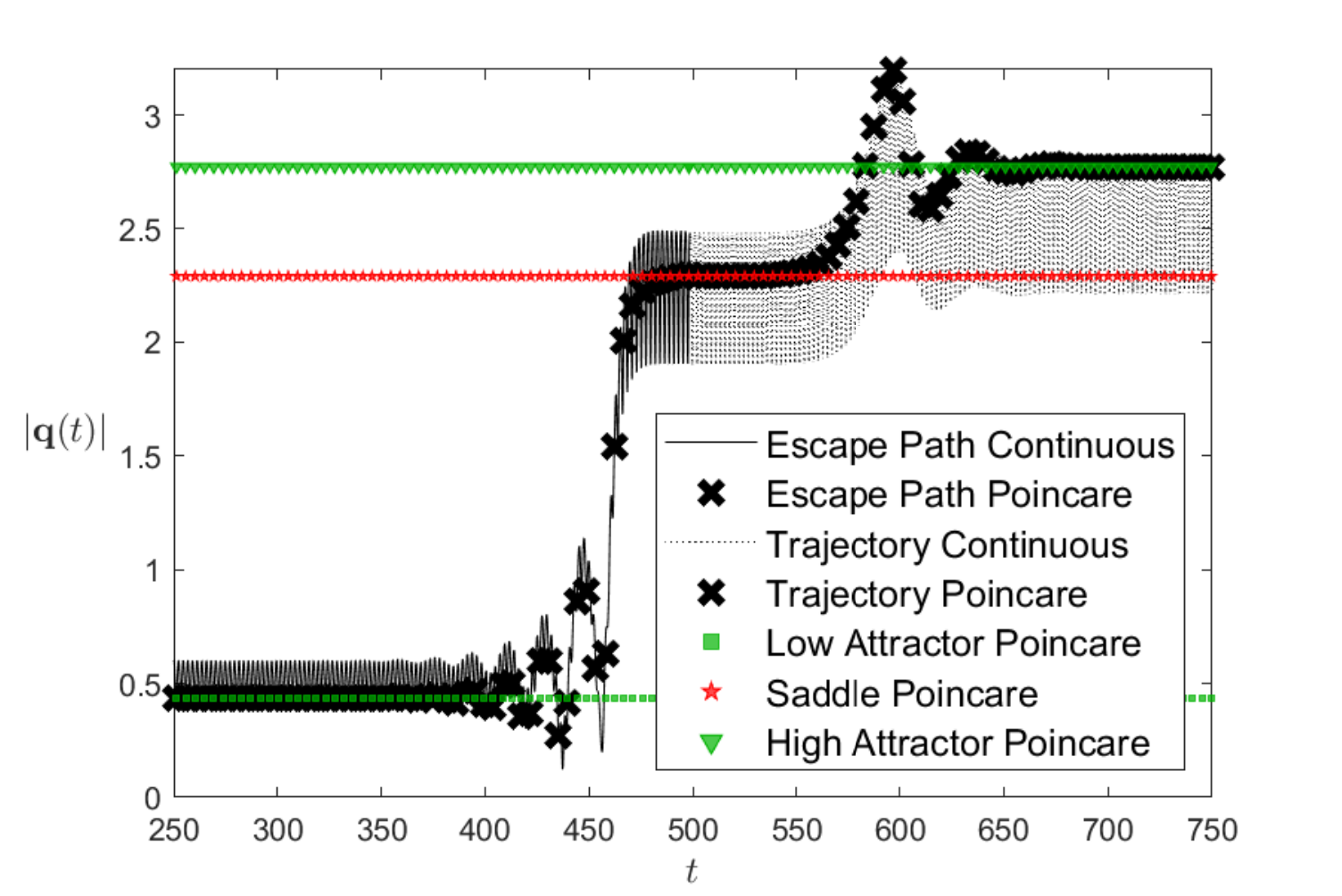}}
\caption{Euclidean norm of a MPEP that escapes the low amplitude attractor (solid) and of an unperturbed trajectory that ascends from the saddle to the high amplitude attractor (dotted).}
\label{OneDuffing_L_L2MPEP}
\end{center}
\end{figure}

\begin{figure}[htbp]
\begin{center}
\centerline{\includegraphics[width=0.5\textwidth]{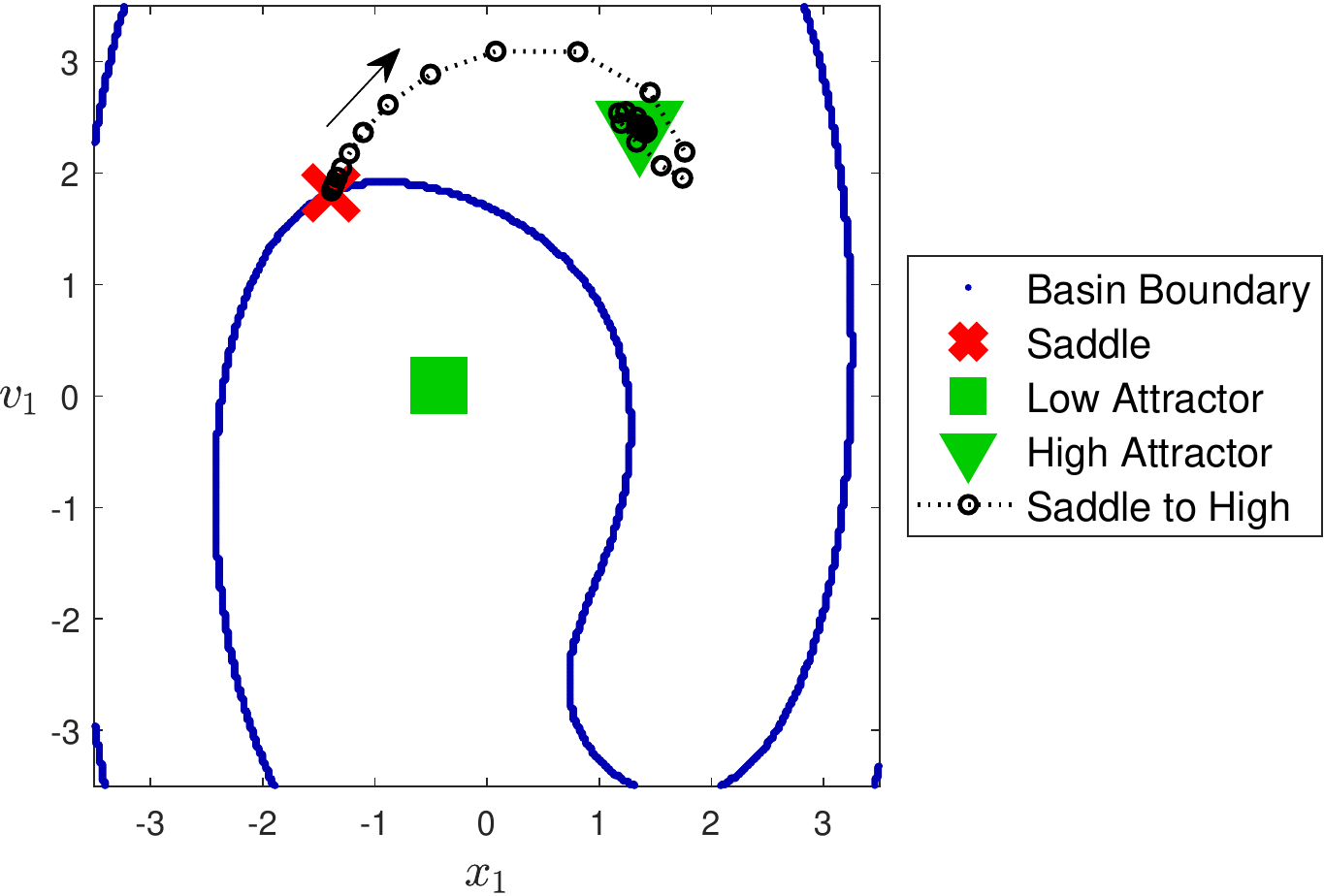}}
\caption{The Poincar\'e section of a trajectory of equation (3) for $N = 1$ that is drawn to the saddle cycle as $t\rightarrow -\infty$, is rotated clockwise in the Poincar\'e section, and is drawn to the high amplitude attractor as $t\rightarrow \infty$.  The rotation of the trajectory in phase space matches the rotation of the escape path of that same basin; this is common in lower dimensional systems such as the bistable unforced Duffing oscillator shown in Fig. 3, and is noteworthy because one could naively expect the most probable escape paths to move along the space of a trajectory but in the opposite direction, and this is not the case.}
\label{OneDuffing_H_SaddleToAttractor}
\end{center}
\end{figure}

For comparison, the Poincar\'e section of a trajectory of \eqref{eq:mainODE} going from the saddle cycle to the high amplitude attractor is visualized in \fref{OneDuffing_H_SaddleToAttractor}. Note that both Poincar\'e sections of the MPEPs from the high amplitude attractor to the saddle cycle and of the trajectory going the other way around are clockwise. 

Another way to visualize a MPEP is to plot the Euclidean norm of its $\mathbf{q}$-components versus time. 
The advantage of this technique is that it allows for visualization of escape paths in a system consisting of an arbitrary number of oscillators.
The Euclidean norm of the MPEP  (the solid line) corresponding to the bottom star in \fref{OneDuffing_ActionPlotWithPhase} is displayed in \fref{OneDuffing_H_L2MPEP}. This MPEP is concatenated with a heteroclinic trajectory (the dotted line) going from the saddle cycle to the low amplitude attractor. A similar visualization of a MPEP going from the low- to the high amplitude attractor is presented in \fref{OneDuffing_L_L2MPEP}.

\begin{figure}[htbp]
\begin{center}
\centerline{\includegraphics[width=0.7\textwidth]{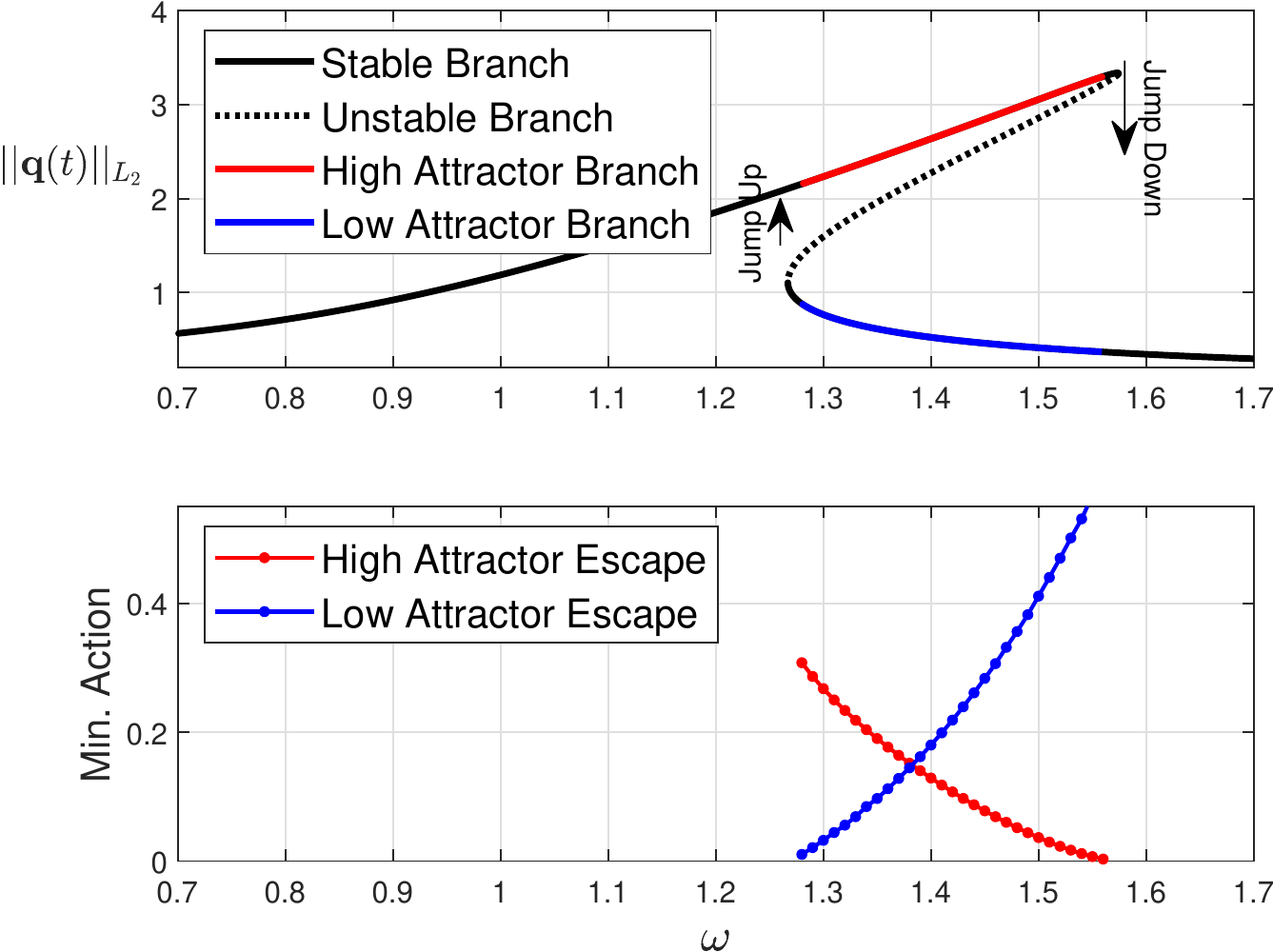}}
\caption{(top)  $L_2$ norm of the frequency response of the forced Duffing oscillator with attractor branches highlighted in color in the hysteresis region.  (bottom) The quasipotential as a function of excitation frequency; action values corresponding to escapes from the low and high amplitude attractors are shown in blue and red, respectively.}
\label{OneDuffing_FrequencyQuasipotential}
\end{center}
\end{figure}

The escape costs from the low- to high amplitude attractor and vice versa; that is, the \emph{quasipotential barriers}, are plotted as functions of the excitation frequency $\omega$ in \fref{OneDuffing_FrequencyQuasipotential}.
The quasipotential barrier to escape the low amplitude attractor approaches zero near the bifurcation point of the deterministic dynamics at $\omega \approx 1.27$. This agrees with experimental and numerical observations in literature \ccite{perkins_effects_2016,alofi_coupled_2021}, wherein noise is more likely to induce a change in stable mode from the low- to the high amplitude attractor near the jump up point of the hysteresis region. Similarly, the quasipotential barrier for the escape from the high-to low amplitude attractor decays to zero as the frequency approaches the second bifurcation point $\omega=\omega_2\approx 1.57$ at which the high amplitude attractor annihilates.
Therefore, the quasipotential barrier serves as a quantifier of the escape difficulty from the basin of the corresponding attractor. At $\omega \approx 1.3825$, the quasipotential barriers for both escapes are the same.


\section{Two Coupled Oscillators} \label{TwoDuffingResults}
The case of two coupled Duffing oscillators is interesting because this is the simplest example that illuminates the effect of coupling and sheds the light on transition mechanisms in larger oscillator arrays. Its dimensionality is high enough to require online detection of basin crossing by Hamiltonian paths, as outlined in Algorithm \ref{alg:CostFunction}. The values of the parameters $\alpha = 1, \, \beta = 0.3, \, \delta = 0.1,$  and $\, F(t) = 0.4 \text{ cos}(\omega t) $ are the same as for the single oscillator in the previous section. The coupling coefficient is set to be $\nu = 0.01$. 
\begin{figure}[htbp]
\begin{center}
\centerline{\includegraphics[width=0.5\textwidth]{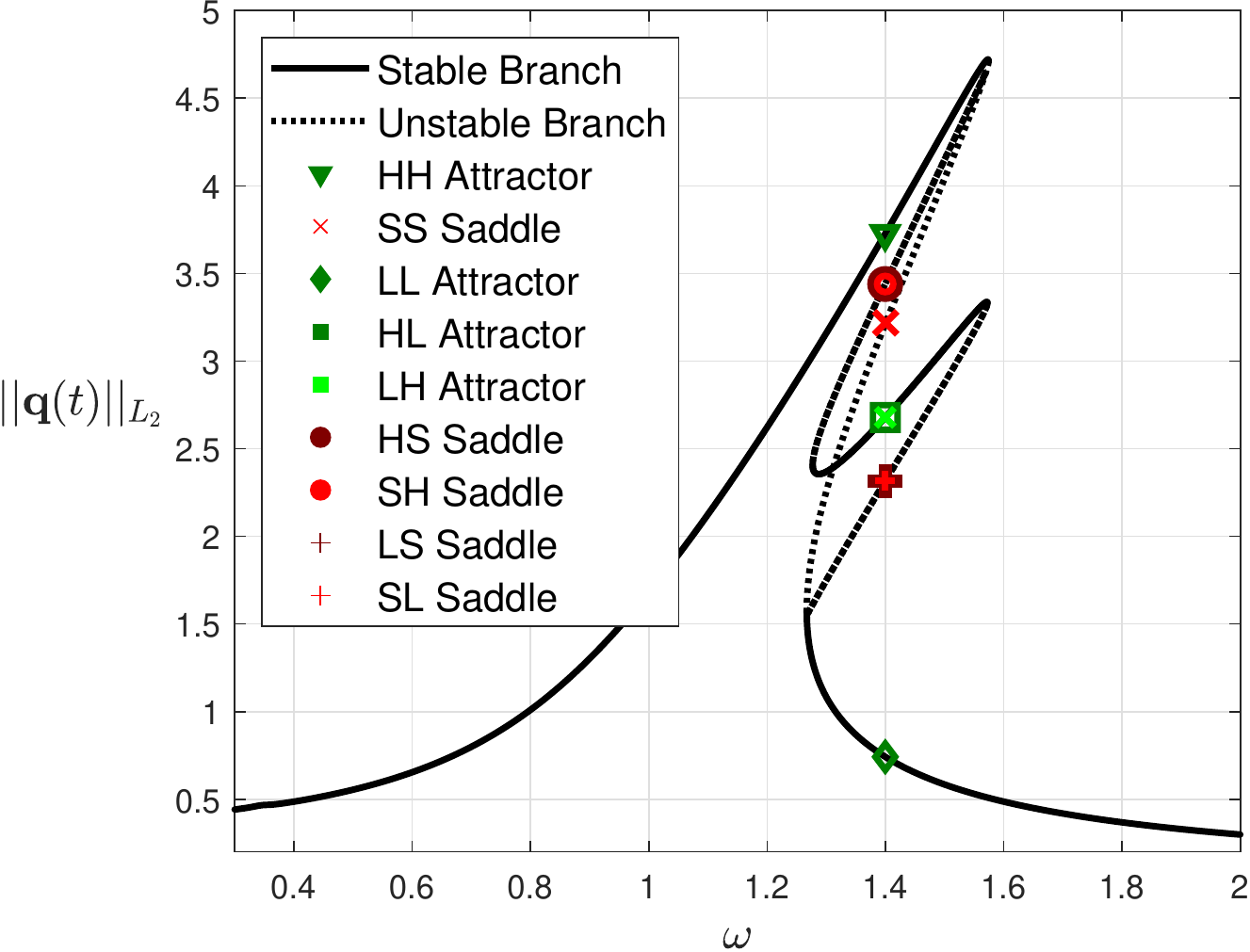}}
\caption{Frequency response of two coupled forced Duffing oscillators as a function of excitation frequency, $\omega$. Attractor and saddle locations are shown at the frequency $\omega = 1.4$. This figure was generated by using the continuation package AUTO2007 \ccite{doedel_auto-07p_2007}.}
\label{TwoDuffing_FrequencyResponse}
\end{center}
\end{figure}

With these parameter values, the deterministic system exhibits the frequency response depicted in \fref{TwoDuffing_FrequencyResponse}.  In the hysteresis region $\omega_1<\omega<\omega_2$, where, $\omega_1 \approx 1.27$ and $\omega_2 \approx 1.57$, there are $9$ solution branches. Solid and dashed curves represent stable and unstable periodic solutions, respectively. As described in Section \ref{IntroductionSection},  letters $H$ and $L$ denote states of individual oscillators moving along a high amplitude orbit and a low amplitude orbit, respectively. The letter $S$ encodes an intermediate state in which an oscillator is close to the saddle orbit of the single oscillator studied in Section \ref{OneDuffingResults}.
The highest (lowest) branch corresponds to the stable periodic solution in which both oscillators vibrate at high (low)  amplitudes, the $HH$  and $LL$ attractors, respectively. 
Two overlapping stable branches (the $HL$ and $LH$ attractors) in the middle represent two symmetric stable periodic solutions where the first oscillator (oscillator 1) has high amplitude while the second oscillator (oscillator 2) has low amplitude and vice versa.  The remaining branches correspond to unstable periodic solutions of the saddle type.

\begin{figure}[htbp]
\begin{center}
\centerline{\includegraphics[width=0.7\textwidth]{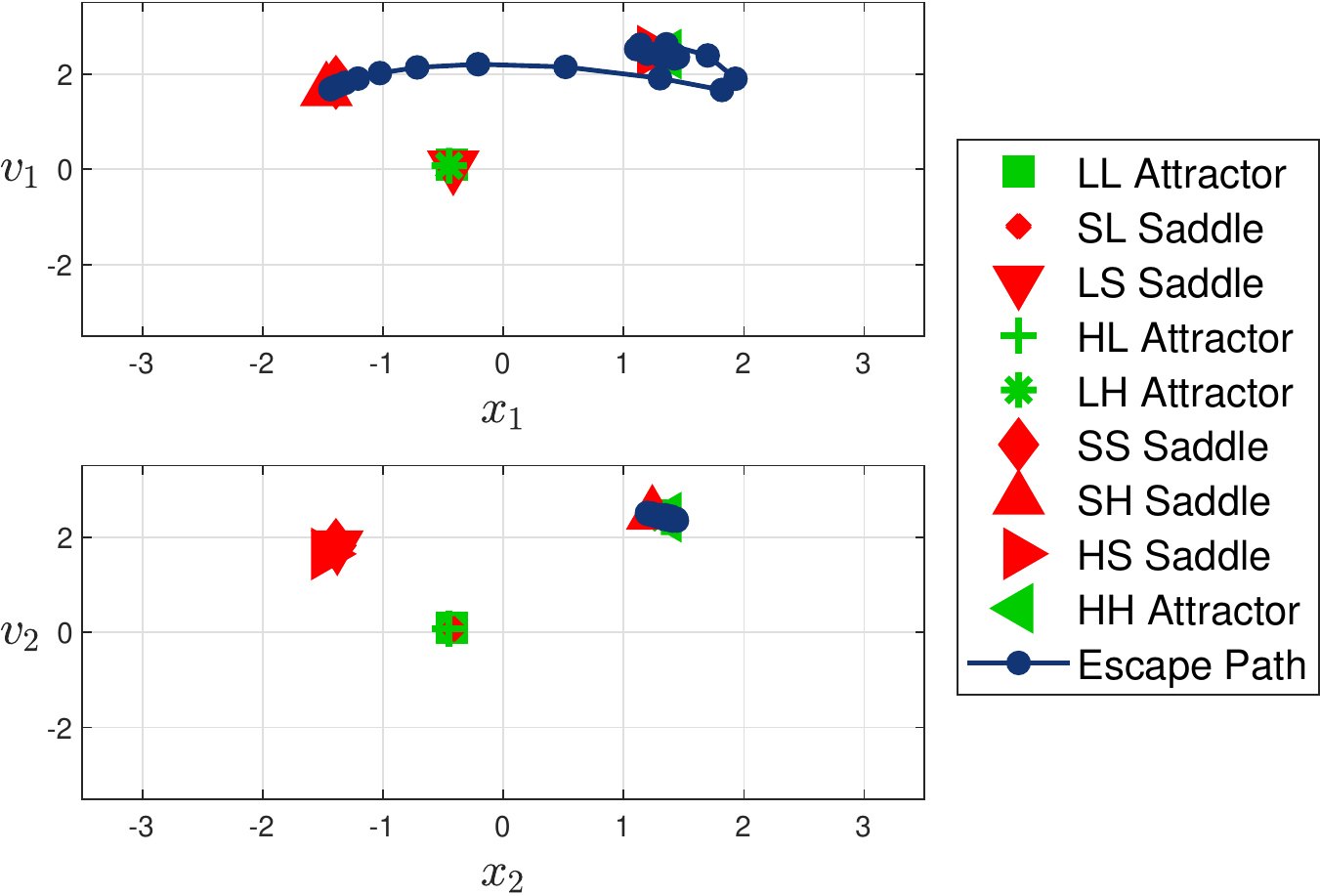}}
\caption{Hyperplanes of a Poincar\'e map of a MAP that escapes the $HH$ attractor.}
\label{TwoDuffing_HH_MPEP}
\end{center}
\end{figure}
\begin{figure}[htbp]
\begin{center}
\centerline{\includegraphics[width=0.7\textwidth]{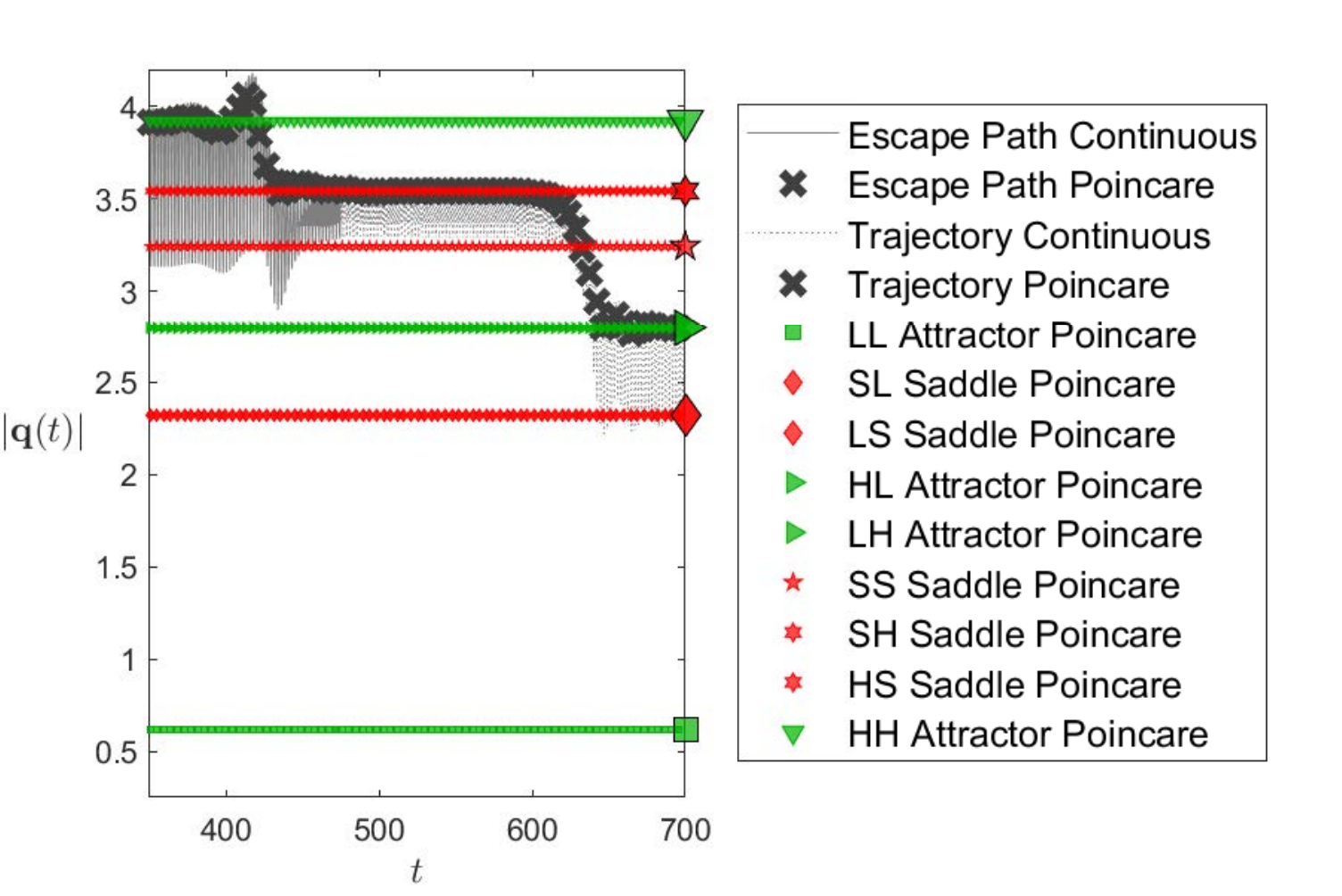}}
\caption{Euclidean norm of a MAP that escapes the $HH$ attractor (solid) and of an unperturbed trajectory that from the $SH$ saddle to the $LH$ attractor (dotted).}
\label{TwoDuffing_HH_L2MPEP}
\end{center}
\end{figure}

\begin{figure}[htbp]
\begin{center}
\centerline{\includegraphics[width=0.7\textwidth]{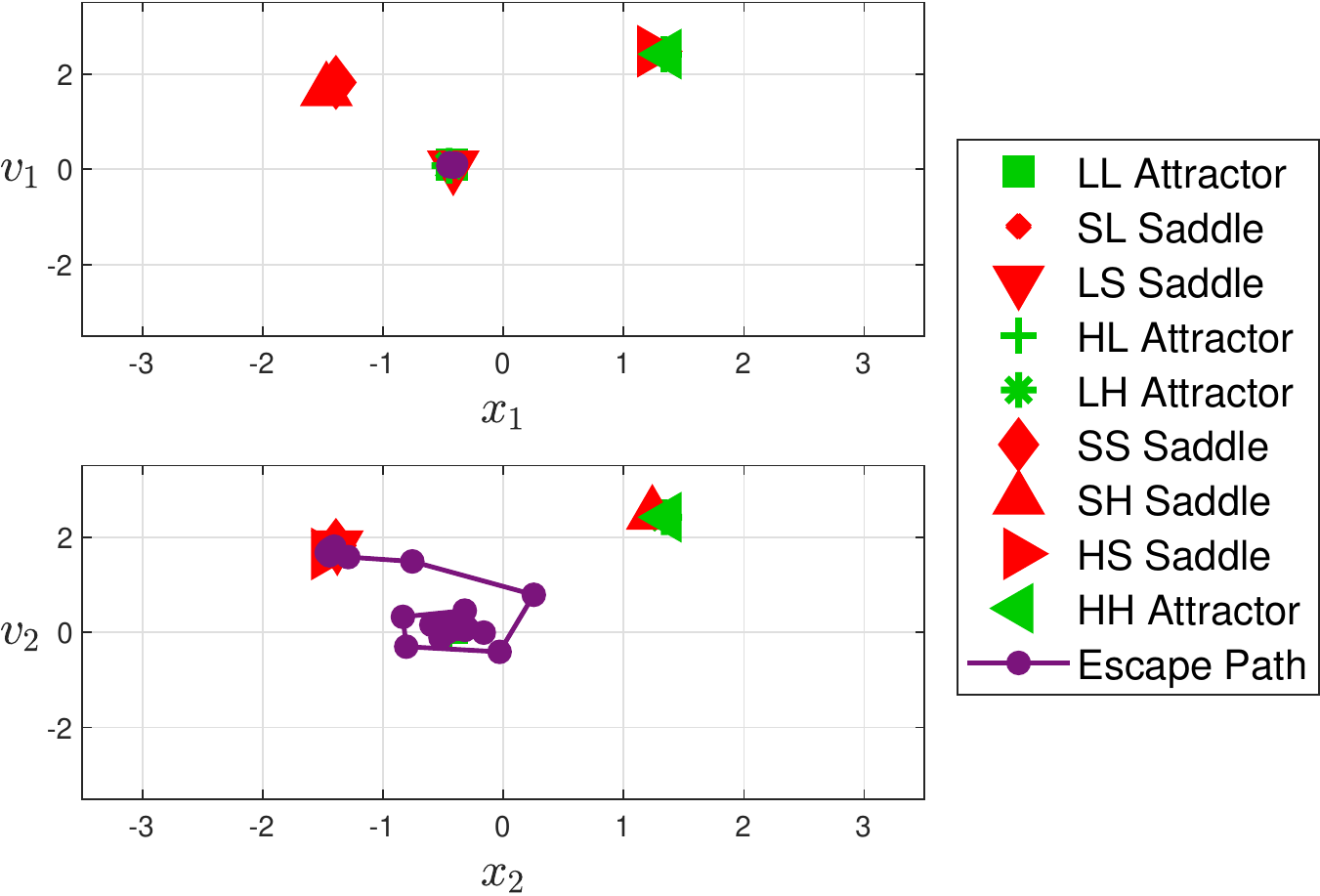}}
\caption{Hyperplanes of a Poincar\'e map of a MPEP that escapes the $LL$ attractor.}
\label{TwoDuffing_LL_MPEP}
\end{center}
\end{figure}

\begin{figure}[htbp]
\begin{center}
\centerline{\includegraphics[width=0.7\textwidth]{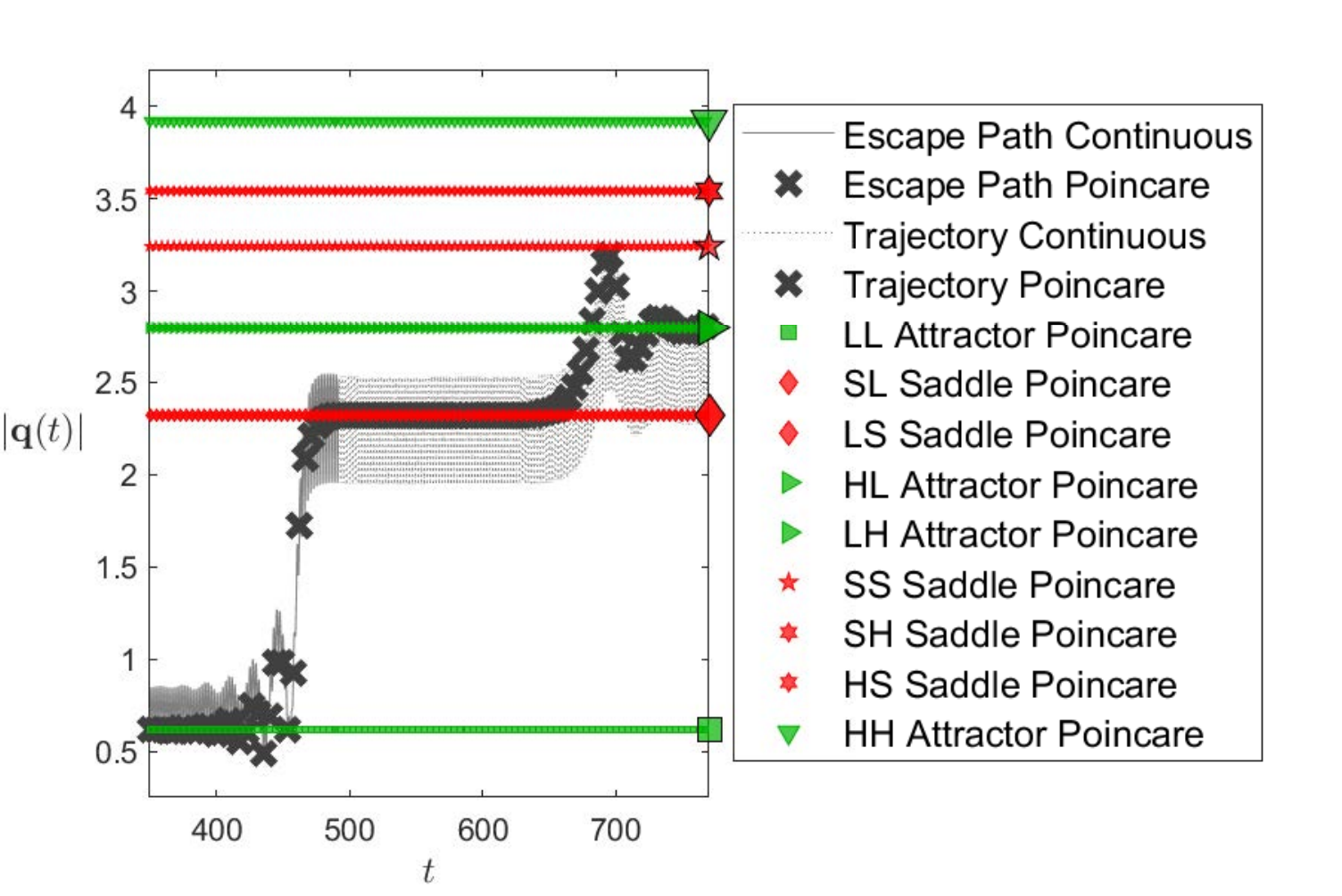}}
\caption{Euclidean norm of a MPEP that escapes the $LL$ attractor (solid) and of an unperturbed trajectory that ascends from the $LS$ saddle to the $LH$ attractor (dotted).}
\label{TwoDuffing_LL_L2MPEP}
\end{center}
\end{figure}

\begin{figure}[htbp]
\begin{center}
\centerline{\includegraphics[width=0.7\textwidth]{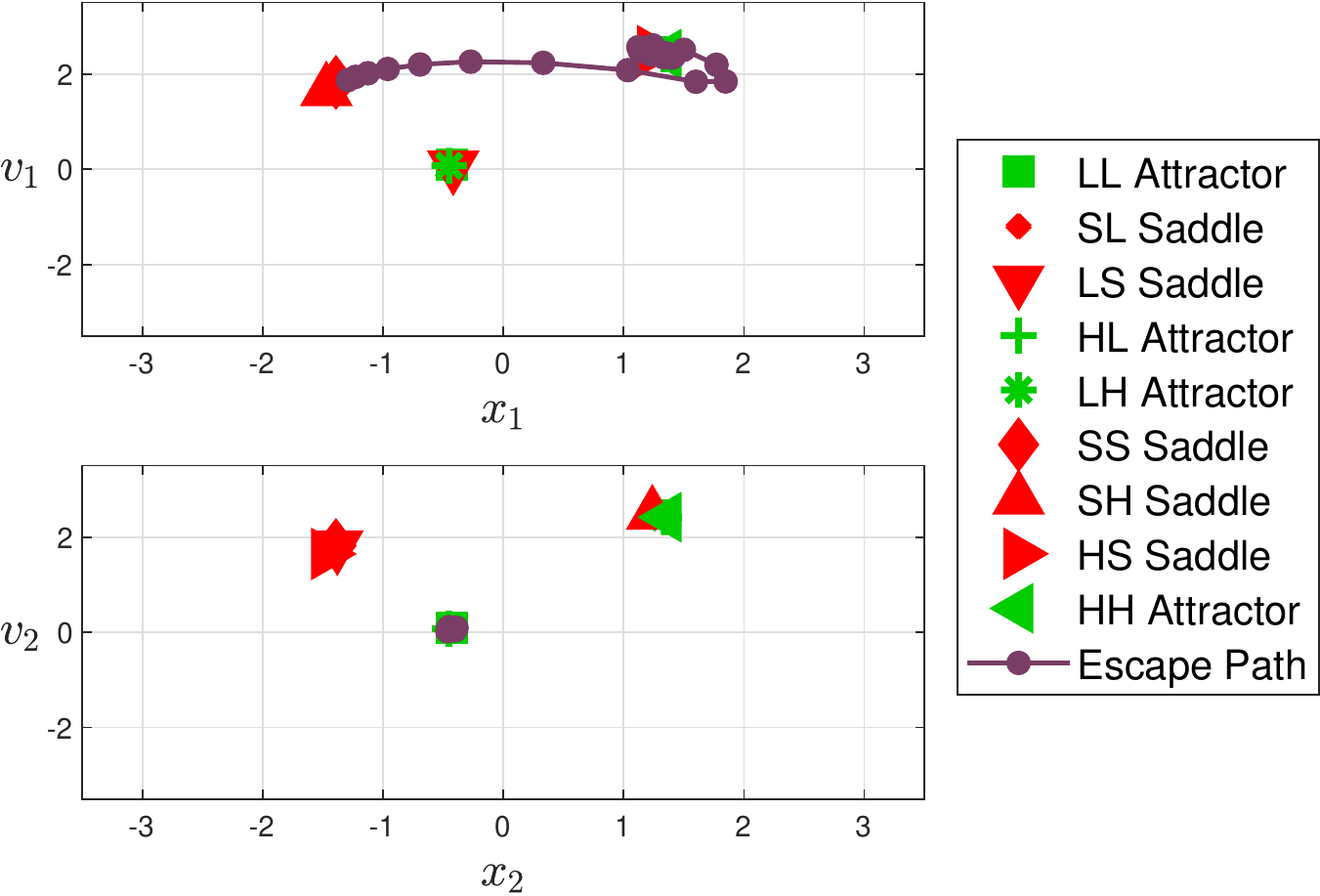}}
\caption{Hyperplanes of a Poincar\'e map of a MPEP that escapes the $HL$ attractor.}
\label{TwoDuffing_HL_MPEP}
\end{center}
\end{figure}

\begin{figure}[htbp]
\begin{center}
\centerline{\includegraphics[width=0.7\textwidth]{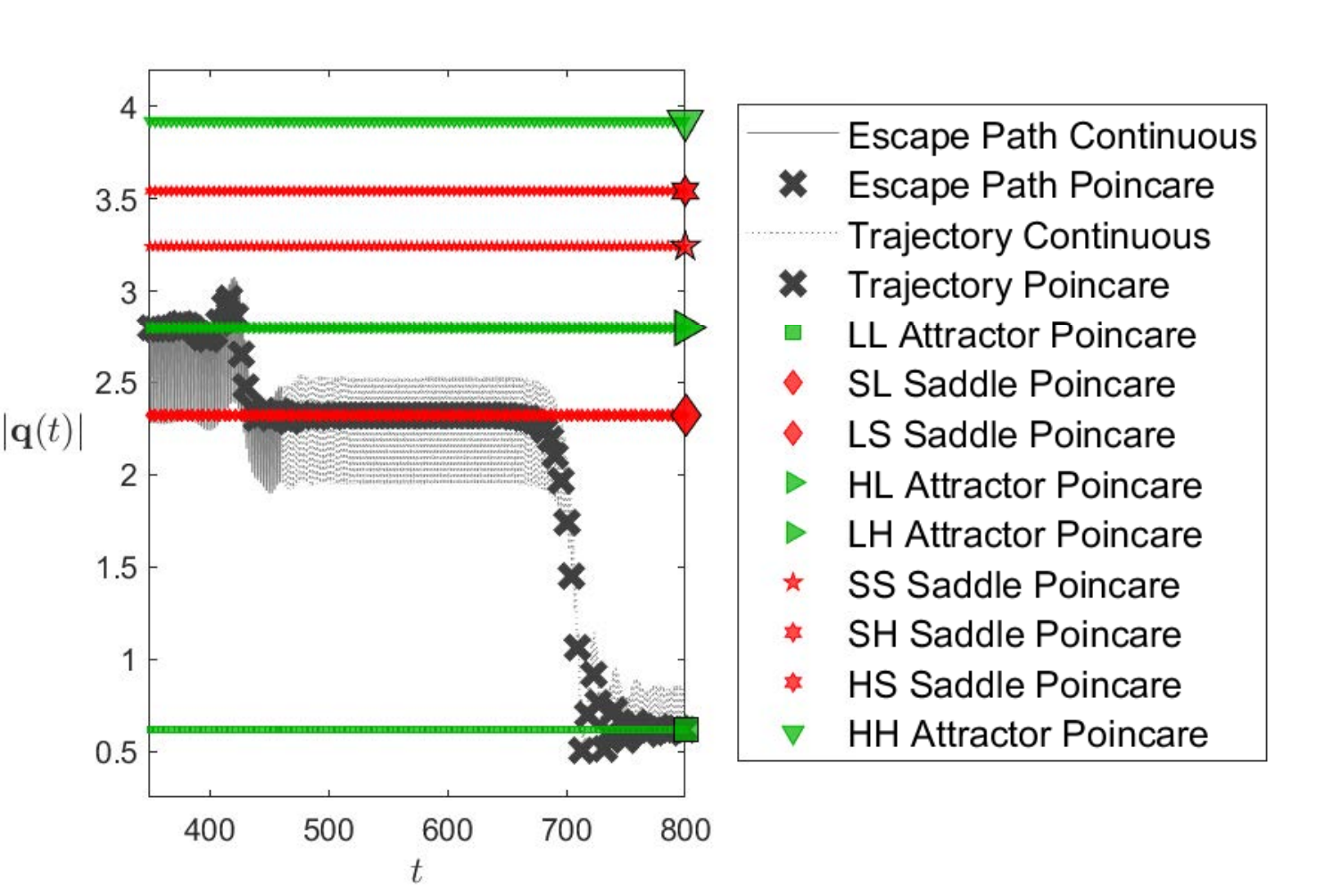}}
\caption{Euclidean norm of a MPEP that escapes the $HL$ attractor (solid) and of an unperturbed trajectory that descends from the $SL$ saddle to the $LL$ attractor (dotted).}
\label{TwoDuffing_HL_L2MPEP}
\end{center}
\end{figure}

For $N=2$ coupled oscillators, the optimization problem for the initial condition for the optimal Hamiltonian path is four-dimensional.   That is, the optimal initial condition $(\bm{q}_{\gamma}^{\ast},\theta_0^{\ast})$ is sought on $\mathbb{S}^{3}_{\gamma}\times \mathbb{S}_T^1$, the direct product of the three-dimensional sphere of radius $\gamma$ embedded into $\mathbb{R}^4$ and the circle of circumference $T$. This optimization problem has been solved to find optimal escape paths from the attractors $HH$, $LL$, and $HL$. Due to symmetry, the solution for the escape from the $LH$ attractor is found for the one for the $HL$ attractor by switching the oscillators. A collection of optimal paths visualized in Figs. \ref{TwoDuffing_HH_MPEP}--\ref{TwoDuffing_HL_L2MPEP} corresponds to the excitation frequency $\omega = 1.4$.

The escape path from the $HH$ attractor is displayed in
Fig. \ref{TwoDuffing_HH_MPEP} by plotting the Poincar\'e sections of its projections onto $(x_1,v_1)$- and $(x_2,v_2)$-planes. After escaping from the basin of the $HH$ attractor along this path, the system arrives at the $SH$ saddle. From there, it can descend either to the $LH$ attractor or back to the $HH$ attractor. \fref{TwoDuffing_HH_MPEP} shows that the first oscillator follows a path similar to the escape path from the high- to low amplitude attractor of a single oscillator in \fref{OneDuffing_H_MPEP_All}, while the motion of the second oscillator experiences a relatively small change throughout this transition. Note the closeness of the projections of the $HH$  and $LH$ attractors onto the $(x_2,v_2)$-plane. The $L_2$-norm of the $\bm{q}(t)$-component of the transition path from the $HH$  to the $LH$ attractor via the $SH$ saddle is plotted in \fref{TwoDuffing_HH_L2MPEP}.

The escape path from the $LL$ attractor  to the $LS$ saddle lying on the boundary separating the $LL$  and $LH$ basins is shown in Figs. \ref{TwoDuffing_LL_MPEP} and \ref{TwoDuffing_LL_L2MPEP}. The path from $LL$ to $SL$ can be found using symmetry.

The escape path from the $HL$ attractor shown in
Figs. \ref{TwoDuffing_HL_MPEP} and \ref{TwoDuffing_HL_L2MPEP} leads to the $SL$ saddle. It has been found that the quasipotential of the $SL$ saddle with respect to the $HL$ attractor is lower than that of the $HS$ saddle. This means that the escape from $HL$ via $HS$ is exponentially less likely than the escape via $SL$.
Furthermore, no direct transition between  the localized modes $HL$ and $LH$ has been found. This suggests that a transition between $HL$ and $LH$ can only be accomplished in the weak noise limit by visiting the $LL$  or the $HH$ mode in-between.


\begin{figure}[htbp]
\begin{center}
\centerline{\includegraphics[width=0.7\textwidth]{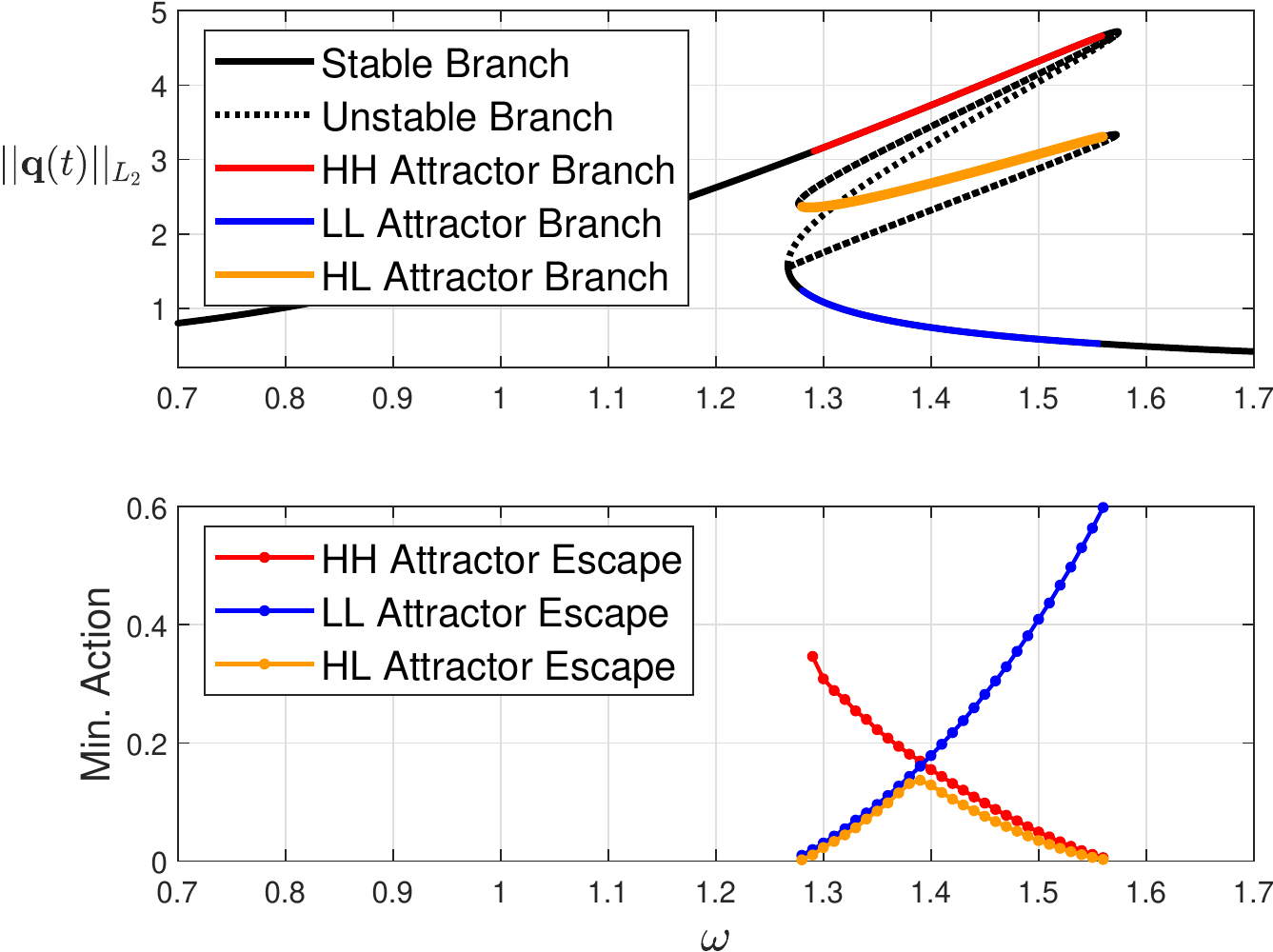}}
\caption{(top) $L_2$ norm of the frequency response of the $N=2$ system with attractor branches highlighted in color in the hysteresis region.  (bottom) The quasipotential as a function of excitation frequency; action values corresponding to escapes from the $LL$, $HL$, $HH$ attractors are shown in blue, yellow, and red respectively.}
\label{TwoDuffing_FrequencyQuasipotential}
\end{center}
\end{figure}

The MPEPs and the corresponding quasipotential barriers have been computed for a set of values of the excitation frequency $\omega$ spanning the hysteresis region $[\omega_1,\omega_2]$.
The quasipotential barriers between each attractor and the corresponding lowest saddle are plotted as functions of $\omega$ in \fref{TwoDuffing_FrequencyQuasipotential}. It is instructive to compare this figure with the similar figure for the single oscillator (\fref{OneDuffing_FrequencyQuasipotential}). It is evident that in the case of weak coupling considered here, the transitions to and from the localized modes in the two-oscillator system are well approximated by the transitions in the single oscillator. The values of the quasipotential barriers for the escapes from the $LL$ , $HL$ , and $HH$ attractors for $\omega = 1.4$ and the coupling coefficient $\nu=0.01$ are given in Table \ref{tab:quasipotentialvalues}.

Furthermore, the following observation can be made: 

\begin{itemize}
    \item Escape from the $LL$ mode requires smaller minimum action near the jump up frequency and larger minimum action near the jump down frequency.
    \item Escape from the $HH$ mode requires larger minimum action near the jump up frequency and goes to zero near the jump down frequency.

 \item The $\Lambda$-shape of the orange curve in \fref{TwoDuffing_FrequencyQuasipotential} is caused by the fact that there are two possible escapes from each localized mode: to $HH$ and to $LL$. For $\omega<1.3825$, the required minimum action for the escape from $HL$ is close but slightly less than the quasipotential barrier for the escape from $LL$. For $\omega > 1.3825$, the  required minimum action for the escape from $HL$ is close but slightly less than the quasipotential barrier for the escape from $HH$. This reduction in quasipotential is due to the coupling: the optimal escape paths for $N=2$ are sought in a larger space than for $N=1$. Hence the quasipotential barriers for $N=2$ can only decrease in comparison to those for the single oscillator. 

\end{itemize}

\section{Circular Oscillator Arrays} \label{NDuffingResults}

\begin{figure}[htbp]
\begin{center}
\centerline{\includegraphics[width=0.5\textwidth]{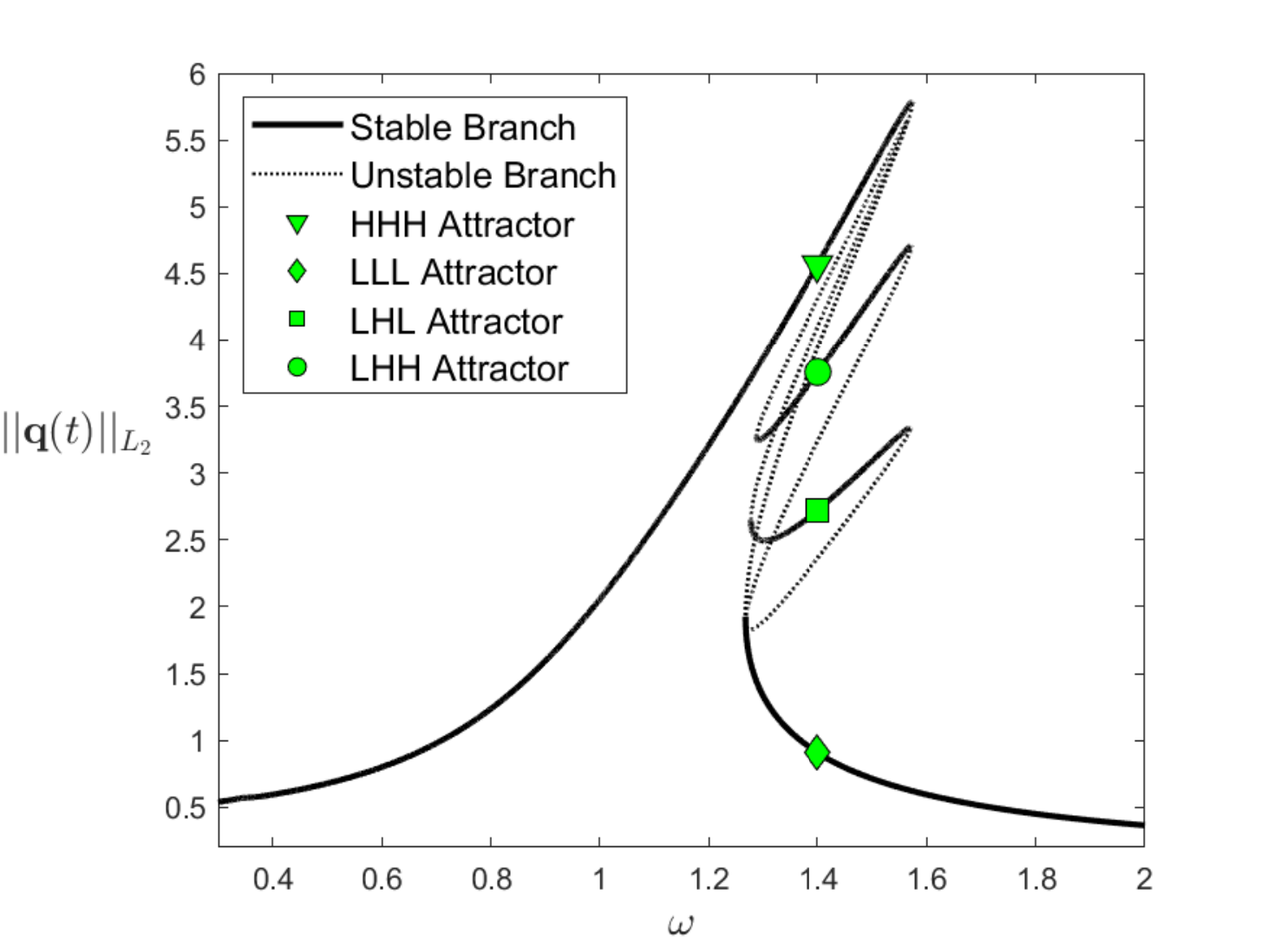}}
\caption{Frequency response of a circular array of $N=3$ coupled forced Duffing oscillators as a function of excitation frequency, $\omega$. Unique attractors are shown at the frequency $\omega = 1.4$. This figure was generated by using the continuation package AUTO2007 \ccite{doedel_auto-07p_2007}.}
\label{ThreeDuffing_FrequencyResponse}
\end{center}
\end{figure}

\begin{figure}[htbp]
\begin{center}
\centerline{\includegraphics[width=0.7\textwidth]{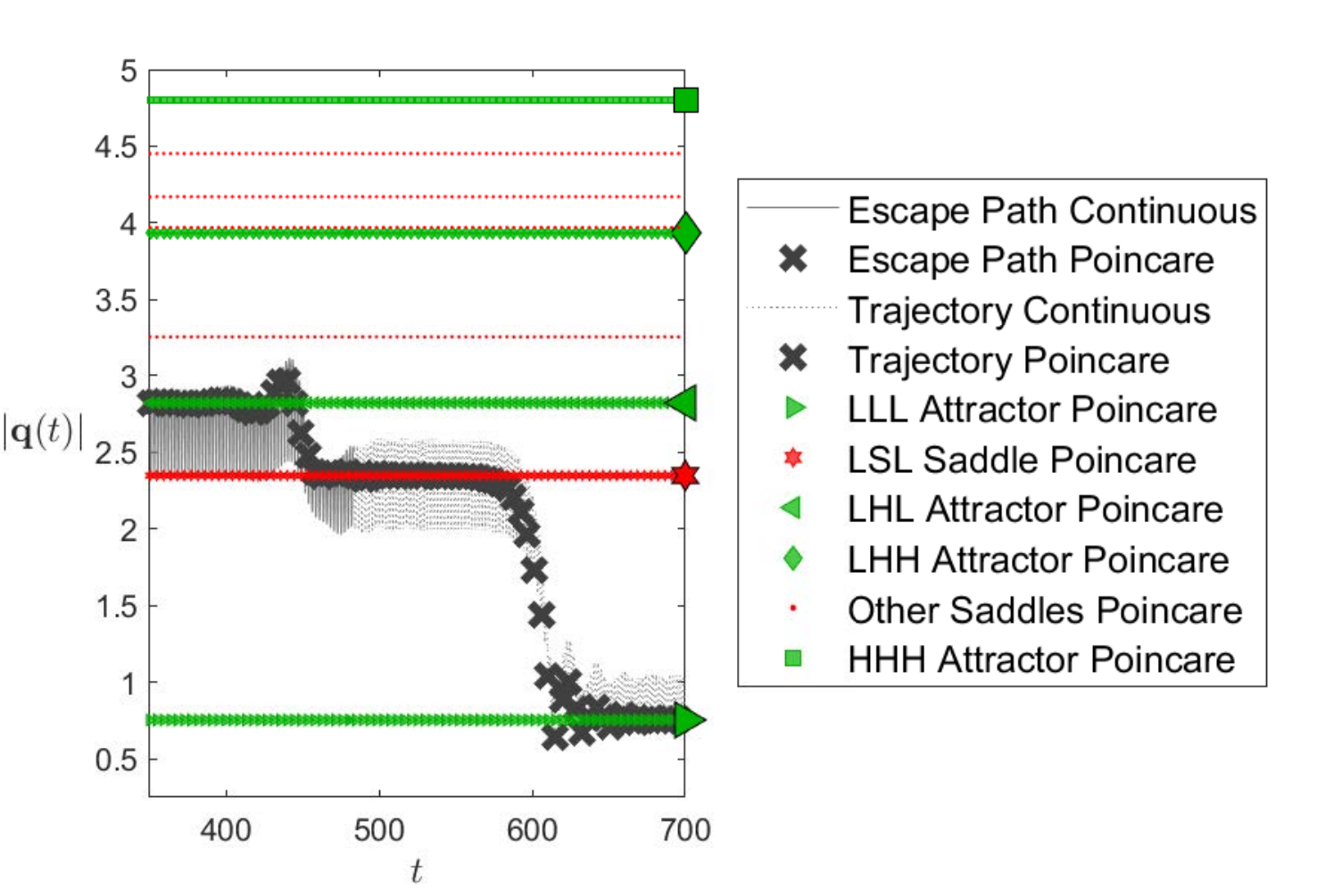}}
\caption{Euclidean norm of a MPEP that escapes the $LHL$ attractor (solid) and of an unperturbed trajectory that descends from the $LSL$ saddle to the $LLL$ attractor (dotted).}
\label{ThreeDuffing_LHL_L2MPEP}
\end{center}
\end{figure}
\begin{figure}[htbp]
\begin{center}
\centerline{\includegraphics[width=0.7\textwidth]{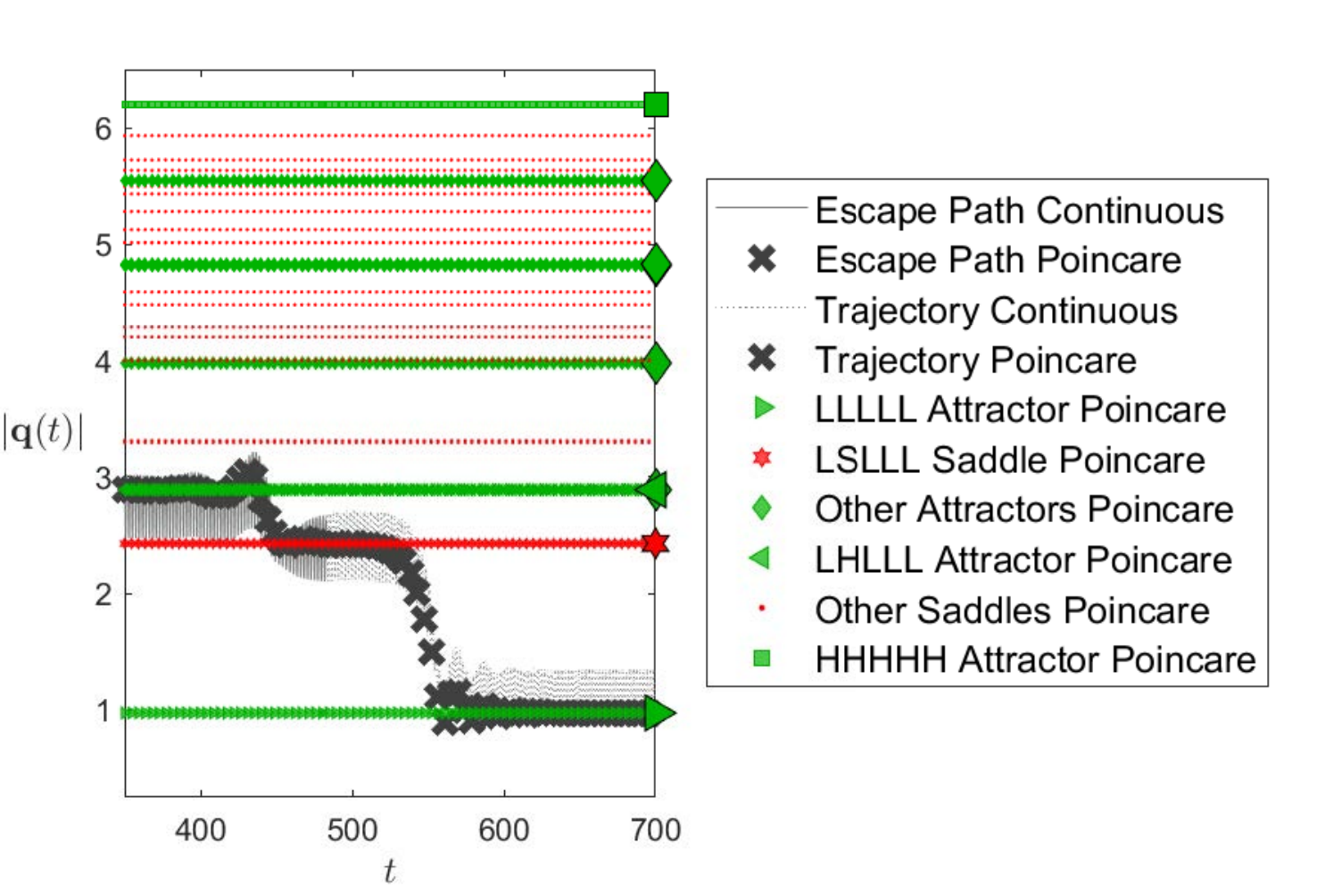}}
\caption{Euclidean norm of a MPEP that escapes the $LHLLL$ attractor (solid) and of an unperturbed trajectory that descends from the $LSLLL$ saddle to the $LLLLL$ attractor (dotted).}
\label{FiveDuffing_LHLLL_L2MPEP}
\end{center}
\end{figure}

The application to circular three- and five-oscillator arrays presented in this section demonstrates the viability of the proposed methodology for high-dimensional systems and sheds light of noise-induced transitions in larger circular arrays. 

The systems under consideration are governed by SDE \eqref{eq:mainSDE} with $N=3$ and $N=5$, i.e., with three and five oscillators, respectively. They are models of vibrations of blades of turbomachinery or cyclically arranged VEHs. The parameter values are the same as for $N=2$. The frequency responses of $N=3$ and $N=5$ systems are shown, respectively, in Figs. \ref{ThreeDuffing_FrequencyResponse} and \ref{FiveDuffing_FrequencyResponse}. Provided that the coupling is weak, the number of stable vibrating modes grows dramatically with $N$. For $N=3$, this number is at least 8, while for $N=5$, it is at least 32, as mentioned in the introduction. Two of the stable modes correspond to all oscillators vibrating at low or high amplitudes, while the rest are localized modes in which some oscillators have high amplitudes while others have low ones. Each localized mode has a number of equivalent localized modes obtained by circular shift and/or changing direction of rotation of oscillator indexing. Moreover, nonequivalent localized modes with equal numbers of high amplitude oscillators have very close time-averaged $L_2$-norms of their $\bm{q}$-components \eqref{L2NormDefinition} resulting in indistinguishable points in the frequency-response diagram. The authors have chosen to focus here on the study of escapes from localized modes for $N=3$ and $N=5$ in which one oscillator has high amplitude while the other $N-1$ oscillators have low amplitudes. The escapes from these modes are the most relevant to noise influenced turbomachinery.

The most probable escape paths from the basins of the $LHL$ and $LHLLL$ attractors of the $N=3$ and $N=5$ systems at frequency $\omega = 1.4$ and coupling coefficient $\nu = 0.01$ are shown in Figs. \ref{ThreeDuffing_LHL_L2MPEP} and \ref{FiveDuffing_LHLLL_L2MPEP}, respectively. After these escapes, the systems land onto the attractors in which all oscillators have low amplitudes. These escape paths are similar to the MPEP leaving the localized mode of the $N=2$ system: the path of the oscillator originally at high amplitude resembles the escape path of the single oscillator, $(N=1)$, from its high attractor, while the motions of the other oscillators are only slightly affected. The quasipotential barriers for these escapes, $U_{LHL}$ and $U_{LHLLL}$ are slightly smaller than the quasipotential barrier $U_H$ for the escape of the single oscillator from its $H$ attractor basin. Furthermore, as expected, the following trend is observed: 
$$
U_{H}>U_{HL}>U_{LHL}\approx U_{LHLLL}.
$$
The values of these barriers are given in Table \ref{tab:quasipotentialvalues}.

\begin{figure}[htbp]
\begin{center}
\centerline{\includegraphics[width=0.7\textwidth]{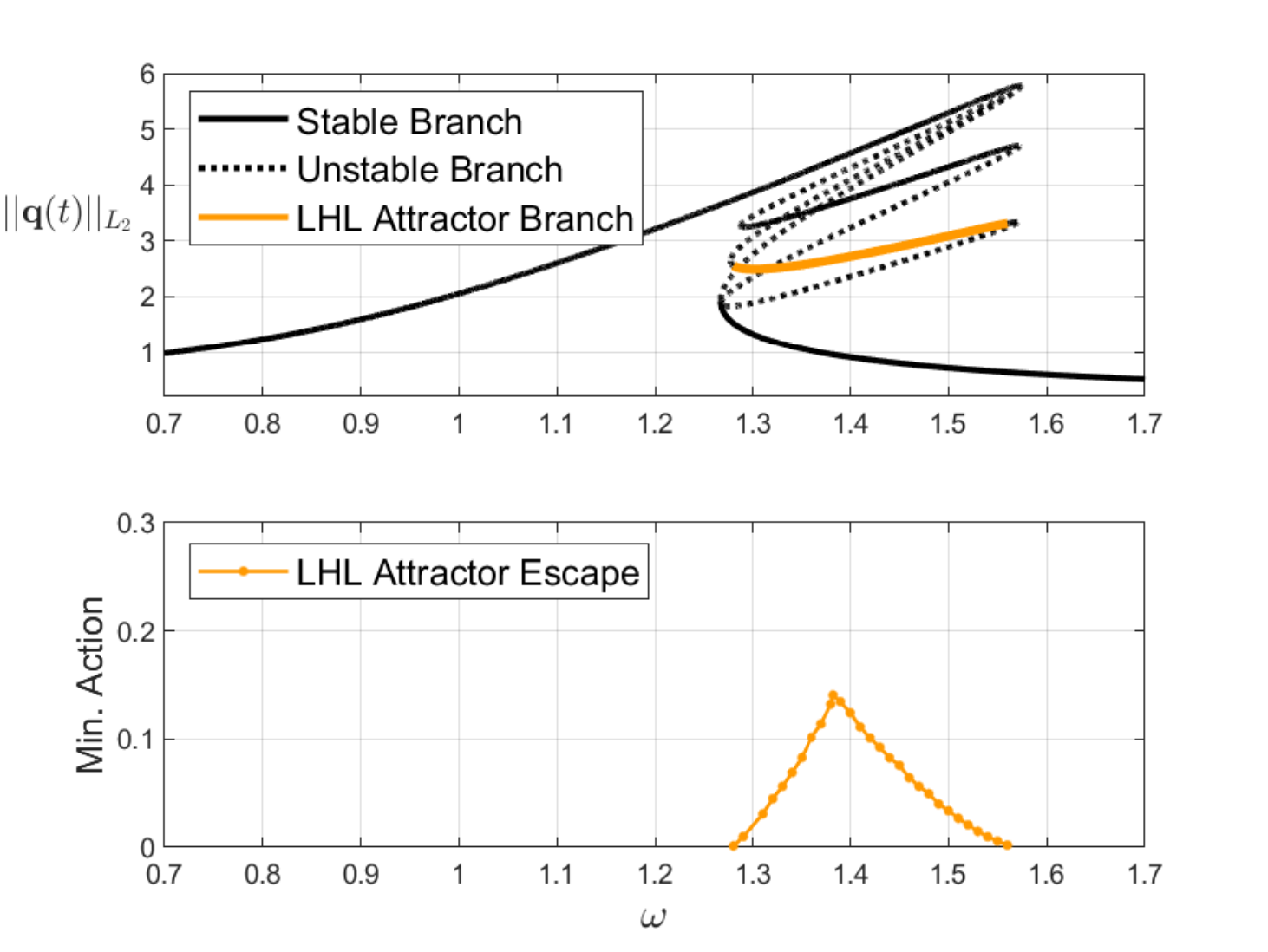}}
\caption{(top) The $L_2$-norm \eqref{L2NormDefinition} of the frequency response of the $N=3$ system with attractor branches highlighted in color in the hysteresis region.  (bottom) The quasipotential corresponding to escapes from the $LHL$ attractor as a function of excitation frequency.}
\label{ThreeDuffing_FrequencyQuasipotential}
\end{center}
\end{figure}

The dependence of the quasipotential barrier for the escape from the $LHL$ attractor on the excitation frequency $\omega$ in the hysteresis region for three oscillators is very similar to the one for the escape from the $HL$ attractor for $N=2$ -- compare Figs.  \ref{ThreeDuffing_FrequencyQuasipotential} and \ref{TwoDuffing_FrequencyQuasipotential}. 

\begin{figure}[htbp]
\begin{center}
\centerline{\includegraphics[width=0.5\textwidth]{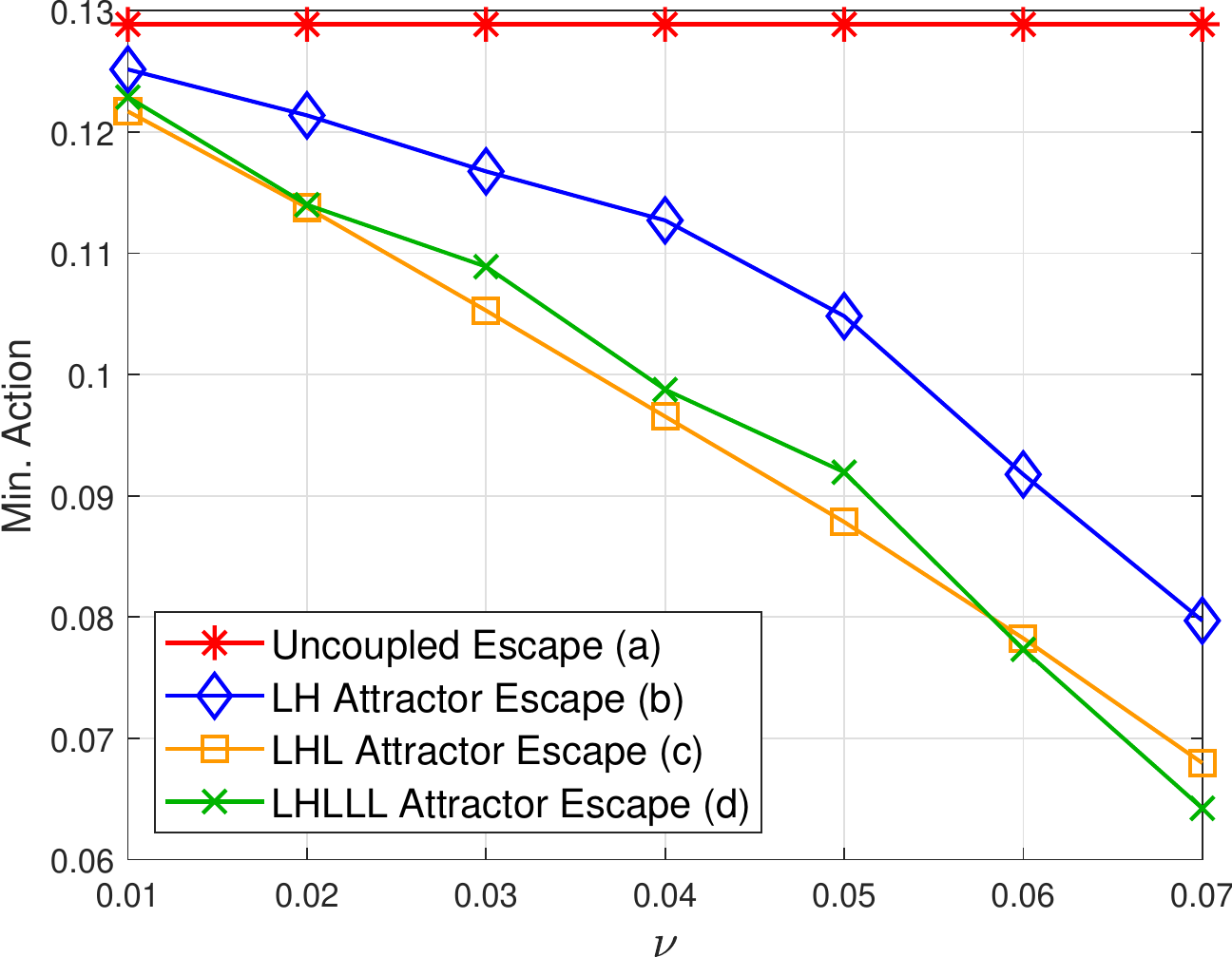}}
\caption{The quasipotential barriers as a function of coupling coefficient $\nu$ that correspond to the following: (a) escapes from the $H$ attractor of the single Duffing oscillator, (b) escapes from the localized mode of the single side coupling ($N=2$) system, and (c) and (d) escapes from the localized mode with a single oscillator being at high amplitude in circular arrays ($N=3$) and ($N=5$), respectively.}
\label{NDuffing_CouplingQuasipotential}
\end{center}
\end{figure}

\begin{table}[t]
\centering
\renewcommand\arraystretch{1.2}
\caption{The quasipotential barriers for the escapes from localized modes at which only one oscillator has high amplitude for $N=1,2,3,5$  and from all modes for $N=2$ at the frequency of $\omega= 1.4$ with $\nu = 0.01$. }
\small
\begin{tabular}{llll}
\\
\hline
$N$ & Initial Attractor & Final Attractor & $S(\varphi)$ \\  \hline
$1$   & $H$ &  $L$ & $0.129$  \\
$2$   & $HL$ & $LL$ & $0.125$  \\ 
$3$   & $LHL$ & $LLL$ & $0.122$  \\
$5$   & $LHLLL$ &  $LLLLL$ & $0.122$  \\ 
\hline{1-4}
$1$   & $L$ & $H$ & $0.180$   \\ 
$2$   & $LL$ & $HL$ & $0.179$ \\ 
\hline{1-4}
$2$   & $HH$ & $HL$ & $0.155$  \\ \hline
\end{tabular}
\label{tab:quasipotentialvalues}
\end{table}

The dependencies of the quasipotential barriers $U_{H}$, $U_{HL}$, $U_{LHL}$, and $U_{LHLLL}$ on the coupling coefficient $\nu$  at $\omega = 1.4$ are plotted in \fref{NDuffing_CouplingQuasipotential}. The barriers for the circular array are lower than those for the single oscillator. In turn, the barriers for $N=3$ and $N=5$ are smaller than for $N=2$. Moreover, the difference between these barriers grows with the increase in the inter-oscillator coupling coefficient $\nu$.  Note that the basins of localized modes get shallower as $\nu$ increases, and as a consequence, it is difficult to find most probable escape paths for $\nu > 0.07$. Furthermore, the localized modes become unstable as $\nu$ reaches $0.08$; hence, localizations do not exist for strong inter-oscillator couplings, which is also noted in prior studies \ccite{balachandran_dynamics_2022}.

Fig. \ref{NDuffing_CouplingQuasipotential} indicates a very important fact that one can predict the qualitative behavior of most probable escape paths and the values of the quasipotential barriers for the escapes from the localized mode where only one oscillator has high amplitude by comparing the results obtained for $N=3$ and $N=5$. It is evident that the differences between the quasipotential barriers $U_{LHL}$ and $U_{LHLLL}$ at the feasible range of $\nu$ are small and comparable with numerical errors. Therefore, for weak inter-oscillator coupling facilitating the existence of response localizations, one can consider the barrier $U_{LHL}$ as an adequate approximation for $U_{LHLLL}$. This suggests that the circular nature of coupling effectively limits the influence of the array on a given oscillator to the influence of its nearest neighbors. As a consequence, the quasipotential barriers for the escapes from the localized modes $LHL\ldots L$ for $N>5$ are close to those for $N=3$ and $N=5$. Meanwhile, the numerical solution is cheaper and more accurate for $N=3$ than for larger $N$.


\section{Discussion} \label{Discussion}
The contribution of this work is two-fold.
\begin{enumerate}
    \item A methodology for computing most probable escape paths and quasipotential barriers from periodic attractors of arrays of nonlinear oscillators with external periodic forcing and small white noise has been developed. 
    \item This methodology has been applied for a case study of a single monostable harmonically driven Duffing oscillator and circular arrays of $N=2$, 3, and 5 oscillators. 
\end{enumerate}
The methodology relies on the large deviations theory \cite{freidlin_random_1998} and Feynman's path integral theory \cite{feynman_1965}, exploits Pontryagin's maximum principle  for the optimal control problems\cite{bardi_optimal_2008} and the Floquet theory for linear ODEs with periodic coefficients \cite{chicone_1999}, and uses the Action Plot Method \cite{beri_solution_2005} as a building block. To the best of the authors' knowledge, this is the first work in which the Action Plot Method was promoted to high dimensions. The main challenges in this promotion are:
\begin{enumerate}
    \item the arising optimization problem for the initial condition for the optimal Hamiltonian path has a discontinuous objective function with multiple local minima located at the bottoms of the ``cliffs";
    \item the final time for a Hamiltonian path is the time when it reaches the appropriate basin boundary, and basin boundaries need to be identified during the optimizations for higher dimensional systems.  
\end{enumerate}
The first challenge has been tackled by the introduction of the Stochastic Unit Vectors (SUV) method, while the second one has been overcome via routine checks in which basin the state of the system is in and refinement of the escape time and location once a basin switch is detected. These techniques have shown themselves to be very robust and suitable for high dimensions. 

The maximal number of coupled oscillators $N=5$ considered in this work is not the limit for the proposed methodology. However, as the number of oscillators $N$ increases, the runtime of the optimization routine grows as $O(N^2)$ (see Appendix A). Furthermore, the behavior of the green curve in Fig. \ref{NDuffing_CouplingQuasipotential} corresponding to $N=5$ suggests that the found MPEP and the corresponding quasipotential barrier become less accurate. Therefore, in order to apply the proposed methodology to larger oscillator arrays it is desirable to make the optimization method more efficient. 

The case study of circular oscillator arrays has shown an important trend: the predictions for the most probable escape paths and quasipotential barriers for response localizations in large arrays can be made based on the results for small arrays. In particular, it has been shown that the results for $N=5$ are well-approximated by those for $N=3$. This is consistent with the results obtained by means of reduced-order modeling in Ref. \cite{balachandran_dynamics_2022}.

The methodology described here has been generalized for different stiffness models. In this case study, the authors chose $K(\bm{x}) = \alpha \bm{x} +\beta \bm{x}^3 +\nu D_N \bm{x}$, but $K(\bm{x}) $ could contain higher order nonlinear stiffness terms,  higher order coupling terms, or inhomogeneous stiffness models, for example. In particular, quantifying quasipotentials for escapes from response localizations subjected to noise in systems that have both cubic stiffness and cubic coupling could be a direction of future work. It is known that such systems exhibit two types of response localizations known as  s-t mode and p-mode, and that the two types of localized modes change stability depending on the cubic stiffness coefficient and the cubic coupling coefficient. Quantifying the quasipotential for those systems would reveal how robust the two types of localized modes are to noise. 

The methodology developed in this work has an important limitation: the noise term in SDE \eqref{eq:mainSDE} is the standard Brownian motion. On each fixed interval $\Delta t$, the increment of the Brownian motion is $N$-dimensional Gaussian random variable with mean 0 and the covariance matrix $I_{N\times N}$, which is unbounded. In physical experiments, only bounded amplitude and bounded bandwidth noise can be realized. Bounded amplitude can be modeled by truncating and renormalizing increments of Brownian motion. This will result in rendering all Hamiltonian paths for which $\|\bm{p}_{\bm{v}}\|$ exceeds the maximum allowed noise magnitude infeasible, i.e., the corresponding action functional value is $+\infty$. Modifying the action functional for noise of limited bandwidth is less straightforward. The extension for more realistic noise models is left for the future.


\section{\label{data availability}Data Availability Statement}

All data used in this work were generated within this work. The authors' codes are available upon request.

\section*{Acknowledgments}
The authors thank Professor Sandra Cerrai for a valuable discussion on the large deviations theory.
This work was partially supported by the Clark Doctoral Fellowship (LC), NSF CAREER Grant DMS-1554907 (MC), AFOSR MURI grant FA9550-20-1-0397 (MC), and NSF grant CMMI-1760366 and associated data science supplements (BB). A part of this research project was conducted by using computational resources available at the Maryland Advanced Research Computing Center (MARCC).

\section{Appendix A: Notes on Computational Cost and Nomenclature} \label{Appendix}

\begin{table*}[b]
\caption{Parameters and computational cost}
\centering
\begin{tabular}{p{0.3\linewidth}p{0.1\linewidth}p{0.1\linewidth}p{0.1\linewidth}p{0.1\linewidth}}
\hline
Description & Symbol  & Example 1 & Example 2 & Example 3 \\ \hline
Number of oscillators   & $N$ & 1 & 2 & 5  \\
Excitation Frequency   &  $\omega$ & 1.4 & 1.4 & 1.4 \\
Coupling Coefficient   &  $\nu$  & - & 0.01 & 0.01 \\ 
Dimension of cost function  &$d$  & 2 & 4 & 10\\
Sphere Magnitude   &  $\gamma$   & $10^{-20}$ & $10^{-15}$ & $10^{-15}$\\
Basin check interval   &  $dT$   & $\pi/\omega$ & $\pi/\omega$ & $\pi/\omega$\\
Min. Escape Radius   & $r_{A_0}$   & 0.1 & 0.2 & 0.5\\
Min. Attracted Radius   & $r_{A_k}$   & 0.1 & 0.2 & 0.5\\
Number of initial samples of cost space   & $n_{init}$   & 300 & 1000 & 1000\\ 
Number of parallel optimizations  &  $n_{s}$   & 10 & 15 & 15\\
Initial SUV step size  &  $\sigma$   & 0.25 & 0.25 & 0.25\\
Maximum optimization iterations  &  $i_{m}$   & 35 & 35 & 35\\ 
Min. number of optimizations to keep  & $n_{m}$   & 5 & 5 & 5\\ 
Number of available processors  &  $c$  & 24 & 24 & 24 \\
\hline{1-5}
Average Processor Efficiency ($\%$)  & $e$  & $54\%$ & $69\%$ & $81\%$ \\ 
Real time cost (seconds)  & $t$  & 160 & 365 & 4942\\ 
\hline
\label{tab:ComputationalCostTable}
\end{tabular}
\end{table*}

This appendix includes an estimate of the computational cost of the optimization algorithm for finding MPEPs and three examples of parameter choices. The purpose of this appendix is to facilitate the reproduction of the results of this work by a third party with reasonable expectations of computational cost. 

To save computational cost, a $2N-1$-dimensional sphere of radius  $r_{A_0}$ embedded in $\mathbb{R}^{2N}$ that surrounds the initial attractor $\bm{A}_0(\theta)$ at a fixed moment of time, $\theta$, is introduced. Here, $\bm{A}_0(\theta)$ describes the $(\bm{x},\bm{v})$ coordinates of the attractor at time $\theta$. The radius $r_{A_0}$ is chosen so that if the path is still within the sphere, the path has not left the current basin yet. Similarly, $r_{A_k}$, is the radius of a $2N-1$-dimensional sphere that surrounds attractor $A_k$. When computing the domains of attraction online, it is assumed that if a trajectory enters the sphere surrounding  attractor, $A_k$ than the trajectory is in the basin of attractor $k$. 

Table \ref{tab:ComputationalCostTable} is a list of the definitions of many of the parameters relevant to the oscillator arrays studied in this work and to the proposed numerical algorithms. Omitted parameters are of lesser influence on the computational cost. $t_i$ is the cost of the initial sample set of the cost function space and $t_o$ is time cost of the parallel optimizations using the SUV algorithm. The order of $t_o$ is loosely bounded by an inequality that depends on the number of optimizations running because the algorithm is allowed to throw out high cost, poor improvement descents while preserving useful descents. This is a gradient free optimization scheme on a cost function with many local minima and discontinuities; finding a global minima is not guaranteed. 

One could estimate the influence of the parameters on the computational costs as 

\begin{equation}
\begin{split}
     O(t_i) &\approx O\left(  \frac{d \times n_{init}}{c\times e \times dT} \right) \\ 
     O\left(  \frac{2 d^2 \times n_m }{c\times e \times dT} \right) &\leq O(t_o) \leq O\left(  \frac{2 d^2 \times n_{s} \times i_m}{c\times e \times dT} \right) \\ 
      t &\approx t_i + t_o
\end{split}
\label{CompCost}
\end{equation}

The principle idea of \eqref{CompCost} is that the computational cost of evaluating the cost function scales with $d \times n/dT$ and that the cost of optimization using SUV scales the number of cost function evaluations by a multiple of $2d$.  With efficient implementations at the excitation frequency of $\omega=1.4$, finding a solution to the $N=5$ system was a 10 dimensional optimization problem that took $80$ minutes to solve with 24 processors in MATLAB. Additional examples are shown in Table \ref{tab:ComputationalCostTable}. These results were produced on the Maryland Advanced Research and Computing Center's (MARCC) Intel Haswell dual socket,  12-core processors, 2.5GHz, 30MB cache, 128GB RAM node. 

The important nomenclature in this work is summarized in Table \ref{tab:NotationTable}. This nomenclature is also described in the work as it is used.

\begin{table*}[b]
\tiny
\caption{Nomenclature}
\centering
\begin{tabular}{p{0.1\linewidth}p{0.12\linewidth}p{0.6\linewidth}}
\hline
 Symbol  & Type & Description \\ \hline
 $N$   & $\mathbb{W}$ & Number of oscillators \\
  $t$   & $\mathbb{R}$ & Time \\
$\bm{x}$   & $\mathbb{R}^N$ & Beam tip displacements \\
$\bm{v}$   & $\mathbb{R}^N$ & Beam tip velocities \\
$\bm{f}(t)$   & $\mathbb{R}^N$ & Deterministic periodic external excitation \\
$\bm{\eta}(t)$   & $\mathbb{R}^N$ & Stochastic external excitation \\
$\delta_s$   & $\mathbb{R}$ & Damping coefficient \\
$\alpha$   & $\mathbb{R}$ & Linear stiffness coefficient \\
$\beta$   & $\mathbb{R}$ & Cubic stiffness coefficient \\
$\nu$   & $\mathbb{R}$ & Linear coupling coefficient \\
$\epsilon$   & $\mathbb{R}$ & Noise intensity/multiplier \\
$\omega$   & $\mathbb{R}$ & Periodic excitation frequency \\
$T$   & $\mathbb{R}$ & Periodic excitation period \\
$\bm{D}_N $  & $\mathbb{R}^{N\times N}$ & Linear Coupling Matrix \\
$\bm{q} $  & $\mathbb{R}^{2N}$ & Vector of displacement states and velocity states $(\bm{x},\bm{v})$ \\
$\bm{p} $  & $\mathbb{R}^{2N}$ & Vector of costates $(\bm{p_x},\bm{p_v})$ \\
$\bm{u} $  & $\mathbb{R}^{N}$ & Vector of deterministic control inputs \\
$||\bm{q}||_{L_2}$  & $\mathbb{R}$ & $L_2$-like norm \\
$\bm{b}(\bm{q})$  & $\mathbb{R}^{2N}$ & Drift vector\\
$\bm{Q}(\bm{q})$  & $\mathbb{R}^{2N\times 2N}$ & Diffusion matrix\\
${S}_\tau(\bm{\varphi})$  & $\mathbb{R}$ & Action functional\\
$\bm{\varphi}$  & $\mathbb{R}^{2N}$ & Candidate escape path\\
$\tau$  & $\mathbb{R}$ & Travel time, final time boundary of $\varphi$, escape time\\
$\mathcal{A}$  & char & Attractor \\
$B$  & char & Basin of attraction \\
$\partial B$  & char & Basin boundary \\
$D$  & char & Domain \\
$\partial D$  & char & Domain boundary \\
$\mathbb{S}_T$  & $\mathbb{S}^1$ & A circle with circumference $T$ \\
$\bm{K}(\bm{x})$  & $\mathbb{R}^N$ & Generalized restoring force function of $\bm{x}$ \\
$\theta$  & $\mathbb{S}_T$ & A periodic time coordinate  \\
$\hat{\bm{q}}$  & $\mathbb{R}^{2N}\times \mathbb{S}_T$ & Augmented state vector for autonomous system, $(\bm{x},\bm{v},\theta)$   \\
$\hat{\bm{p}}$  & $\mathbb{R}^{2N+1}$ & Augmented costate vector for autonomous system, $(\bm{p_x},\bm{p_v}, p_\theta)$   \\
$n_a$  & $\mathbb{W}$ & Number of attractors \\
$k$  & $\{0,1,...,n_a\}$ & Index of $n_a$ attractors \\
$k = 0$  &  & Index of the initial attractor \\
$A_k$  & char & The attractor with index k\\
$B_k$  & char & Basin of attraction of $A_k$ \\
$\partial B_k$  & char & Basin boundary of $A_k$\\
$\partial B_{kl}$  & char & Intersection of Basin boundary of $A_k$ and $A_l$\\
$\bm{A}_k(\theta)$  & $\mathbb{R}^{2N}$ & The location in coordinates $(\bm{x}(\theta),\bm{v}(\theta))$ of attractor with index $k$ at time $\theta$  \\
$\bm{A}_0$  & $\mathbb{R}^{2N}$ & The location of the initial attractor\\
$\hat{\bm{A}}_k$  & $\mathbb{R}^{2N}\times\mathbb{S}_T$ & The location in coordinates $(\bm{x},\bm{v},\theta)$ of attractor with index $k$   \\
$\bar{\bm{A}}_k$  & $\mathbb{R}^{4N+1}\times\mathbb{S}_T$ & Augmented $\bar{\bm{A}}_k$ in Hamiltonian coordinates, $(\bm{x},\bm{v},\theta,\bm{0},\bm{0},0)$ \\
$\tilde{\bm{A}}_k(\theta)$  & $\mathbb{R}^{4N}$ & Augmented $\bm{A}_k(\theta)$ in Hamiltonian coords. w/o explicit $\theta$ and $p_\theta$, $(\bm{x}(\theta),\bm{v}(\theta),\bm{0},\bm{0})$ \\
$J$  & $\mathbb{R}^{4N\times4N}$ & Jacobian of Hamiltonian system \\
$\bm{J_K}$  & $\mathbb{R}^{N\times N}$ & Jacobian of vector field $\bm{K}(\bm{x})$ \\
$\bm{z}$  & $\mathbb{R}^{4N}$ & Coordinates of linearized system \\
$\bm{\Psi}(t)$  & $\mathbb{R}^{4N \times 4N}$ & Fundamental solution matrix \\
$\bm{\Phi}(t)^s_{\theta_0}$  & $\mathbb{R}^{4N \times 2N}$ & Stable eigenvectors of Monodromy matrix \\
$\bm{\Phi}(t)^u_{\theta_0}$  & $\mathbb{R}^{4N \times 2N}$ & Untable eigenvectors of Monodromy matrix \\
$\gamma$  & $\mathbb{R}$ & Small radius of a sphere that surrounds the initial attractor at fixed time \\
$\mathbb{S}_\gamma^{2N-1}$  & $\mathbb{S}^{2N-1}$ & A 2N sphere with radius $\gamma$  \\
$\bm{q}_\gamma$  & $\mathbb{S}_\gamma^{2N-1}$ & A value of $\bm{q}$ with euclidean norm equal to $\gamma$  \\
$dT$  & $\mathbb{R}$ & The amount of time to numerically map the path forward in time prior to checking if the path has entered a new basin of attraction  \\
$L$  & char & "Low amplitude" \\
$H$  & char & "High amplitude" \\
\hline
\label{tab:NotationTable}
\end{tabular}
\end{table*}

\section*{References}

\begin{thebibliography}{10}

\bibitem{duffing_erzwungene_1918}
G.~Duffing, ``Erzwungene schwingungen bei veränderlicher eigenfrequenz,'' {\em
  Vieweg u. Sohn, Braunschweig}, vol.~7, 1918.

\bibitem{papangelo_multistability_2019}
A.~Papangelo, F.~Fontanela, A.~Grolet, M.~Ciavarella, and N.~Hoffmann,
  ``Multistability and localization in forced cyclic symmetric structures
  modelled by weakly-coupled {Duffing} oscillators,'' {\em Journal of Sound and
  Vibration}, vol.~440, pp.~202--211, Feb. 2019.

\bibitem{grolet_free_2012}
A.~Grolet and F.~Thouverez, ``Free and forced vibration analysis of a nonlinear
  system with cyclic symmetry: {Application} to a simplified model,'' {\em
  Journal of Sound and Vibration}, vol.~331, pp.~2911--2928, June 2012.

\bibitem{dick_intrinsic_2008}
A.~Dick, B.~Balachandran, and c.~Mote, ``Intrinsic localized modes in
  microresonator arrays and their relationship to nonlinear vibration modes,''
  {\em Nonlinear Dynamics}, vol.~54, pp.~13--29, Jan. 2008.

\bibitem{ikeda_intrinsic_2013}
T.~Ikeda, Y.~Harata, and K.~Nishimura, ``Intrinsic localized modes of harmonic
  oscillations in nonlinear oscillator arrays,'' {\em Journal of Computational
  and Nonlinear Dynamics}, vol.~8, no.~4, 2013.
\newblock Publisher: American Society of Mechanical Engineers Digital
  Collection.

\bibitem{balachandran_response_2015}
B.~Balachandran, E.~Perkins, and T.~Fitzgerald, ``Response localization in
  micro-scale oscillator arrays: influence of cubic coupling nonlinearities,''
  {\em International Journal of Dynamics and Control}, vol.~3, no.~2,
  pp.~183--188, 2015.
\newblock Publisher: Springer.

\bibitem{doedel_auto-07p_2007}
E.~J. Doedel, T.~F. Fairgrieve, B.~Sandstede, A.~R. Champneys, Y.~A. Kuznetsov,
  and X.~Wang, ``{AUTO}-{07P}: {Continuation} and bifurcation software for
  ordinary differential equations,'' tech. rep., 2007.

\bibitem{emad_experimental_2000}
J.~Emad, A.~F. Vakakis, and N.~Miller, ``Experimental nonlinear localization in
  a periodically forced repetitive system of coupled magnetoelastic beams,''
  {\em Physica D: Nonlinear Phenomena}, vol.~137, pp.~192--201, Mar. 2000.

\bibitem{kimura_coupled_2009}
M.~Kimura and T.~Hikihara, ``Coupled cantilever array with tunable on-site
  nonlinearity and observation of localized oscillations,'' {\em Physics
  Letters A}, vol.~373, pp.~1257--1260, Mar. 2009.

\bibitem{kimura_experimental_2012}
M.~Kimura and T.~Hikihara, ``Experimental manipulation of intrinsic localized
  modes in macro-mechanical system,'' {\em Nonlinear Theory and Its
  Applications, IEICE}, vol.~3, no.~2, pp.~233--245, 2012.

\bibitem{perkins_effects_2016}
E.~Perkins, M.~Kimura, T.~Hikihara, and B.~Balachandran, ``Effects of noise on
  symmetric intrinsic localized modes,'' {\em Nonlinear Dynamics}, vol.~85,
  no.~1, pp.~333--341, 2016.
\newblock ISBN: 0924-090X Publisher: Springer.

\bibitem{niedergesas_experimental_2021}
B.~Niedergesäß, A.~Papangelo, A.~Grolet, A.~Vizzaccaro, F.~Fontanela,
  L.~Salles, A.~Sievers, and N.~Hoffmann, ``Experimental observations of
  nonlinear vibration localization in a cyclic chain of weakly coupled
  nonlinear oscillators,'' {\em Journal of Sound and Vibration}, vol.~497,
  p.~115952, 2021.
\newblock Publisher: Elsevier.

\bibitem{kenyon_forced_2002}
J.~A. Kenyon and J.~H. Griffin, ``Forced {Response} of {Turbine} {Engine}
  {Bladed} {Disks} and {Sensitivity} to {Harmonic} {Mistuning},'' {\em Journal
  of Engineering for Gas Turbines and Power}, vol.~125, pp.~113--120, Dec.
  2002.

\bibitem{srinivasan_flutter_1997}
A.~V. Srinivasan, ``Flutter and {Resonant} {Vibration} {Characteristics} of
  {Engine} {Blades},'' {\em Journal of Engineering for Gas Turbines and Power},
  vol.~119, pp.~742--775, Oct. 1997.

\bibitem{castanier_modeling_2006}
M.~P. Castanier and C.~Pierre, ``Modeling and {Analysis} of {Mistuned} {Bladed}
  {Disk} {Vibration}: {Current} {Status} and {Emerging} {Directions},'' {\em
  Journal of Propulsion and Power}, vol.~22, pp.~384--396, Mar. 2006.

\bibitem{sever_experimental_2004}
I.~A. Sever, {\em Experimental validation of turbomachinery blade vibration
  predictions}.
\newblock PhD thesis, Imperial College London (University of London), 2004.

\bibitem{gardonio_vibration_2014}
P.~Gardonio and M.~Zilletti, ``Vibration energy harvesting based on mode
  localization,'' {\em ISMA 2014)(Katholieke Universiteit Leuven, Belgium,
  15–17 September 2014)}, 2014.

\bibitem{lefeuvre_comparison_2006}
E.~Lefeuvre, A.~Badel, C.~Richard, L.~Petit, and D.~Guyomar, ``A comparison
  between several vibration-powered piezoelectric generators for standalone
  systems,'' {\em Sensors and Actuators A: Physical}, vol.~126, pp.~405--416,
  Feb. 2006.

\bibitem{ewins_effects_1969}
D.~J. Ewins, ``The effects of detuning upon the forced vibrations of bladed
  disks,'' {\em Journal of Sound and Vibration}, vol.~9, no.~1, pp.~65--79,
  1969.
\newblock Publisher: Elsevier.

\bibitem{yan_vibration_2008}
Y.~Yan, P.~Cui, and H.~Hao, ``Vibration mechanism of a mistuned bladed-disk,''
  {\em Journal of sound and vibration}, vol.~317, no.~1-2, pp.~294--307, 2008.
\newblock Publisher: Elsevier.

\bibitem{freidlin_random_1998}
M.~I. Freidlin and A.~D. Wentzell, ``Random perturbations,'' in {\em Random
  perturbations of dynamical systems}, pp.~15--43, Springer, 1998.

\bibitem{bardi_optimal_2008}
M.~Bardi and I.~Capuzzo-Dolcetta, {\em Optimal control and viscosity solutions
  of {Hamilton}-{Jacobi}-{Bellman} equations}.
\newblock Springer Science \& Business Media, 2008.

\bibitem{beri_solution_2005}
S.~Beri, R.~Mannella, D.~G. Luchinsky, A.~Silchenko, and P.~V. McClintock,
  ``Solution of the boundary value problem for optimal escape in continuous
  stochastic systems and maps,'' {\em Physical Review E}, vol.~72, no.~3,
  p.~036131, 2005.
\newblock Publisher: APS.

\bibitem{lin_quasi-potential_2019}
L.~Lin, H.~Yu, and X.~Zhou, ``Quasi-potential calculation and minimum action
  method for limit cycle,'' {\em Journal of Nonlinear Science}, vol.~29, no.~3,
  pp.~961--991, 2019.
\newblock Publisher: Springer.

\bibitem{chicone_1999}
C.~Chicone, {\em Ordinary Differential Equations with Applications}.
\newblock Springer-Verlag, New York, 1999.

\bibitem{chen_noise_2016}
Z.~Chen, Y.~Li, and X.~Liu, ``Noise induced escape from a nonhyperbolic chaotic
  attractor of a periodically driven nonlinear oscillator,'' {\em Chaos: An
  Interdisciplinary Journal of Nonlinear Science}, vol.~26, no.~6, p.~063112,
  2016.
\newblock Publisher: AIP Publishing LLC.

\bibitem{von_wagner_calculation_2000}
U.~von Wagner and W.~V. Wedig, ``On the {Calculation} of {Stationary}
  {Solutions} of {Multi}-{Dimensional} {Fokker}–{Planck} {Equations} by
  {Orthogonal} {Functions},'' {\em Nonlinear Dynamics}, vol.~21, pp.~289--306,
  Mar. 2000.

\bibitem{von_wagner_double_2002}
U.~Von~Wagner, ``On double crater-like probability density functions of a
  duffing oscillator subjected to harmonic and stochastic excitation,'' {\em
  Nonlinear Dynamics}, vol.~28, no.~3-4, pp.~343--355, 2002.

\bibitem{martens_calculation_2012}
W.~Martens, U.~von Wagner, and V.~Mehrmann, ``Calculation of high-dimensional
  probability density functions of stochastically excited nonlinear mechanical
  systems,'' {\em Nonlinear Dynamics}, vol.~67, pp.~2089--2099, Feb. 2012.

\bibitem{forster_approximate_2018}
A.~Förster, L.~Panning-von Scheidt, and J.~Wallaschek, ``Approximate
  {Solution} of the {Fokker}–{Planck} {Equation} for a {Multidegree} of
  {Freedom} {Frictionally} {Damped} {Bladed} {Disk} {Under} {Random}
  {Excitation},'' {\em Journal of Engineering for Gas Turbines and Power},
  vol.~141, Sept. 2018.

\bibitem{dykman_optimal_1992}
M.~I. Dykman, P.~V.~E. McClintock, V.~N. Smelyanski, N.~D. Stein, and N.~G.
  Stocks, ``Optimal paths and the prehistory problem for large fluctuations in
  noise-driven systems,'' {\em Phys. Rev. Lett.}, vol.~68, pp.~2718--2721, May
  1992.
\newblock Publisher: American Physical Society.

\bibitem{luchinsky_optimal_2002}
D.~G. Luchinsky, S.~Beri, R.~Mannella, P.~V. McClintock, and I.~Khovanov,
  ``Optimal fluctuations and the control of chaos,'' {\em International Journal
  of Bifurcation and Chaos}, vol.~12, no.~03, pp.~583--604, 2002.
\newblock Publisher: World Scientific.

\bibitem{ramakrishnan_energy_2010}
S.~Ramakrishnan and B.~Balachandran, ``Energy localization and white
  noise-induced enhancement of response in a micro-scale oscillator array,''
  {\em Nonlinear dynamics}, vol.~62, no.~1, pp.~1--16, 2010.
\newblock Publisher: Springer.

\bibitem{perkins_noise-enhanced_2012}
E.~Perkins and B.~Balachandran, ``Noise-enhanced response of nonlinear
  oscillators,'' {\em Procedia Iutam}, vol.~5, pp.~59--68, 2012.
\newblock Publisher: Elsevier.

\bibitem{perkins_noise-influenced_2013}
E.~Perkins, C.~Chabalko, and B.~Balachandran, ``Noise-influenced transient
  energy localization in an oscillator array,'' {\em Nonlinear Theory and Its
  Applications, IEICE}, vol.~4, no.~3, pp.~232--243, 2013.
\newblock ISBN: 2185-4106 Publisher: The Institute of Electronics, Information
  and Communication Engineers.

\bibitem{xu_global_2003}
W.~Xu, Q.~He, T.~Fang, and H.~Rong, ``Global analysis of stochastic bifurcation
  in {Duffing} system,'' {\em International Journal of Bifurcation and Chaos},
  vol.~13, no.~10, pp.~3115--3123, 2003.
\newblock Publisher: World Scientific.

\bibitem{xu_stochastic_2004}
W.~Xu, Q.~He, T.~Fang, and H.~Rong, ``Stochastic bifurcation in {Duffing}
  system subject to harmonic excitation and in presence of random noise,'' {\em
  International Journal of Non-Linear Mechanics}, vol.~39, no.~9,
  pp.~1473--1479, 2004.
\newblock ISBN: 0020-7462 Publisher: Elsevier.

\bibitem{yu_numerical_2004}
J.~S. Yu and Y.~K. Lin, ``Numerical path integration of a non-homogeneous
  {Markov} process,'' {\em International Journal of Non-Linear Mechanics},
  vol.~39, no.~9, pp.~1493--1500, 2004.

\bibitem{kumar_modified_2010}
P.~Kumar and S.~Narayanan, ``Modified path integral solution of
  {Fokker}–{Planck} equation: response and bifurcation of nonlinear
  systems,'' {\em Journal of computational and nonlinear dynamics}, vol.~5,
  no.~1, 2010.

\bibitem{narayanan_numerical_2012}
S.~Narayanan and P.~Kumar, ``Numerical solutions of {Fokker}–{Planck}
  equation of nonlinear systems subjected to random and harmonic excitations,''
  {\em Probabilistic Engineering Mechanics}, vol.~27, no.~1, pp.~35--46, 2012.
\newblock ISBN: 0266-8920 Publisher: Elsevier.

\bibitem{cilenti_transient_2021}
L.~Cilenti and B.~Balachandran, ``Transient probability in basins of noise
  influenced responses of mono and coupled {Duffing} oscillators,'' {\em Chaos:
  An Interdisciplinary Journal of Nonlinear Science}, vol.~31, p.~063117, June
  2021.
\newblock Publisher: American Institute of Physics.

\bibitem{balachandran_dynamics_2022}
B.~Balachandran, T.~Breunung, G.~D. Acar, A.~Alofi, and J.~A. Yorke, ``Dynamics
  of circular oscillator arrays subjected to noise,'' {\em Nonlinear Dynamics},
  pp.~1--14, 2022.
\newblock Publisher: Springer.

\bibitem{agarwal_influence_2018}
V.~Agarwal, X.~Zheng, and B.~Balachandran, ``Influence of noise on frequency
  responses of softening {Duffing} oscillators,'' {\em Physics Letters A},
  vol.~382, no.~46, pp.~3355--3364, 2018.
\newblock ISBN: 0375-9601 Publisher: Elsevier.

\bibitem{alofi_coupled_2021}
A.~M. Alofi, {\em Coupled {Oscillator} {Arrays}: {Dynamics} and {Influence} of
  {Noise}}.
\newblock {PhD} {Thesis}, 2021.

\bibitem{chao_tao_2021}
Y.~Chao and M.~Tao, ``Parametric resonance for enhancing the rate of metastable
  transition.''.

\bibitem{dembo_1998}
A.~Dembo and O.~Zeitouni, {\em Large Deviations Techniques and Applications}.
\newblock Springer, New York, 1998.

\bibitem{chen_smoluchowskikramers_2005}
Z.~Chen and M.~Freidlin, ``Smoluchowski–{Kramers} approximation and exit
  problems,'' {\em Stochastics and Dynamics}, vol.~5, no.~04, pp.~569--585,
  2005.
\newblock Publisher: World Scientific.

\bibitem{freidlin_2004}
M.~Freidlin, ``Some remarks on the smoluchowski–kramers approximation,'' {\em
  Journal of Statistical Physics}, vol.~117, no.~3/4, pp.~617--634, 2004.

\bibitem{dykman_1979}
M.~I. Dykman and M.~A. Krivoglaz, ``Theory of fluctuational transitions between
  stable states of a nonlinear oscillator,'' {\em Sov. Phys. JETP}, vol.~50,
  no.~1, pp.~30--37, 1979.

\bibitem{feynman_1965}
R.~P. Feynman and A.~R. Hibbs, {\em Quantum mechanics and path integrals}.
\newblock McGraw-Hill, New York, 1965.

\bibitem{kirk_optimal_2004}
D.~E. Kirk, {\em Optimal control theory: an introduction}.
\newblock Courier Corporation, 2004.

\bibitem{arnold_1978}
V.~I. Arnold, {\em Mathematical Methods of Classical Mechanics}.
\newblock Springer-Verlag, New York, 1978.

\end{thebibliography}
\bibliographystyle{ieeetr}

\end{document}